\documentstyle[11pt]{article}
\title{	The Effects of Flavor Changing Neutral Current \\
	and \\
	Cabibbo-Kobayashi-Maskawa Matrix\footnote[2]{
	This paper is Master Thesis at Faculty of Science,
	Hiroshima University.}}
\author{L.T. Handoko\thanks{
	E-mail address : handoko@theo.phys.sci.hiroshima-u.ac.jp}}
\date{	\it Department of Physics, Hiroshima University \\
	1-3-1 Kagamiyama, Higashi Hiroshima - 724, Japan}
\input{epsf}

\begin{document}

\setlength{\baselineskip}{24pt}

\maketitle

\setlength{\baselineskip}{8mm}

\thispagestyle{empty}
\begin{abstract}
	We study the effects of tree-level flavor changing neutral
	current and its present status with Cabibbo-Kobayashi-Maskawa matrix. 
	Especially, we remark the effects of flavor changing neutral 
	current on the unitary triangle of Cabibbo-Kobayashi-Maskawa matrix.
	In the case when the unitarity is nearly conserved, 
	we find that the allowed FCNC in the $db$ sector is 
	$\left| {{z_d}^{db}}/{{V_{CKM}}^{td}
	{{V_{CKM}}^{tb}}^\ast} \right| \sim O({10}^{-2})$,    
	from the evaluation of the present bounds for 
	the mixing size of neutral $B$ meson ($x_d$), and the
	$CP$ violation parameter in the neutral $K$ meson ($\epsilon$).	
	Although the size is small, the new physics due to the
	tree-level FCNC in the $db$ sector is expected to be
	visible in the $B \rightarrow X_d \, l^+ \, l^- $ 
	and the $CP$ asymmetries in the neutral $B$ meson system.
	In the case when the unitarity is violated and the 
	quadrangle is conserved, the FCNC is large enough, i.e. 
	$\left| {{z_d}^{db}}/{{V_{CKM}}^{td}
	{{V_{CKM}}^{tb}}^\ast} \right| \sim O(1)$. 
	Then, a significant contribution is also expected in the 
	inclusive $B \rightarrow X_d \, \gamma$ decay.	
	As a typical model with this charateristic, we consider 
	Vetor-like Quark Model, where the vector-like 
	quarks are added into the three standard generations of quarks. 
	An overview of the flavor physics in the SM is also given. 
\end{abstract}

\clearpage
\tableofcontents

\clearpage
\section{INTRODUCTION}

In the present, within the Standard Model (henceforth SM)\footnote{
Throughout this paper, the phrase Standard Model (SM) point to
{\it Minimal Electroweak} Standard Model. {\it Minimal} means 
that the Higgs sector contains only one Higgs doublet.} 
the most of present high energy physics experiments can be
explained relatively well. However, many phenomenas are
still unexplained, e.g. 
fermion mass problem, $CP$ violation, strong {\it CP} problem,
Electric Dipole Moment and so on.
In order to explain them, many models beyond the SM are proposed (refs.
\cite{nelson}, \cite{barr} and \cite{branco}).
Especially, in the sense of the existence of tree-level
flavor changing neutral current
(henceforth called FCNC), one can divide them into two types, i.e.
\begin{enumerate}
  	\item Models with tree-level FCNC
		(Vector-like Quark Model, etc.). 
	\item Models without tree-level FCNC (some Two Higgs Doublet
		Model, etc.).
\end{enumerate}
In the SM, FCNC occurs only in the one loop and higher
orders, but does not emerge in the tree-level.

The existence of FCNC's in the SM have consequences in the
rare decays of mesons, e.g. $K$ and $B$ mesons. As will be
reviewed shortly in the first section, these rare decays have an important 
role to investigate the Cabibbo-Kobayashi-Maskawa matrix 
(henceforth called CKM matrix).
The study of CKM matrix will lead us to have a better understanding
about Yukawa sector in the SM, as well as new physics
in the beyond SM. 
So in another word one can say that through study of the rare
meson decays, we may learn some clues about Yukawa-sector of the electroweak 
theory which has not been understood well. In this meaning, it is important 
to test the various models beyond the SM.

In this paper, we study the effects of FCNC's which occur in the tree-level
by employing Vector-like Quark model (henceforth called VQM). This
model is a possible minimum extension of the SM. 
In addition to the three standard generations of quarks, 
the vector-like quarks are intrduced (see for example 
ref. \cite{branco}.
In this model, there are two types of tree-level FCNC's, i.e.,
\begin{enumerate}
  	\item Tree-level FCNC's among the ordinary quarks and
	\item Tree-level FCNC's between the ordinary and the vector-like quarks.
\end{enumerate}
They can be tested separately as discussed in refs. \cite{handoko2},
\cite{handoko3} and references therein.
By applying this model in the FCNC processes and rare decays, we investigate
the effects of these FCNC's especially on the triangle of
CKM matrix that appears from the unitarity of CKM matrix.
In this paper we concentrate on the $db$ sector of
CKM matrix. In this paper, some neutral $K$ and $B$ mesons 
processes will be evaluated.

This paper is organized as follows. First, we make an overview of the
SM and the induced flavor physics phenomenas, that is, unitarity of
CKM matrix and the experimental bounds for the CKM matrix elements from
{\it CP} violation and rare decays in $K$ and $B$ mesons
system. According to the processes, we discuss some
experimental constraints on the unitarity
of CKM matrix. Next, we give a brief introduction of the
VQM. We show that in the model tree-level FCNC's are
appearing, and the sizes of FCNC's are indicated as
$z^{\alpha \beta}$. The upper-bounds for $z^{\alpha \beta}$ will 
be found from the $K$ and $B$ mesons processes, under an
assumption that the tree-level FCNC's are dominant.
Further we evaluate the constraints for two cases,
\begin{enumerate}
	\item Constraint when the unitarity is nearly 
		conserved. 
	\item Constraint when the unitarity is violated and 
		the quadrangle is conserved.
\end{enumerate}
In the case (1), small FCNC is allowed from the
evaluation of the experimental bounds of $B^0 - \bar{B}^0$
mixing and the $CP$ violation in the neutral $K$ meson. 
On the other hand, large FCNC is allowed in the case (2).
Under the bounds for each case, the predictions 
for the neutral $B$ meson processes will be presented. 
The predictions for $B \rightarrow X_d \, l^+ \, l^- $ 
and $CP$ asymmetries in the neutral $B$ meson system 
show that significant contributions are expected in the
model, although the FCNC is small.
For large FCNC, a significant contribution is expected in
the inclusive $B \rightarrow X_d \, \gamma$ decay.
Additionally, we also confirm contribution of the new physics due to 
the VQM in the electroweak oblique correction parameters,
$S,T,U$ (ref. \cite{lavoura}), and find that the
contributions are negligible.
Apart from the experimental constraints, we will also present 
theoretical studies on the FCNC's, as a consequence of 
the behaviour of the mass matrices in the model.

All of the contents will be discussed in the main paper, and the detail
calculations and explanations can be seen in the appendices.

\clearpage
\section{FLAVOR PHYSICS IN THE SM : AN OVERVIEW}
\label{sec:fpsm}

In this section, we will give an overview of the SM, and its implications
on the flavor physics (see for example refs. \cite{cheng}, \cite{muta}
and references therein).

\subsection{The model}
\label{subsec:msm}

SM is a model which is based on the $SU(2) \otimes U(1)$
gauge group. In the SM, left-handed particles are introduced
as $SU(2)$ doublet, but right-handed particles are as $U(1)$ singlet. 
The contents of the particles in this group are denoted as below,
\begin{itemize}
	\item Fermions :
		\begin{eqnarray}
			{\rm Quarks} & : & \left(
				\begin{array}{c}
					u^i \\
					d^i
				\end{array}
				\right)_L \: , \:
				{u_R}^i \: , \: {d_R}^i \\
			{\rm Leptons} & : & \left(
				\begin{array}{c}
					{\nu_l}^i \\
					l^i
				\end{array}
				\right)_L \: , \:
				{l_R}^i \: .
		\end{eqnarray}
	\item Gauge bosons :
		\begin{equation}
			\left(
			\begin{array}{c}
				{W_\mu}^1 \\
				{W_\mu}^2 \\
				{W_\mu}^3
			\end{array}
			\right) \: , \:
			B_\mu \: .
		\end{equation}
	\item Scalars :
		\begin{equation}
			\phi \equiv \left(
			\begin{array}{c}
				\chi^+ \\
				\frac{1}{\sqrt{2}} \left( v + \phi^0 \right)
			\end{array}
			\right) \: .
			\label{eq:scalar}
		\end{equation}
\end{itemize}
where $u^i$ and $d^i$ represent up-type ($u \, , \, c \, , \, t$) and
down-type ($d \, , \, s \, , \, b$) quarks, with generation
indices $i = 1,2,3$, and same for leptons $l^i$ ($e \, , \, \mu \, , \, \nu$)
and neutrino $\nu^i$ ($\nu_e \, , \, \nu_\mu \, , \, \nu_\tau$).
Throughout this paper, we use the following notations for
chirality, $L \equiv \frac{1}{2} (1 - \gamma_5)$ and
$R \equiv \frac{1}{2} (1 + \gamma_5)$.
In the Minimal SM, we consider only one Higgs doublet to generate the
fermion masses.

The various terms in the lagrangian can be written by demanding
$SU(2) \otimes U(1)$ gauge invariance and lepton-quark universality
of the electroweak interactions. It will reproduced
in part in Appendix \ref{app:sml}. In relation with $K$ and $B$
physics, the main interest lies in the study of the flavor changing 
transitions, involving fermions, gauge bosons and Higgs fields.

As the results of Appendix \ref{app:sml}, we can write the
related interactions for physics in the meson decays briefly as,
\begin{eqnarray}
	{\cal L}_{W^{\pm}} & = & \frac{g}{\sqrt{2}} \left(
		{V_{CKM}}^{ij} \bar{u}^i \gamma^{\mu} \, L \, d^j
		+ \bar{\nu_l}^i \gamma^{\mu} \, L \, l^i \right) \, W_{\mu}^+
		+ h.c. \: ,
		\label{eq:lwsm} \\
	{\cal L}_{\chi^{\pm}} & = & \frac{g}{\sqrt{2} M_W} \left[
		{V_{CKM}}^{ij} \bar{u}^i \left(
		m_{u^i} L - m_{d^j} R \right) d^j
		+ m_l \bar{\nu_l}^i \, L \, l^i \right] \,
		\chi^+ + h.c. \: , \\
	{\cal L}_A & = & \frac{e}{3} \left(
		2 \bar{u}^i \gamma^{\mu} u^i
		- \bar{d}^i \gamma^{\mu} d^i
		- 3 \bar{l}^i \gamma^{\mu} l^i
		\right) \, A_{\mu} \: , \\
	{\cal L}_Z & = & \frac{g}{2\cos\theta_W}
		\left\{ \bar{u}^i \gamma^{\mu} \left[\left(
		1 - \frac{4}{3} \sin^2\theta_W \right) L -
		\frac{4}{3} \sin^2\theta_W R \right] u^i \right.
		\nonumber \\
	& &	+ \left. \bar{d}^i \gamma^{\mu}
		\left[\left( \frac{2}{3} \sin^2\theta_W - 1 \right) L +
		\frac{2}{3}\sin^2\theta_W R \right] d^i \right.
		\nonumber \\
	& &	+ \left. \bar{\nu_l}^i \gamma^{\mu} \left[\left(
		1 - 4 \sin^2\theta_W \right) L -
		4 \sin^2\theta_W R \right] {\nu_l}^i \right.
		\nonumber \\
	& &	+ \left. \bar{l}^i \gamma^{\mu}
		\left[\left( 2 \sin^2\theta_W - 1 \right) L +
		2 \sin^2\theta_W R \right] l^i
	  	\right\} \, Z_{\mu} \: ,
\end{eqnarray}
and the related tripple gauge interactions are,
\begin{eqnarray}
 	{\cal L}_{W^{\pm} W^{\mp} A} & = & i e \, \left[
		\left( \partial^\mu A^\nu - \partial^\nu A^\mu \right)
		{W_\mu}^+ {W_\nu}^- + A^\nu \left(
		\partial^\mu {W^\nu}^+ - \partial^\nu {W^\mu}^+ \right)
		{W_\mu}^- \right. \nonumber \\
	& & 	\left. - A^\nu \left( \partial_\mu {W_\nu}^-
		- \partial_\nu {w_\mu}^- \right) {W^\mu}^+
		+ A^\mu \left( {W_\mu}^+ \partial^\mu
		- \partial^\mu {W_\mu}^+ \right) {W_\mu}^- \right] \: , \\
	{\cal L}_{\chi^{\pm} \chi^{\mp} A} & = & i e \,
		A_\mu \chi^- \partial^\mu \chi^+ + h.c. \: .
\end{eqnarray}
Note here that the last two terms in the ${\cal L}_{W^{\pm} W^{\mp} A}$ comes
out from the gauge fixing terms (Eq. (\ref{eq:gf})).

The main goal of this section is to investigate the CKM matrix, $V_{CKM}$,
which appears in the above charge current interaction terms. It can be
achieved by examining the {\it CP} violations and rare decays in the SM.
This will be the aim of Sec. \ref{subsec:ckm} and \ref{subsec:ecckm}. In the
Sec. \ref{subsec:ckm}, we will restudy the behaviour of CKM matrix in the SM,
and extract all of considerable processes that should be used to test
the CKM matrix.

\subsection{CKM matrix}
\label{subsec:ckm}

From the discussion in Appendix \ref{app:yukawasm}, it is clear that
the flavor changing transition in the SM emerge from the diagonalization
of the quark mass matrices after spontaneous symmetry breaking (SSB).
The Yukawa couplings in the SM are arbitrary complex numbers, and
therefore there is no hope of understanding their origin within the SM.
The origin of the differences of fermion masses is one of
the outstanding problems in the particle physics. Perhaps, 
in a more fundamental framework these couplings
may be derived from some deeper dynamics, for example this puzzle may be
able to be worked out in the VQM framework as will be done in
Sec. \ref{subsubsec:nsfcnc}. Since the interpretation of
the experimental results in the SM framework does not require a prior
knowledge of this connection, we shall assume the validity of CKM
matrix as a consistent framework and study the consequences of this
assumption in weak decays.

The matrix elements $V_{CKM}$ are determined by charged current coupling
to the $W^{\pm}$ bosons and/or $\chi^{\pm}$. Symbolically
it can be written as,
\begin{equation}
	V_{CKM} \equiv \left(
	\begin{array}{ccc}
		{V_{CKM}}^{ud}	& {V_{CKM}}^{us}	& {V_{CKM}}^{ub} \\
		{V_{CKM}}^{cd}	& {V_{CKM}}^{cs}	& {V_{CKM}}^{cb} \\
		{V_{CKM}}^{td}	& {V_{CKM}}^{ts}	& {V_{CKM}}^{tb}
	\end{array}
	\right) \: .
	\label{eq:ckm}
\end{equation}
All of these elements have to be determined experimentally via
$u^i d^i W$ and $u^i d^i \chi$ interactions. This is the
principal task of experimental flavor physics.

Before going to the next section, let us give a parametrization
that is usually used in relation with the experiment results. 
As done in Appendix \ref{app:kobayashi}, in the $3 \times 3$ CKM matrix,
after making any transformations, one phase is always remained.
This phase leads to $CP$ violation in the flavor physics.
However, under this fact and the condition that the CKM
matrix is unitary, we can parametrize it in any arbitrary
ways (Appendix \ref{app:kobayashi} and
\ref{app:wolfenstein}). Here we employ an approximation but very
popular form of the matrix $V_{CKM}$ due to Wolfenstein, that is
\begin{equation}
	V_{CKM} \cong \left(
	\begin{array}{ccc}
		1 - \frac{1}{2} \lambda^2 	& \lambda	&
			A \lambda^3 (\rho - i \eta +
			\frac{i}{2} \eta \lambda^2) \\
		- \lambda	& 1 - \frac{1}{2} \lambda^2 -
			i \eta A^2 \lambda^4	&
			A \lambda^2 \left( 1 + i \eta \lambda^2
			\right) \\
		A \lambda^3 (1 - \rho - i \eta)	& -A \lambda^2 	& 1
	\end{array} \right) \: .
	\label{eq:ckmw}
\end{equation}
This Wolfenstein parametrization still has three real parameters and one phase.
This parametrization is an approximation form of the $V_{CKM}$,
but has advantages especially in evaluating the CKM matrix
elements phenomenologically.

Of the two generations part of $V_{CKM}$ in Eq. (\ref{eq:ckm}),
four involving the $u,d,c,s$ quarks (Cabibbo angle) were already known
before the discovery of bottom and top quarks. Two of the matrix
elements involving the bottom quark, ${V_{CKM}}^{ub}$ and ${V_{CKM}}^{cb}$,
have been measured in the decays of $B$ hadrons (ref. \cite{data}). The
remaining three elements, ${V_{CKM}}^{td}$, ${V_{CKM}}^{ts}$ and
${V_{CKM}}^{tb}$, are also accessible in $B$ decays. These will be
discussed in Sec. \ref{subsec:ecckm}.

\subsection{Unitarity of CKM matrix}
\label{subsec:uckm}

As stated before, the CKM matrix elements obey unitary constraints,
$V_{CKM} \, {V_{CKM}}^\dagger = 1$. This has implications that any 
pair of rows or columns of CKM matrix are orthogonal, then leads 
to six orthogonal conditions, i.e.,
\begin{enumerate}
  	\item Orthogonality conditions on the columns :
	\begin{eqnarray}
		ds & : & \sum_{i = u,c,t} {V_{CKM}}^{id} {{V_{CKM}}^{is}}^\ast
			= 0 \nonumber \: , \\
		db & : & \sum_{i = u,c,t} {V_{CKM}}^{id} {{V_{CKM}}^{ib}}^\ast
			= 0
			\label{eq:occol} \: , \\
		sb & : & \sum_{i = u,c,t} {V_{CKM}}^{is} {{V_{CKM}}^{ib}}^\ast
			= 0 \: .  \nonumber
	\end{eqnarray}
  	\item Orthogonality conditions on the rows :
	\begin{eqnarray}
		uc & : & \sum_{i = d,s,b} {V_{CKM}}^{ui} {{V_{CKM}}^{ci}}^\ast
			= 0 \: , \nonumber \\
		ut & : & \sum_{i = d,s,b} {V_{CKM}}^{ui} {{V_{CKM}}^{ti}}^\ast
			= 0
			\label{eq:ocrow} \: , \\
		ct & : & \sum_{i = d,s,b} {V_{CKM}}^{ci} {{V_{CKM}}^{ti}}^\ast
			= 0 \: . \nonumber
	\end{eqnarray}
\end{enumerate}
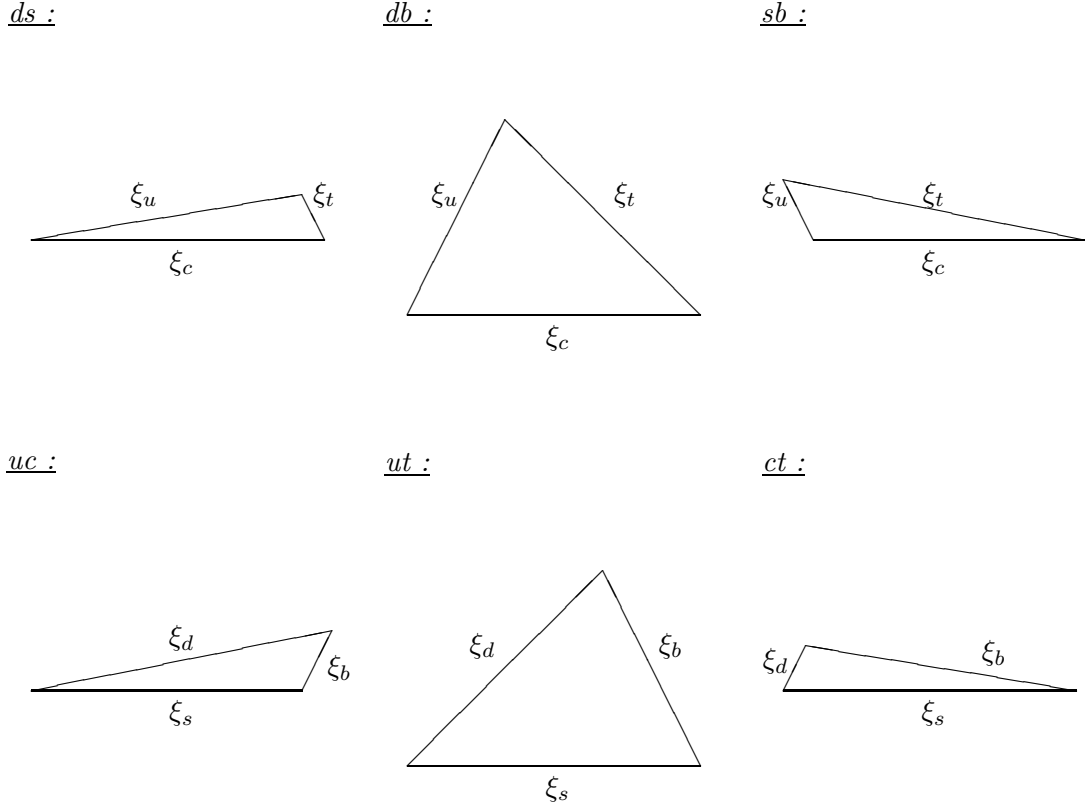
\begin{figure}[t]
	\unitlength 1mm
	\begin{center}
	\begin{picture}(140,110)
		\put(0,110){\makebox(0,0){\underline{\it ds :}}}
		\put(0,80){\line(1,0){39}}
		\put(0,80){\line(6,1){36}}
		\put(39,80){\line(-1,2){3}}
		\put(15,86){\makebox(0,0){$ \xi_u $}}
		\put(39,86){\makebox(0,0){$ \xi_t $}}
		\put(20,77){\makebox(0,0){$ \xi_c $}}
		\put(50,110){\makebox(0,0){\underline{\it db :}}}
		\put(50,70){\line(1,0){39}}
		\put(50,70){\line(1,2){13}}
		\put(89,70){\line(-1,1){26}}
		\put(55,86){\makebox(0,0){$ \xi_u $}}
		\put(79,86){\makebox(0,0){$ \xi_t $}}
		\put(70,67){\makebox(0,0){$ \xi_c $}}
		\put(100,110){\makebox(0,0){\underline{\it sb :}}}
		\put(104,80){\line(1,0){36}}
		\put(140,80){\line(-5,1){40}}
		\put(104,80){\line(-1,2){4}}
		\put(99,86){\makebox(0,0){$ \xi_u $}}
		\put(120,86){\makebox(0,0){$ \xi_t $}}
		\put(120,77){\makebox(0,0){$ \xi_c $}}
		\put(0,50){\makebox(0,0){\underline{\it uc :}}}
		\put(0,20){\line(1,0){36}}
		\put(0,20){\line(5,1){40}}
		\put(36,20){\line(1,2){4}}
		\put(20,27){\makebox(0,0){$ \xi_d $}}
		\put(20,17){\makebox(0,0){$ \xi_s $}}
		\put(41,23){\makebox(0,0){$ \xi_b $}}
		\put(50,50){\makebox(0,0){\underline{\it ut :}}}
		\put(50,10){\line(1,0){39}}
		\put(50,10){\line(1,1){26}}
		\put(89,10){\line(-1,2){13}}
		\put(60,26){\makebox(0,0){$ \xi_d $}}
		\put(85,26){\makebox(0,0){$ \xi_b $}}
		\put(70,7){\makebox(0,0){$ \xi_s $}}
		\put(100,50){\makebox(0,0){\underline{\it ct :}}}
		\put(100,20){\line(1,0){39}}
		\put(139,20){\line(-6,1){36}}
		\put(100,20){\line(1,2){3}}
		\put(99,24){\makebox(0,0){$ \xi_d $}}
		\put(128,25){\makebox(0,0){$ \xi_b $}}
		\put(120,17){\makebox(0,0){$ \xi_s $}}
	\end{picture}
	\caption{The CKM unitarity triangles following from the orthogonality
		of CKM matrix, where in the $ab$ sector,
		$\xi_i \equiv {V_{CKM}}^{ia} {{V_{CKM}}^{ib}}^\ast$ for the
		upper graphs, and
		$\xi_i \equiv {V_{CKM}}^{ai} {{V_{CKM}}^{bi}}^\ast$
		for the lower ones.}
	\label{fig:tri}
	\end{center}
\end{figure}

By using Wolfenstein parametrization, we can depict each of orthogonality
condition in Eqs. (\ref{eq:occol}) and (\ref{eq:ocrow}) as a triangle.
Thus, we obtain six triangles for them as depicted in Fig. (\ref{fig:tri}).
This will be derived quantitavely in the next section. We will discuss 
these triangles and the processes which determine them in 
Sec. \ref{subsec:ecckm}. 
In general, the triangles are a consequence of the unitarity
of CKM matrix, can be showed in a simple way (ref. \cite{sanda}).
From Eqs. (\ref{eq:occol}) and (\ref{eq:ocrow}),
\begin{equation}
	{V_{CKM}}^{1i} {{V_{CKM}}^{1j}}^\ast + {V_{CKM}}^{2i}
		{{V_{CKM}}^{2j}}^\ast
		+ {V_{CKM}}^{3i} {{V_{CKM}}^{3j}}^\ast = 0
\end{equation}
for $i \neq j$. Now multiply it by ${{V_{CKM}}^{3i}}^\ast {V_{CKM}}^{3j}$ and
take the imaginary part, we obtain
\begin{equation}
	{\rm Im} \left[ {{V_{CKM}}^{3i}}^\ast {V_{CKM}}^{3j}
			{V_{CKM}}^{1i} {{V_{CKM}}^{1j}}^\ast \right]
	+ {\rm Im} \left[ {{V_{CKM}}^{3i}}^\ast {V_{CKM}}^{3j}
			{V_{CKM}}^{2i} {{V_{CKM}}^{2j}}^\ast \right] = 0 \: .
\end{equation}
Then,
\begin{equation}
	{\rm Im} \left[ {{V_{CKM}}^{3i}}^\ast {V_{CKM}}^{3j}
			{V_{CKM}}^{1i} {{V_{CKM}}^{1j}}^\ast \right]
	= - {\rm Im} \left[ {{V_{CKM}}^{3i}}^\ast {V_{CKM}}^{3j}
			{V_{CKM}}^{2i} {{V_{CKM}}^{2j}}^\ast \right] \: .
\end{equation}
In the same way, for $i \neq j \neq k \neq l$,
\begin{eqnarray}
	{\rm A} & \equiv & \pm \: {\rm Im} \left[ {{V_{CKM}}^{ki}}^\ast
		{V_{CKM}}^{kj} {V_{CKM}}^{li} {{V_{CKM}}^{lj}}^\ast \right]
		\nonumber \\
	& = & \pm \: {\rm Im} \left[ \left| {V_{CKM}}^{ki}
		{{V_{CKM}}^{kj}}^\ast \right| e^{i \phi_1} \times
		\left| {V_{CKM}}^{li} {{V_{CKM}}^{lj}}^\ast \right|
		e^{-i \phi_2} \right] \nonumber \\
	& = & \pm \: \left| {{V_{CKM}}^{ki}}^\ast {V_{CKM}}^{kj} \right|
		\left| {V_{CKM}}^{li} {{V_{CKM}}^{lj}}^\ast \right|
		\sin \left(\phi_1 - \phi_2 \right)
	\label{eq:ickm}
\end{eqnarray}
is the area of triangle, where $\phi_1 - \phi_2 $ is the angle between
${{V_{CKM}}^{ki}}^\ast {V_{CKM}}^{kj} $ and
$ {V_{CKM}}^{li} {{V_{CKM}}^{lj}}^\ast $.
Eq. (\ref{eq:ickm}) also indicates that A is invariant under the
phase redefinition (Appendices \ref{app:kobayashi} and
\ref{app:wolfenstein}).

Finally, it is clear that a test of each triangle
conservation should be a test of
the unitarity of CKM matrix itself. This could be done by examining some
processes which determine the CKM matrix elements. This will be the aim
of the next subsection.

\subsection{Experimental constraints on the CKM matrix}
\label{subsec:ecckm}

Unitarity constraints on the CKM matrix elements in Sec. \ref{subsec:uckm}
provide very powerful and interesting relations. We ought to test all
of the triangles in Fig. (\ref{fig:tri}) in order to test the Yukawa sector
in the SM. In general, the sides of the triangles can be measured by
FCCC and FCNC processes, and the angles are by $CP$ violations.

As stated in Sec. \ref{subsec:ckm}, apart from the already
measured six CKM matrix elements, there are still three
unknown elements. All of them are the elements in the third row of
CKM matrix, and involving bottom and top
quarks. Of course, it could be measured directly in the FCCC decays in
the high energy experiment in the future, i.e. top decay processes.
However, since the mass of top quark is very large, according to
ref. \cite{cdf}, $m_t \cong 174 \pm 10^{+13}_{-12}$(GeV),
we have difficulty realizing it in the present. 
This problem can be cleared
in some $K$ and $B$ mesons processes. The reason is that, the
$t \, d^i \, W$ and/or $t \, d^i \, \chi $ interactions
appear in the one-loop order, and the top quark contributions in its 
internal lines are dominant.

In our opinion, the most interesting triangle is the $db$ sector.
The reason is, the $db$ sector triangle will be measured 
confidently in the near future, especially when the $B$ factory and
LHC are started. So, here we will concentrate on the triangle of $db$
sector and do not go on discussing the other sectors.

\subsubsection{Experimental constraints from FCCC processes}
\label{subsubsec:eccs}

We briefly give the experimental results (refs. \cite{data}, \cite{ali}
and \cite{nir2}) of first and second rows of $V_{CKM}$ in Eq. (\ref{eq:ckm}).
All of them are determined in the direct FCCC decays.
The measurements of two generations part (Cabibbo matrix part) are 
accurate. On the other hand, $\left| {V_{CKM}}^{ub} \right| $ and
$\left| {V_{CKM}}^{cd} \right| $ are less accurate, since there are
theoretical uncertainties due to the puzzle of hadron matrix elements.
However, because our interest is the FCNC processes, we will not discuss
the FCCC processes in detail.

Here, we list the experiment results and the processes to determine
them,
\begin{enumerate}
  	\item $\left| {V_{CKM}}^{ud} \right| $ : \\
  		It is determined by nuclear beta decays, when compared
  		to muon decay. One obtains,
  		\begin{equation}
			\left| {V_{CKM}}^{ud} \right| = 0.9744 \pm 0.0010 \: .
			\label{eq:vud}
		\end{equation}
	\item $\left| {V_{CKM}}^{us} \right| $ : \\
		There are mainly two ways to determine it, that is via
		$K_{e3}$ decays ($K^+ \rightarrow \pi^0 e^+ \nu_e$ and
		$K^0_L \rightarrow \pi^- e^+ \nu_e$), and via semileptonic
		hyperon decays. One averages these two results to obtain,
  		\begin{equation}
			\left| {V_{CKM}}^{us} \right| = 0.2205 \pm 0.0018 \: .
			\label{eq:vus}
		\end{equation}
	\item $\left| {V_{CKM}}^{cd} \right| $ : \\
		The magnitude of $\left| {V_{CKM}}^{cd} \right| $ may be
		deduced from neutrino and antineutrino production of charm
		off valence $d$ quarks, and semileptonic strangeless $D$
		decays ($D^0 \rightarrow \pi^- e^+ \nu_e$). But the second
		one is less accurate since the theoretical uncertainties
		in the form factor. From the first process,
  		\begin{equation}
			\left| {V_{CKM}}^{cd} \right| = 0.204 \pm 0.017 \: .
			\label{eq:vcd}
		\end{equation}
	\item $\left| {V_{CKM}}^{cs} \right| $ : \\
		This is deduced from $D_{e3}$ decays
		($D^0 \rightarrow K^- e^+ \nu_e$), analogous to
		$\left| {V_{CKM}}^{us} \right| $. It follows that,
  		\begin{equation}
			\left| {V_{CKM}}^{cs} \right| = 1.02 \pm 0.18 \: .
			\label{eq:vcs}
		\end{equation}
	\item $\left| {V_{CKM}}^{cb} \right| $ : \\
		$\left| {V_{CKM}}^{cb} \right| $ can be gained from the
		inclusive semileptonic $B$ decays, that is
		$B \rightarrow X_c l \nu_l$
		($B^0 \rightarrow {D^\ast}^+ l^- \nu_l$,
		$B^- \rightarrow {D^\ast}^0 l^- \nu_l$).
		The present data have an accuracy of $\pm 12 \% $,
  		\begin{equation}
			\left| {V_{CKM}}^{cb} \right| = 0.042 \pm 0.005 \: .
			\label{eq:vcb}
		\end{equation}
	\item $\left| {V_{CKM}}^{ub} \right| $ : \\
		Due to experimental progress, the fact that
		$\left| {V_{CKM}}^{ub} \right| $ can be shown to be non-zero
		as required to explain $CP$ violation, has been addressed by
		many groups. $\left| {V_{CKM}}^{ub} \right| $ is usually
		determined in the ratio $\left| {{V_{CKM}}^{ub}}/{
		{V_{CKM}}^{cb}} \right| $,
  		\begin{equation}
			\left| \frac{{V_{CKM}}^{ub}}{{V_{CKM}}^{cb}} \right|
			= 0.08 \pm 0.02 \: .
			\label{eq:vub}
		\end{equation}
\end{enumerate}
Note that, accurately measured $\left| {V_{CKM}}^{us} \right| $, have 
an important role in the Wolfenstein parametrization 
in Eq. (\ref{eq:ckmw}), since the CKM matrix has been expanded in term of
$\left| {V_{CKM}}^{us} \right| \equiv \lambda$.

On using Eq. (\ref{eq:ckmw}), we can translate the above list to find
the experimental constraints for the parameters in the Wolfenstein
parametrized CKM matrix. The bounds for them can be summarized as,
\begin{enumerate}
  	\item $ \lambda $: \\
    		\begin{equation}
			\left| {V_{CKM}}^{us} \right| = \lambda
			\longrightarrow \lambda = 0.2205 \pm 0.0018 \: .
			\label{eq:lambda}
		\end{equation}
	\item $ A$ : \\
    		\begin{equation}
			\left| {V_{CKM}}^{cb} \right| = A \lambda^2
			\longrightarrow A = 0.86 \pm 0.10 \: .
			\label{eq:a}
		\end{equation}
	\item $ \rho $ and $\eta $ : \\
    		\begin{equation}
			\left| \frac{{V_{CKM}}^{ub}}{{V_{CKM}}^{cb}} \right|
			= \lambda \sqrt{\rho^2 + \eta^2} \longrightarrow
			\sqrt{\rho^2 + \eta^2} = 0.36 \pm 0.09 \: .
			\label{eq:cavcb}
		\end{equation}
		To get bounds for $\rho $ and $\eta $ separately, we need
		additional input from FCNC processes.
\end{enumerate}

\subsubsection{Experimental constraints from FCNC processes}
\label{subsubsec:ecdbs}

In this section, we discuss the remaining CKM matrix elements
which could be determined by FCNC processes.
From Eq. (\ref{eq:ckmw}), we should normalize each side of the triangle
in Fig. (\ref{fig:tri}) by ${\left| {V_{CKM}}^{cd} {{V_{CKM}}^{cb}}^\ast
\right|}$ like below,
\begin{eqnarray}
	\frac{\left| {V_{CKM}}^{ud} {{V_{CKM}}^{ub}}^\ast \right|}
		{\left| {V_{CKM}}^{cd} {{V_{CKM}}^{cb}}^\ast \right|}
		& \cong & \sqrt{\rho^2 + \eta^2} \equiv
		\left| {\rm CA} \right| \, ,
		\label{eq:ca} \\
	\frac{\left| {V_{CKM}}^{td} {{V_{CKM}}^{tb}}^\ast \right|}
		{\left| {V_{CKM}}^{cd} {{V_{CKM}}^{cb}}^\ast \right|}
		& \cong & \sqrt{(1 - \rho)^2 + \eta^2}
		\equiv \left| {\rm AB} \right| \, .
		\label{eq:ab}
\end{eqnarray}
Therefore, in the complex plane $\left( \rho \, , \, \eta \right)$,
the triangle can be written as Fig. (\ref{fig:dbs}), by normalizing
$\left| {V_{CKM}}^{cd} {{V_{CKM}}^{cb}}^\ast \right| \cong A \lambda^3
\equiv 1$.
Moreover, the side $\left| {\rm AB} \right|$ can be rewritten in term of
$\left| {\rm CA} \right|$ by using Eqs. (\ref{eq:ca}) and (\ref{eq:ab}) ,
\begin{eqnarray}
	\left|{\rm AB} \right| & = & \sqrt{1 - 2 \rho + (\rho^2 + \eta^2)}
		\nonumber \\
	& = & \sqrt{1 + \left| {\rm CA} \right|^2 - 2 \left| {\rm CA} \right|
		\cos \gamma} \: .
	\label{eq:abca}
\end{eqnarray}
Each angle in Fig. (\ref{fig:dbs}) can also be expressed in terms of
$\rho$ and $\eta$ by using simple trigonometry (ref. \cite{buras}),
\begin{eqnarray}
	\sin 2 \alpha & = & \frac{2 \eta (\eta^2 + \rho^2 - \rho)}{
		(\rho^2 + \eta^2)((1 - \rho)^2 + \eta^2)}
		= \frac{2 \sin \gamma (\left| {\rm CA} \right| - \cos \gamma)}{
		\left| {\rm AB} \right|^2} \: ,
	\label{eq:sina} \\
	\sin 2 \beta & = & \frac{2 \eta (1 - \rho)}{(1 - \rho)^2 + \eta^2}
		= \frac{2 \left| {\rm CA} \right| \sin \gamma
		(1 - \left| {\rm CA} \right| \cos \gamma)}{
		\left| {\rm AB} \right|^2} \: ,
	\label{eq:sinb} \\
	\sin 2 \gamma & = & \frac{2 \rho \eta}{\rho^2 + \eta^2}
		= \frac{2 \rho \eta}{\left| {\rm CA} \right|^2} \: .
	\label{eq:sing}
\end{eqnarray}
In addition to the above relations, there is also a constraint from
the feature of triangle, i.e.
\begin{equation}
	\alpha + \beta + \gamma = 180^0 \: .
	\label{eq:angle}
\end{equation}
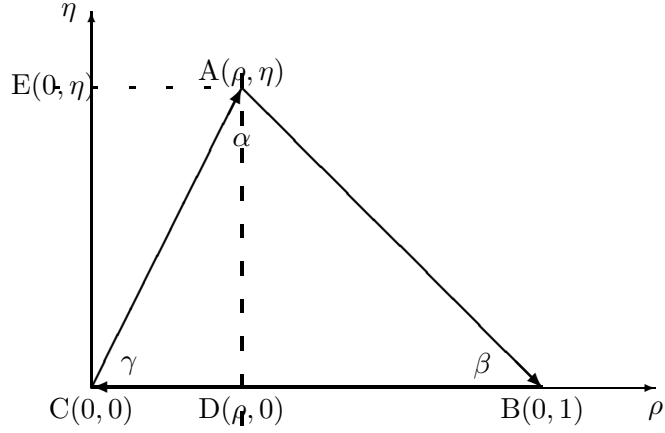
\begin{figure}[t]
	\unitlength 1mm
	\begin{center}
	\begin{picture}(80,55)
		\put(5,5){\vector(0,1){50}}
		\put(5,5){\vector(1,0){75}}
		\thicklines
		\put(65,5.1){\vector(-1,0){60}}
		\put(5,5){\vector(1,2){20}}
		\put(25,45){\vector(1,-1){40}}
		\multiput(25,0)(0,5){10}{\line(0,1){1.8}}
		\multiput(0,45)(5,0){5}{\line(1,0){0.8}}
		\put(2,55){\makebox(0,0){$\eta$}}
		\put(80,2){\makebox(0,0){$\rho$}}
		\put(5,2){\makebox(0,0){C$(0,0)$}}
		\put(65,2){\makebox(0,0){B$(0,1)$}}
		\put(25,47){\makebox(0,0){A$(\rho,\eta)$}}
		\put(25,38){\makebox(0,0){$\alpha$}}
		\put(57,8){\makebox(0,0){$\beta$}}
		\put(10,8){\makebox(0,0){$\gamma$}}
		\put(25,2){\makebox(0,0){D$(\rho,0)$}}
		\put(0,45){\makebox(0,0){E$(0,\eta)$}}
	\end{picture}
	\caption{The CKM unitarity triangle in the $db$ sector.}
	\label{fig:dbs}
	\end{center}
\end{figure}

The next question is, what the processes can be used to measured
the sides and the angles of triangle in Fig. (\ref{fig:dbs}).
Here, we list the processes which should determine the sides and
the angles.
\begin{itemize}
  	\item $\left| {\rm CA} \right|$ : $ B \rightarrow X_c l \nu_l $,
  		$\cdots $
  	\item $\left| {\rm AB} \right|$ : ${B_d}^0 - \bar{B_d}^0$ mixing,
  		$B \rightarrow X_d l^+ l^- $,
  		$B \rightarrow X_d \, \gamma $,
  		${B_d}^0 \rightarrow l^+ l^- $, $\cdots $
  	\item $\left| {\rm AD} \right|$ : ${\epsilon^\prime}/\epsilon$,
  		$K_L \rightarrow \pi^0 e^+ e^-$,
  		$K_L \rightarrow \pi^0 \nu \bar{\nu} $, $\cdots $
  	\item $\left| {\rm AE} \right|$ : $\left( K^0_L \rightarrow \mu \,
  		\bar{\mu} \right)_{SD} $, $\cdots $
  	\item $\alpha$ : ${B_d}^0 \rightarrow \pi^+ \pi^- \, , \,
  			\rho^{\pm} \pi^{\mp} $ , $\cdots $
  	\item $\beta$ : ${B_d}^0 \rightarrow \psi K_s \, , \,
  			D^{\pm} D^{\mp} $ , $\cdots $
  	\item $\gamma$ : ${B_s}^0 \rightarrow \rho^0 K_s \,$ , $\cdots $
  	\item The sign of $\rho$ and $\eta$ : $\epsilon$.
\end{itemize}
The effects of tree-level FCNC's will be applyed in the most reliable
processes from the above list, that is ${B_d}^0 - \bar{B_d}^0$ mixing,
$CP$ violations in the ${B_d}^0 \rightarrow \pi^+ \pi^-, \:
\psi K_s$ and ${B_s}^0 \rightarrow \rho^0 K_s$ decays and the parameter 
$\epsilon$ in the neutral $K$ meson system. The reason that, the processes 
will be determined well in such experiments. So we are not
discussing the others in either cases, SM and VQM, in this paper.

\noindent
\medskip \\
1) {\it $B \rightarrow X_q \, \gamma$ decay}

\begin{figure}[t]
	\epsfxsize=16cm
	\epsfysize=4cm
	\epsffile{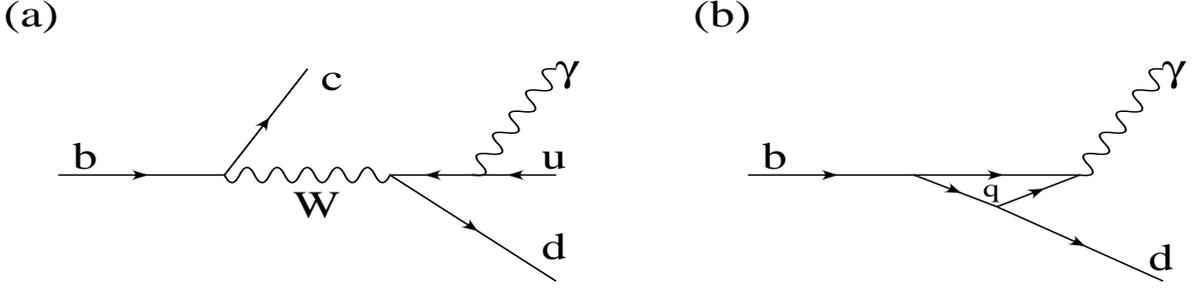}
	\caption{(a) Tree-level and (b) one-loop penguin diagram
		for the inclusive $b \rightarrow d \,
		\gamma$ decay, where $q$ denotes up-quarks.}
	\label{fig:bsgto}
\end{figure}

Approximately, the short distance contribution of this decay can be
expressed in the quark model by inclusive $b \rightarrow q \, \gamma$ decay,
since the bottom quark is heavy compared with the QCD scale.
In relation to the side $\left| {\rm AB} \right|$ of triangle, we should
observe the 
one-loop penguin diagram of $b \rightarrow d \gamma$ which has been
calculated in Appendix \ref{app:bqgsm}.
Unfortunately, in the inclusive $b \rightarrow d \gamma$ decay, contribution
from the penguin diagrams (Fig. (\ref{fig:bsgto})b) is
smaller than from the tree-level diagram (Fig. (\ref{fig:bsgto})a),
that is with order 
$\left| {{V_{CKM}}^{td}}/{V_{CKM}}^{cb}{} \right| \propto O(\lambda)$.
In the $b \rightarrow s \gamma$ decay, the situation is different, since
the tree-level is suppressed by order $O(\lambda^3)$ due to
the CKM matrix element ${V_{CKM}}^{ub}$.

In order to make the magnetic moment type $b \rightarrow d \gamma$ to
be dominance, we exclude the tree-level contribution by improving the
following condition (ref. \cite{grin})
\begin{equation}
	E_\gamma > \frac{{m_B}^2 - {m_D}^2}{2 m_B} = 2.3({\rm GeV}) \: .
\end{equation}
Thus, the penguin diagrams are to be the main contribution in the decay.

The SM prediction for the amplitude of $b \rightarrow d \gamma$ decay
is given in Appendix \ref{app:bqgsm}. From the result, the branching
ratio is,
\begin{eqnarray}
	{Br(b \rightarrow d \, \gamma)}^{SM} & \cong &
		\frac{\alpha {G_F}^2}{128 \pi^4} {m_{B_d}}^5 \, \tau_{B_d}
		\, {Q_u}^2
		\left| {V_{CKM}}^{td} {{V_{CKM}}^{tb}}^\ast \right|^2
		\, \left| {F_{bd\gamma}}^{SM} (m_b) \right|^2 \nonumber \\
	& = &	0.230 \times A^2 \, \lambda^6 \,
		\left[ \left( 1 - \rho \right)^2 + \eta^2 \right] \: ,
\end{eqnarray}
where we neglect the down quark mass $m_d$, because of ${m_d}^2 \ll {m_b}^2$.
Here, we used $m_t = 174$(GeV) and $ m_{B_d} = 5373
\pm 4.2 $(MeV), and ${F_{bd\gamma}}^{SM} (m_b)$ is given in Appendix
\ref{app:bqgsm}.

On the other hand, experimentally $b \rightarrow d \, \gamma$ has been
observed, but only the upper-bound have been obtained by CLEO collaboration
(ref. \cite{payne}). The value is the sum of the branching ratio of
the exclusive decays $B^- \rightarrow \rho^- \, \gamma$,
$B \rightarrow \rho^0 \, \gamma$ and $B^0 \rightarrow \omega \, \gamma$,
\begin{equation}
	{Br(b \rightarrow d \, \gamma)}^{exp} < 6.3 \times 10^{-5} \: .
\end{equation}
This result yields the upper-bound for the side $\left| {\rm AB} \right|$
of triangle,
\begin{equation}
	\sqrt{\left( 1 - \rho \right)^2 + \eta^2} =
		\left| {\rm AB} \right|^{SM} < 2.082 \: .
	\label{eq:abbdg}
\end{equation}
Thus, it is clear that $b \rightarrow d \gamma$ itself is rather useless
to determine the side $\left| {\rm AB} \right|$, since it gives only
the upper-bound. So it would better to determine it by 
${B_d}^0 - \bar{B_d}^0$ mixing. However, this decay should be a
powerful test for the sizes of FCNC's in the future.

Apart from the above evaluation, it is also usual to
consider a less model-dependent ratio of $b \rightarrow q \, \gamma$
decay, i.e.
\begin{equation}
	\left( \frac{Br(b \rightarrow d \gamma)}{
		Br(b \rightarrow s \gamma)} \right)^{SM}  =
		\left| \frac{{V_{CKM}}^{td}{{V_{CKM}}^{tb}}^\ast}{
		{V_{CKM}}^{ts} {{V_{CKM}}^{tb}}^\ast} \right|^2
		\cong \lambda^2 \,
		\left[ \left( 1 - \rho \right)^2 + \eta^2 \right]
		\: .
	\label{eq:bqgratio}
\end{equation}
Since the experimental result for $b \rightarrow s \, \gamma$ is,
\begin{equation}
	{Br(b \rightarrow s \, \gamma)}^{exp} = \left(
		2.32 \pm 0.51 \pm 0.29 \pm 0.32 \right) \times 10^{-4}
\end{equation}
from CLEO-II group (ref. \cite{payne}), one finds
\begin{equation}
	\left| {\rm AB} \right|^{SM} < 1.957 \: .
\end{equation}
This bound is better than the bound in Eq. (\ref{eq:abbdg}).

Let us give a comment here about $b \rightarrow d \gamma$
decay. Present, experimentalists have difficulty measuring
the decay (ref. \cite{payne}). But, here we hope that they
can make a progress to measure it in the future. At any cost, the decay
is important to give a confident value for the side
$\left| {\rm AB} \right|$ together with $x_d$.

\noindent
\medskip \\
2) {\it $K^0 - \bar{K}^0$ mixing}

The small $CP$ violation in the neutral $K$ meson system can give
some constraints to determine the CKM matrix. The violation has been
observed, for example in the neutral $K$'s that decay into
$(2 \pi)_{I=0}$. One of them is described in the well-measured parameter
$\epsilon$. Definition of the parameter can be seen
in Appendix \ref{app:k0k0sm}, and is given in Eq. (\ref{eq:epsilonsm}) as,
\begin{eqnarray}
	{\epsilon}^{SM} & \cong & \frac{{G_F}^2}{6 \sqrt{2} \pi^2 \Delta M}
		m_K \, {f_K}^2 B_K \, {M_W}^2 \, {\rm Im}
		\left( {M_{12}}^{SM} \right) e^{i \pi/4} \nonumber \\
	& = & e^{i \pi/4} \, \frac{{G_F}^2}{6 \sqrt{2} \pi^2 \Delta M} m_K
		\, {f_K}^2 B_K \, {M_W}^2 \, A^2 \, \lambda^6 \, \eta \,
		\nonumber \\
	& &	\left[ -\eta_c \, {F_{\Delta S=2}}^{SM}(x_c)
		+ A^2 \, \lambda^4 (1 - \rho) \, \eta_t \,
		{F_{\Delta S=2}}^{SM}(x_t) + \eta_{ct} \,
		{F_{\Delta S=2}}^{SM}(x_c,x_t) \right] \: ,
\end{eqnarray}
by using Eq. (\ref{eq:ckmw}) and dropping the second term in Eq.
(\ref{eq:epsilonsm}), since
$2 {\rm Re} \, M_{12} \cong \Delta M^{exp} = 3.5 \times 10^{-15}$ is tiny.
Thus, knowing $\epsilon$ experimentally will give the prediction for
the CKM matrix elements. However, there is still theoretical uncertainties
in the bag parameter $B_K$.

Experimentally,
\begin{equation}
	{\epsilon}^{exp} = e^{i \pi/4} \left( 2.258 \pm 0.018 \right)
		\times 10^{-3} \: .
\end{equation}
After putting the values for each parameter, the SM gives the prediction for
$\epsilon$ as,
\begin{equation}
	{\epsilon}^{SM} = e^{i \pi/4} \left( 3.82 \times 10^4 \right)
 		B_K \, A^2 \, \lambda^6
		\, \eta \left[ \left( 7.84 \times 10^{-4} \right)
		+ 1.46 \, A^2 \lambda^4 (1 - \rho) \right] \: ,
		\label{eq:hyperbola}
\end{equation}
for $f_K = 161$(MeV), $\eta_c = 0.85$, $\eta_{ct} = 0.36$, $\eta_t = 0.57$
and $m_t = 174$(GeV). In Fig. (\ref{fig:trism}), we plot this equation with
$B_K = 0.82 \pm 0.10$ from the recent lattice calculations. This
constraint is important to determined the position of point A in the triangle,
or the signs of $\rho$ and $\eta$ in the $(\rho,\eta)$ plane.

\noindent
\medskip \\
3) {\it ${B_q}^0 - \bar{B_q}^0$ mixing}

Here, we discuss the ${B_q}^0 - \bar{B_q}^0$ mixing, with $q = d,s$.
Mixing in the ${B_q}^0 - \bar{B_q}^0$ mixing involves the third generations
and there is no reason to assume a small phase between $M_{12}$ and
$\Gamma_{12}$ (Appendix \ref{app:b0b0}). However, the discussion is
simplified because of Eq. (\ref{eq:gm12}),
$ {\Gamma_{12}}^{SM} \ll {M_{12}}^{SM} $.
The SM predictions for the size of mixing is given in Eq. (\ref{eq:xq}),
\begin{equation}
	{x_q}^{SM} = \frac{{G_F}^2}{6 \pi^2} m_{B_q} {M_W}^2
		\tau_{B_q} \eta_{B_q} {f_{B_q}}^2 B_{B_q}
		\left| {V_{CKM}}^{tq} {{V_{CKM}}^{tb}}^\ast \right|^2 \,
		\left| {F_{\Delta B = 2}}^{SM}(x_t) \right| \: .
		\label{eq:xqsm}
\end{equation}
On using Eq. (\ref{eq:ckmw}), it yields,
\begin{eqnarray}
	q = d 	& : & \left| {V_{CKM}}^{td} {{V_{CKM}}^{tb}}^\ast \right|
			\cong A^2 \, \lambda^6 \left[ \left(
			1 - \rho \right)^2 + \eta^2 \right] \: , \\
	q = s 	& : & \left| {V_{CKM}}^{ts} {{V_{CKM}}^{tb}}^\ast \right|
			\cong A^2 \, \lambda^4 \: .
\end{eqnarray}
Since the parameters $A$ and $\lambda $ have been determined in Eqs.
(\ref{eq:lambda}) and (\ref{eq:a}), one is able to predict $x_s$ more
confidentally than $x_d$. But, experimentally $x_d$ is
precisely measured yet.

The results for $x_d$ from the four experiments ARGUS, CLEO, ALEPH
and DELPHI (refs. \cite{data} and references therein) gives,
\begin{equation}
	{x_d}^{exp} = 0.71 \pm 0.07
	\label{eq:xdexp}
\end{equation}
by using $\tau_{B_d} = 1.44 \pm 0.15$(ps). Unfortunately, this precision is
not matched by the theory, which is uncertain due to the imprecise
knowledge of $f_{B_d}$ and $B_{B_d} $. The prediction of the SM is
expressed as,
\begin{equation}
	{x_d}^{SM} = \left( 2.51 \times 10^5 \right) \times
		{f_{B_d}}^2 B_{B_d} \, A^2 \, \lambda^6 \,
		\left[ \left( 1 - \rho \right)^2 + \eta^2 \right] \: .
		\label{eq:circle}
\end{equation}
This equation determines the sides $\left| {\rm AB} \right|$, and is
depicted in Fig. (\ref{fig:trism}) including the uncertainties,
where $ m_{B_d} = m_{B_s} = 5$(GeV),
$ \eta_{B_d} = \eta_{B_s} = 0.55$ and $ m_t = 174$(GeV).
In the Fig. (\ref{fig:trism}) we adopt $160 \le \sqrt{{f_{B_d}}^2 B_{B_d}}
\le 240$(MeV) from the recent lattice calculations and improved QCD sum rules.
It yields,
\begin{equation}
	0.631 \le \left| {\rm AB} \right|^{SM} \le 1.386 \: .
	\label{eq:abxd}
\end{equation}

Meanwhile, experimentally the mixing parameter $x_s$ is not determined
yet, and only the lower-bound has been known (ref. \cite{data}),
\begin{equation}
	{x_s}^{exp} > 1.5 \: .
	\label{eq:xsexp}
\end{equation}
Thus, in the meaning of the triangle sides,  $x_s$ itself does not work
an important
role. However, knowing $x_d$ and $x_s$ experimentally could reduce
the theoretical parameters and obtain a less model-dependent quantity,
\begin{equation}
	\left( \frac{x_d}{x_s} \right)^{SM} = \left| \frac{{V_{CKM}}^{td}}{
		{V_{CKM}}^{ts}} \right|^2
	= \lambda^2 \left[ \left( 1 - \rho \right)^2 + \eta^2 \right] \: ,
\end{equation}
by assuming $SU(3)$ symmetry breaking is zero, i.e.
${f_{B_q}}^2 B_{B_q}$ are same for $q = d, s$ . Then, the bound for
$\left| {\rm AB} \right|$ is,
\begin{equation}
	\left| {\rm AB} \right|^{SM} \le 3.122 \: .
\end{equation}
So, the bound from single quantity $x_d$ is more reliable. But we expect a
better constraint from this {\it clean} quantity when $x_s$
could be measured precisely in the future experiment.

\noindent
\medskip \\
4) {\it $CP$ violations in ${B_q}^0 - \bar{B_q}^0$ mixing}

We consider a neutral ${B_q}^0$ meson and its antiparticle
$\bar{{B_q}^0}$. As final states $f$, we consider
$f : \psi K_s, \: \pi^+  \pi^-, \: \rho K_s$ correspond to the angles
$\alpha, \: \beta, \: \gamma$ of the triangles (Appendix \ref{app:b0b0}).
All of $f$'s are considered as $CP$ even final states.
We are interested in the neutral $B$'s decays into a $CP$ eigenstate $f$.
Then, we shall use time-dependent asymmetry of each decay as given in
Eq. (\ref{eq:asb}),
\begin{equation}
	a_f(t) \cong -\sin (\Delta M t) \, \sin \phi \: ,
\end{equation}
and the time-integrated asymmetry is,
\begin{equation}
	a_f \equiv \int_0^1 dt \, a_f(t) = -\frac{x_q}{1 +
		{x_q}^2} \sin \phi \: .
\end{equation}
The detailed definition for $ \sin \phi $ is given in Eq. (\ref{eq:sinp}).
\begin{figure}[t]
	\epsfxsize=16cm
	\epsfysize=10cm
	\epsffile{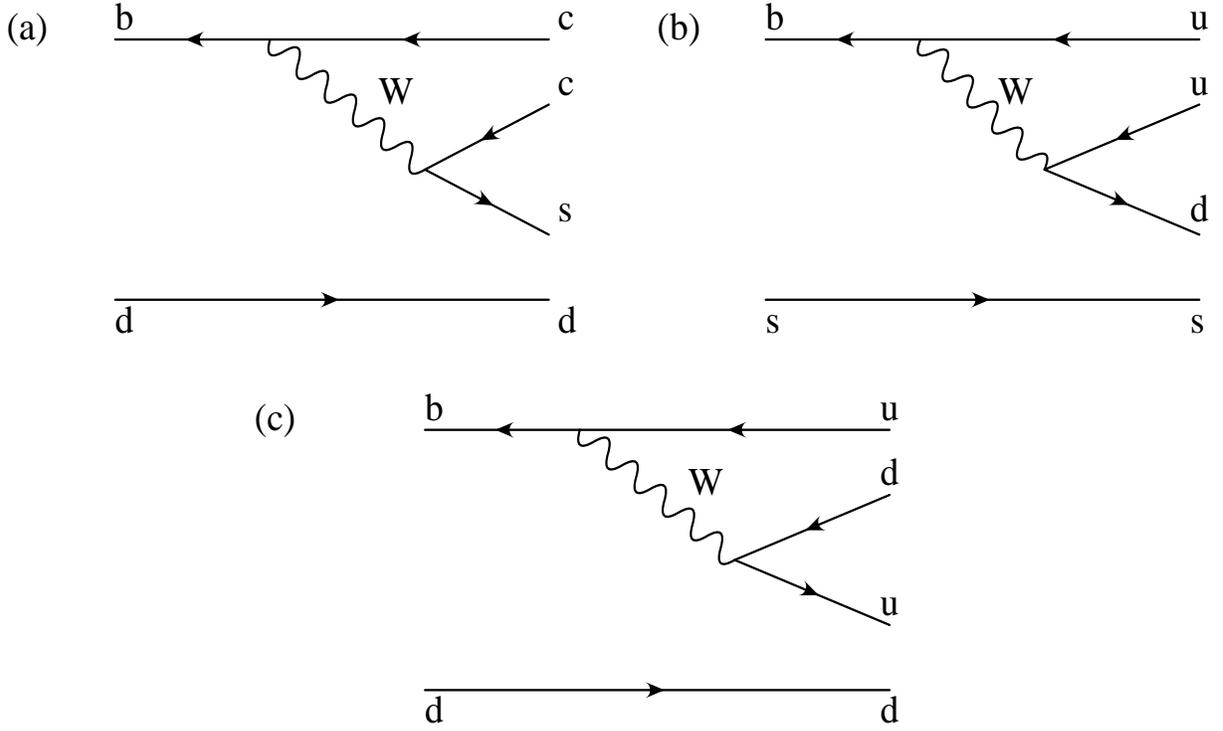}
	\caption{Diagrams which responsible for
	(a) $B_d^0 \rightarrow \psi K_s $, (b) $B_s^0 \rightarrow \rho K_s $
	and (c) $B_d^0 \rightarrow \pi^+ \pi^- $ decays.}
	\label{fig:cpb0b0}
\end{figure}

We are now ready to give three explicit quantities for asymmetries that
measure the angles  $\alpha, \: \beta, \: \gamma$.
\begin{description}
  	\item[(a)] ${B_d}^0 \rightarrow \psi K_s$ : \\
 		The mixing phase in the ${B_d}^0 - \bar{B_d}^0$ mixing is
 		given in Eq. (\ref{eq:qpbq}) for $q = d$.
 		The other phase is due to single final kaon $K_s$. Then
 		we have to take into account the mixing phase in
 		the $K^0 - \bar{K}^0$ mixing in Eq. (\ref{eq:qpk0}).
 		From the amplitude of ${B_d}^0 \rightarrow \psi K_s$ and its
 		anti particle decays as depicted in Fig.
 		(\ref{fig:cpb0b0}a) and Eq. (\ref{eq:aa}), the mixing
 		phase is,
 		\begin{equation}
			\left( \frac{\bar{A}}{A} \right)_{\psi K_s}^{SM} =
			\frac{\left( {V_{CKM}}^{cs} {{V_{CKM}}^{cb}}^\ast
 			\right)^\ast}{{V_{CKM}}^{cs} {{V_{CKM}}^{cb}}^\ast}\: .
		\end{equation}
		Combining all of the phases into Eq. (\ref{eq:sinp}),
 		\begin{equation}
			\sin \phi_{\psi K_s}
			= - \sin \left[ 2 \: {\rm arg} \left( \frac{
			-{V_{CKM}}^{cd} {{V_{CKM}}^{cb}}^\ast}{
 		     	{V_{CKM}}^{td} {{V_{CKM}}^{tb}}^\ast} \right)
 		     	\right]\: .
		\end{equation}
		This gives the size of the angle between $\vec{\rm AB}$ and
		$\vec{\rm BC}$, that is $\phi_{\psi K_s} = - 2 \beta$.
		Experimentally, $\psi K_s$ mode is the only $CP$ eigenstates
		that has been observed so far.
  	\item[(b)] ${B_d}^0 \rightarrow \pi^+ \pi^-$ : \\
		The mixing phase for this mode is,
 		\begin{equation}
			\left( \frac{\bar{A}}{A} \right)_{\pi^+ \pi^-}^{SM} =
			\frac{\left( {V_{CKM}}^{ud} {{V_{CKM}}^{ub}}^\ast
 			\right)^\ast}{{V_{CKM}}^{ud}
 			{{V_{CKM}}^{ub}}^\ast} \: .
		\end{equation}
		Similar with the $\psi K_s$ mode,
 		\begin{equation}
			\sin \phi_{\pi^+ \pi^-}
			= - \sin \left[ 2 \: {\rm arg} \left( \frac{
			-{V_{CKM}}^{td} {{V_{CKM}}^{tb}}^\ast}{
 			{V_{CKM}}^{ud} {{V_{CKM}}^{ub}}^\ast} \right)
 			\right] \: .
		\end{equation}
		It yields, $\phi_{\pi^+ \pi^-} = - 2 \alpha$.
  	\item[(c)] ${B_s}^0 \rightarrow \rho K_s$ : \\
		The mixing phase for this mode is same with $\pi^+ \pi^-$,
		since the quark subprocess is same, but the mixing from
		the neutral $B$ mixing is given in Eq. (\ref{eq:qpbq}) for
		$q = s$. After taking into account the phase mixing
		from $K_s$ in Eq. (\ref{eq:qpk0}), we obtain
 		\begin{equation}
			\sin \phi_{\rho K_s}
			= - \sin \left[ 2 \: {\rm arg} \left( \frac{
			-{V_{CKM}}^{ud} {{V_{CKM}}^{ub}}^\ast}{
 			{V_{CKM}}^{cd} {{V_{CKM}}^{cb}}^\ast} \right)
 			\right] \: .
		\end{equation}
		Then, this gives $\phi_{\rho K_s} = - 2 \gamma$.
\end{description}

However, all of these asymmetries, which give the sizes of angles,
could be determined in the near future experiments, $B$ factory
etc.

\subsection{The triangle in the SM}

\begin{figure}[t]
	\begin{center}
		\input{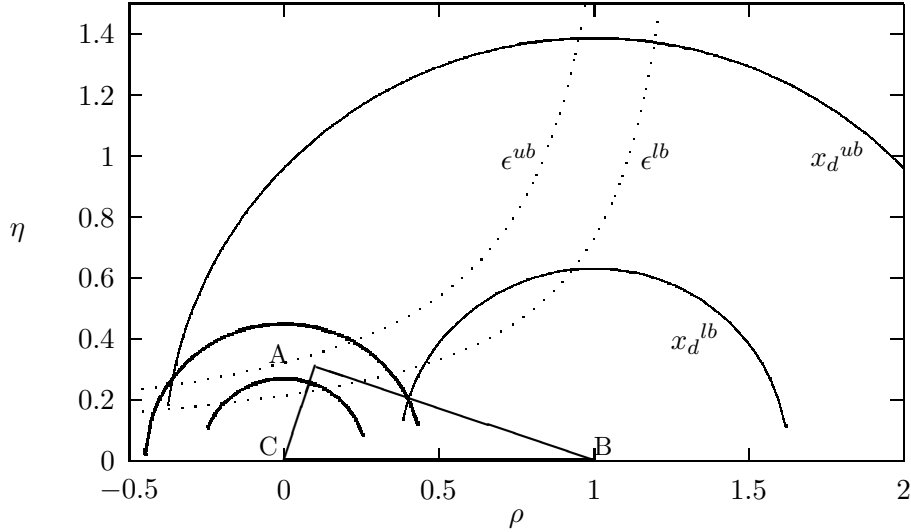}
	\end{center}
	\caption{The experimental bounds for the CKM unitarity triangle of
		the $db$ sector, with $ub$ and
		$lb$ denote upper-bound and lower-bound respectively.}
	\label{fig:trism}
\end{figure}
In the preceding sections, we have reevaluated the constraints
from some direct CC processes for two generations elements,
and the NC processes for the others.
The last task is to describe the experimental constraints on
the sides and angles of the triangle. We depicted them in Fig. (\ref{fig:tri}).

\begin{figure}[t]
	\begin{center}
		\input{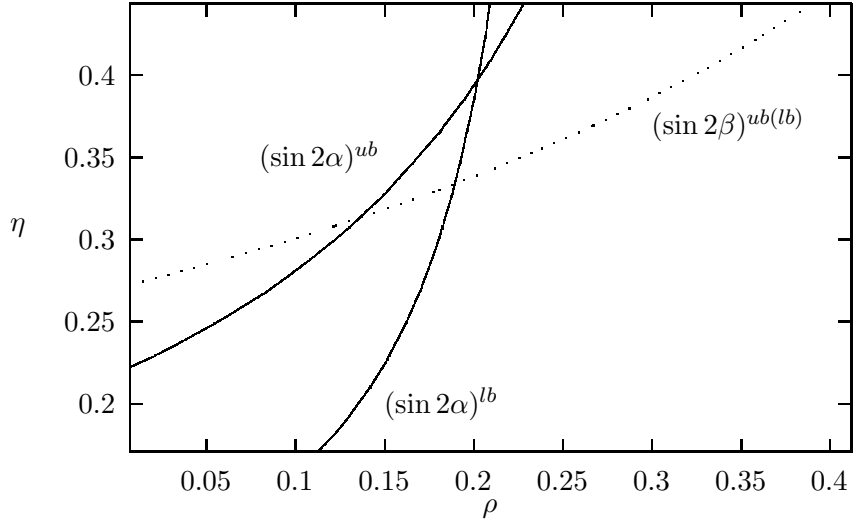}
	\end{center}
	\caption{The bounds for $\sin 2 \alpha$ and
		$\sin 2 \beta$ for the allowed 
		region of the pairs of $(\rho,\eta)$.}
	\label{fig:sinab}
\end{figure}
From Eqs. (\ref{eq:cavcb}) and (\ref{eq:abxd}), the bound
for the sides $\left| {\rm AB} \right|$ 
and $\left| {\rm CA} \right|$ in the SM are 
\begin{eqnarray}
	0.631 & \le \left| {\rm AB} \right|^{SM} < & 1.386 \: , 
	\label{eq:abtotal} \\
	0.25 & \le \left| {\rm CA} \right|^{SM} < & 0.45 \: 
\end{eqnarray}
respectively. the bound of $\left| {\rm CA} \right|$ is depicted as 
circles with centre point $(0,0)$ in Fig. (\ref{fig:trism}). 
Constraints from the mixing size $x_d$ is plotted in Fig. (\ref{fig:trism})
as circles with center point $(1,0)$ by using
the bound in Eq. (\ref{eq:abtotal}).
On the other hand, $\epsilon$ does not have a direct influence
to the sides or angles, but it has determined where the point A
lies. The bound has been depicted as hyperbolas according to
Eq. (\ref{eq:hyperbola}). From Fig. (\ref{fig:trism}), point
A must lie in the above $\rho$ axis. We note here that the wide 
allowed region in the bounds are mostly due to the theoretical 
uncertainties appear in the hadronic matrix elements.

On the other hand, the determination of the angles are still difficult.
However, the knowledge of $\rho$ and $\eta$ will give predictions for 
the sizes of the angles, by assuming that the triangle is perfectly satisfied.
Here we display them in the $(\rho,\eta)$ plane by employing the 
relations in Eqs. (\ref{eq:abca}) $\sim$ (\ref{eq:angle}). 
The relations yield,
\begin{eqnarray}
	\sin 2 \alpha & \cong & \frac{2 \, \eta \left( 1 -
		\frac{\rho}{\left| {\rm CA} \right|} \right)}{
		\left| {\rm AB} \right|^2 \left| {\rm CA} \right|} \: ,
		\label{eq:sinasm} \\
	\sin 2 \beta & \cong & \frac{2 \, \eta \left( 1
		- \rho \right)}{\left| {\rm AB} \right|^2 } \: .
		\label{eq:sinbsm}
\end{eqnarray}
by eliminating $\sin 2 \gamma$. 
In Fig. (\ref{fig:sinab}), the equations are shown with its dependence 
on the angles for the reliable values of $\rho$ and $\eta$.
The maximum allowed regions for $\rho$ and $\eta$ have been found
from the bounds in the Fig. (\ref{fig:trism}). The uncertainties in 
the figure are due to the bounds of the sides $\left| {\rm AB} \right|$ and 
$\left| {\rm CA} \right|$. We can comment that the figures are 
not sensitive for the uncertainty of $\left| {\rm AB} \right|$, 
but rather sensitive
for $\left| {\rm CA} \right|$. This has implication that
the difference between upper-bound and lower-bound in
$\sin 2 \beta$ is not visible. This leads that a better
prediction for $\left| {\rm CA} \right|$, i.e. ${V_{CKM}}^{ub}$,
must be expected in the future.

\clearpage
\section{FLAVOR PHYSICS WITH TREE-LEVEL FCNC}

As stated in Sec. \ref{sec:fpsm}, in the SM FCNC processes occur
through flavor changing in the CC interactions as the consequences
of CKM matrix. It means that in the SM, FCNC processes have only
appeared at one-loop and/or higher orders. This is usually called as
GIM mechanism, which naturally brings to the realization of the suppression
of FCNC processes.
However, in some models beyond the SM, the tree-level FCNC is possibly
realized, in contrast with the models which forbid the tree-level FCNC.

In this paper, we employ a model, which realizes the tree-level FCNC
in a simple way, that is called as VQM. In fact, the calculations and
results are not available for the world of VQM only, but also for the
other models with similar features. We will do similar procedures 
in Sec. \ref{sec:fpsm}.

\subsection{The model}
\label{subsec:sqm}

The model is a simplest extension of the SM, and has been
proposed for example in ref. \cite{branco}. In addition to the three standard
generations of quarks, we introduce one down-type and one up-type
quarks, and let
the particle content in the other sectors are unaltered.
Then the particle content in the quark sector becomes,
\begin{equation}
	\left(
	\begin{array}{c}
		u^i \\
		d^i
	\end{array}
	\right)_L  \: , \:
	{t_L}^\prime \: , \: {b_L}^\prime \: , \:
	{u_R}^{\alpha} \: , \: {d_R}^{\alpha}
\end{equation}
where $\alpha$ and $i$ are generation indices with $\alpha = 1,2,3,4$ and
$i = 1,2,3$.
Here, $t^\prime$ and $b^\prime$ are fourth generation up-type
($Q_{t^\prime} = 2/3$) and down-type ($Q_{b^\prime} = -1/3$) vector-like quarks.

The existence of vector-like quarks will change the NC interactions.
In addition to the interactions in Sec. \ref{subsec:msm},
there are new interactions in the $Z$ boson, photon and
neutral Higgs sectors as below (Appendix \ref{app:sqml}), 
\begin{eqnarray}
	{\cal L}_A & = & \frac{e}{3} \left(
		2 \bar{u}^{\alpha} \gamma^{\mu} u^{\alpha}
		- \bar{d}^{\alpha} \gamma^{\mu} d^{\alpha}
		\right) \, A_{\mu} , \\
	{\cal L}_Z & = & \frac{g}{2\cos\theta_W}
		\left\{ \bar{u}^{\alpha} \gamma^{\mu} \left[\left(
		{z_u}^{\alpha\beta} - \frac{4}{3} \sin^2\theta_W
		\delta^{\alpha\beta}
		\right) L - \frac{4}{3} \sin^2\theta_W \delta^{\alpha\beta} R
		\right] u^{\beta} \right. \nonumber \\
	& &	+ \left. \bar{d}^{\alpha} \gamma^{\mu}
		\left[\left( \frac{2}{3} \sin^2\theta_W \delta^{\alpha\beta} -
		{z_d}^{\alpha\beta}\right) L +
		\frac{2}{3}\sin^2\theta_W \delta^{\alpha\beta}
		R\right] d^{\beta}\right\} \, Z_{\mu} ,
		\label{eq:lzsqm} \\
	{\cal L}_H & = & \frac{-g}{2 M_W} \left[
		{z_u}^{\alpha\beta} \bar{u}^{\alpha}
		\left(m_{u^{\alpha}} L + m_{u^{\beta}} R\right) u^{\beta}
		+ {z_d}^{\alpha\beta} \bar{d}^{\alpha}
		\left(m_{d^{\alpha}} L + m_{d^{\beta}} R\right) d^{\beta}
		\right] H \, ,  \\
	{\cal L}_{\chi^0} & = & \frac{-ig}{2 M_W} \left[
		{z_u}^{\alpha\beta} \bar{u}^{\alpha}
		\left(m_{u^{\alpha}} L - m_{u^{\beta}} R\right) u^{\beta}
		- {z_d}^{\alpha\beta} \bar{d}^{\alpha}
		\left(m_{d^{\alpha}} L - m_{d^{\beta}} R\right) d^{\beta}
		\right] \chi^0 \, .
\end{eqnarray}
where,
\begin{eqnarray}
	{z_d}^{\alpha\beta} & \equiv & \sum_{i=d,s,b} V^{\alpha i}
		{V^{\beta i}}^{\ast} = \delta^{\alpha\beta} -
		V^{\alpha b^\prime} {V^{\beta b^\prime}}^{\ast} \: ,
	\label{eq:zd} \\
	{z_u}^{\alpha\beta} & \equiv & \sum_{i=d,s,b} U^{\alpha i}
		{U^{\beta i}}^{\ast} = \delta^{\alpha\beta} -
		U^{\alpha b^\prime} {U^{\beta b^\prime}}^{\ast} \: .
	\label{eq:zu}
\end{eqnarray}

From the above lagrangians, it is clear that the tree-level FCNC's have
been appearing. The tree-level FCNC's occur through the couplings
$z_d$ and $z_u$, which have similar role with $V_{CKM}$
in the SM to realize flavor changing among the quarks. The difference is,
$z_{u(d)}$ are among the quarks with same charge, and
$V_{CKM}$ is among the quarks with different charges.

Since the main interest of the paper is the unitarity of CKM matrix,
especially in the $db$ sector, we simplify the discussion by ignoring
the up-type vector-like quark, and considering only one down-type vector-like
quark. However, the up-type vector-like quark has an important role in
connection with the effects of tree-level FCNC's on the orthogonal
conditions in Eq. (\ref{eq:ocrow}). The problem is the up-quarks
contained meson, like $D$ meson, processes must be
considered. Then, it is expected that, for example in the $D^0 - \bar{D}^0$
mixing, the processes are small in the SM, since the quarks that
appear in the internal line are light. The heaviest one is
the bottom quark, but it is still not heavy
enough compared with the top quark that appears in the internal line
in the $K$ and $B$ mesons processes. So, in the next subsection,
we will discuss the CKM matrix in the VQM with only one down-type vector-like
quark (henceforth called ODVQM).

\subsection{CKM matrix}

Since the vector-like quark is belong to the $SU(2)$ singlet group,
there is no interaction like $u^i b^\prime W$ and/or
$u^i \, b^\prime \, \chi$  in the bare lagrangian.
But, redefining the quark fields will generate the interaction
which are mediated by the modified CKM matrix as in
Eq. (\ref{eq:ckmsqm}),
\begin{eqnarray}
	{V_{CKM}}^{j \beta} & \equiv & \sum_{i=1}^{3}
		U^{ij} {V^{i \beta}}^{\ast} \nonumber \\
	& \equiv & \left(
	\begin{array}{cccc}
		{V_{CKM}}^{ud} 	& {V_{CKM}}^{us} 	& {V_{CKM}}^{ub}
			& {V_{CKM}}^{ub^\prime} \\
		{V_{CKM}}^{cd} 	& {V_{CKM}}^{cs} 	& {V_{CKM}}^{cb}
			& {V_{CKM}}^{cb^\prime} \\
		{V_{CKM}}^{td} 	& {V_{CKM}}^{ts} 	& {V_{CKM}}^{tb}
			& {V_{CKM}}^{tb^\prime}
	\end{array}
	\right) \: .
	\label{eq:ckmodsqm}
\end{eqnarray}

The interesting feature of Eq. (\ref{eq:ckmodsqm}) is the unitarity 
violation of CKM matrix. This modified CKM matrix gives the following relation,
\begin{equation}
	\left( V_{CKM} {V_{CKM}}^\dagger \right)^{\alpha \beta} = 
		\sum_{i=1}^{3} V^{\alpha i} {V^{\beta i}}^{\ast}
	= {z_d}^{\alpha\beta} \: ,
	\label{eq:zv}
\end{equation}
by using Eq. (\ref{eq:zd}), and the unitarity $U U^\dagger= 1$. 
Our interest is the $db$ sector, where the relation becomes,
\begin{equation}
	{z_d}^{db} = \sum_{i = u,c,t} {V_{CKM}}^{id} {{V_{CKM}}^{ib}}^\ast \: .
	\label{eq:zdbv}
\end{equation}
Compared with Eq. (\ref{eq:occol}), it can be said that ${z_d}^{db}$
indicates the unitarity violation of CKM matrix.

\subsection{Experimental constraints on the FCNC's }
\label{subsec:fp}

We are discussing the main content of the paper. The employed 
experiments are same with ones in the SM case in the 
preceding section. In addition to the standard $K$ and $B$ mesons
processes, the constraints from $S,T,U$ parameters are also 
confirmed briefly.

\subsubsection{Tree-level FCNC dominance}
\label{subsubsec:tlfcnc}

Before making observation of the effects of tree-level FCNC's
on the CKM matrix, let us impose upper-bounds for
the mixings in the ODVQM. The constraints can be made by assuming
that the processes are dominated by $Z$ exchange tree-level FCNC
diagrams. 

The effective four Fermi interactions for the $Z$ and $W^\pm$ exchanges
tree-level processes can be generated from the interactions in Eqs.
(\ref{eq:lwsm}) and (\ref{eq:lzsqm}) and their mass terms, that is
\begin{eqnarray}
	{{{\cal L}_{eff}}^Z}^{ODVQM} & = &
		\sqrt{2} \, G_F \, {z_d}^{\alpha \beta} \,
		\left( \bar{d}^\alpha \, \gamma_\mu \, L \, d^\beta \right)
		\left[ \bar{\nu}^i \, \gamma_\mu \, L \, \nu^i 
		+ \bar{l}^i \, \gamma_\mu \, L \, l^i 
		- \frac{1}{2} \, {z_d}^{\eta \lambda} \,
		\bar{d}^\eta \, \gamma_\mu \, L \, d^\lambda
		\right] \: ,
		\label{eq:lzeff} \\
	{{\cal L}_{eff}}^{W^\pm} & = & 2 \sqrt{2} \, G_F \,
		{V_{CKM}}^{\alpha \beta}
		\left( \bar{u}^\alpha \, \gamma_\mu \, L \, d^\beta \right)
		\left( \bar{\nu_l}^i \, \gamma_\mu \, L \, l^i \right)
		+ h.c.
		\label{eq:lweff} \: \: .
\end{eqnarray}
We also give four fermi interaction for $W^\pm$ here, since most of
the FCNC processes that are considered below are normalized by FCCC
processes.

\noindent
\medskip \\
1) {\it Upper-bound for $\left| {z_d}^{ds} \right|$}

The upper-bound can be found from $K^+ \rightarrow \pi^+ \, \nu \, \bar{\nu}$
decay. This decay can be expressed in the quark subprocess by
$s \rightarrow d \, \nu \, \bar{\nu}$. This decay is usually determined
in the ratio
\begin{equation}
	\frac{{Br \left( K^+ \rightarrow \pi^+ \, \nu \,
		\bar{\nu} \right)}^{ODVQM}}{ Br \left( K^+ \rightarrow
		\pi^0 \, e^+ \, \nu_e \right) } =
	\frac{\langle \pi^+ \left| {\cal L}_{eff} \right| K^+ \rangle }{
		\langle \pi^0 \left| {{\cal L}_{eff}}^{ODVQM}
		\right| K^+ \rangle }
	= \frac{3}{2} \left(\frac{{z_d}^{ds}}{{V_{CKM}}^{us}} \right)^2 \: .
\end{equation}
for all three neutrinos in the final state. Here, we have used Eqs.
(\ref{eq:lzeff}) and (\ref{eq:lweff}), and the $SU(2)$ Clebsh-Gordan
coefficients to evaluate the hadron matrix elements.

The recent experiment results give,
\begin{eqnarray}
	{Br \left( K^+ \rightarrow \pi^+ \, \nu \, \bar{\nu} \right)}^{exp}
		& \le & 5.2 \times 10^{-9} \: , \\
	{Br \left( K^+ \rightarrow \pi^0 \, e^+ \, \nu_e \right)}^{exp}
		& = & 4.82 \times 10^{-2} \: .
\end{eqnarray}
These yield
\begin{equation}
	\left| \frac{{z_d}^{ds}}{{V_{CKM}}^{us}} \right| \le 2.7
		\times 10^{-4} \: .
		\label{eq:zsdus}
\end{equation}
Then, for $\left| {V_{CKM}}^{us} \right| $ in Eq. (\ref{eq:vus}),
the upper-bound for $\left| {z_d}^{ds} \right|$ is,
\begin{equation}
	\left| {z_d}^{ds} \right| \le 6.0 \times 10^{-5} \: .
		\label{eq:zdds}
\end{equation}

\noindent
\medskip \\
2) {\it Upper-bound for $\left| {z_d}^{qb} \right|$}

We work with $B \rightarrow X_q \, l^+ \, l^-$ decay to get the upper-bound
of $\left| {z_q}^{db} \right|$ with $q = d,s$. Because the matrix elements
of $B$ meson is not evaluated well, let us consider the decay in the
quark model of $b \rightarrow q \, l^+ \, l^- $. In order to reduce the
theoretical uncertainties from the $m_{B_q}$, the following ratio has
been usually used,
\begin{eqnarray}
	\frac{{Br \left( B \rightarrow X_q \, l^+ \, l^- \right)}^{ODVQM}}{
		Br \left( B \rightarrow X_c \, l \, \nu_l \right)}
	& \propto &
	\frac{{Br \left( b \rightarrow q \, l^+ \, l^- \right)}^{ODVQM}}{
		Br \left( b \rightarrow c \, l \, \nu_l \right)}
		\nonumber \\
	& \cong & \frac{1}{4 P({m_c}^2)}
		\left| \frac{{z_d}^{qb}}{{V_{CKM}}^{cb}} \right|^2
		\left[ \left( 1 - 2 \sin^2 \theta_W \right)^2
		+ 4 \sin^2 \theta_W \right] \: ,
	\label{eq:bqmm}
\end{eqnarray}
where $P({m_c}^2) \cong 0.55$ is the phase factor.
The detailed calculation can be seen in ref. \cite{mannel}.

Experimentally the final state has not been tagged, so we can think
that the upper-bounds for $q = d,s$ are same, i.e.
\begin{equation}
	{Br \left( B \rightarrow X_q \, \l^+ \, \l^- \right)}^{exp}
		\le 5 \times 10^{-5} \: .
\end{equation}
Because of
\begin{equation}
	{Br \left( B \rightarrow X_c \, l \, \nu \right)}^{exp} = 0.12 \: ,
\end{equation}
we obtain
\begin{equation}
	\left| \frac{{z_d}^{qb}}{{V_{CKM}}^{cb}} \right| \le 0.043 \: ,
		\label{eq:zqbcb}
\end{equation}
and
\begin{equation}
	\left| {z_d}^{qb} \right| \le 2.0 \times 10^{-3} \: ,
\end{equation}
for $\left| {V_{CKM}}^{cb} \right| $ in Eq. (\ref{eq:vcb}).

Anyway, there is also a constraint from ${B_q}^0 - \bar{B_q}^0$ mixing.
Especially from ${B_d}^0 - \bar{B_d}^0$ mixing we can expect more
precise bound, since experimentally $x_d$ has been measured precisely.
The mixing in the $Z$ tree diagram dominance is,
\begin{equation}
	{x_q}^{ODVQM} = \frac{2 \sqrt{2} G_F}{3} m_{B_q}
		\tau_{B_q} \eta_{B_q} {f_{B_q}}^2 B_{B_q}
		\left| {z_d}^{qb} \right|^2  \: .
		\label{eq:xqtl}
\end{equation}
Using the experiment results in Eqs. (\ref{eq:xdexp}) and
(\ref{eq:xsexp}), one finds
\begin{eqnarray}
	\left| {z_d}^{sb} \right| & > & 1.1 \times 10^{-3} \: ,
		\label{eq:zdsb} \\
	\left| {z_d}^{db} \right| & = & 7 \times 10^{-4} \: .
		\label{eq:zddb}
\end{eqnarray}
However, one should not trust in the bounds, since we need $CP$
violations occur in the neutral $K$ and $B$ mesons system. So we must not 
ignore the CC diagrams or take $\left| {V_{CKM}}^{tq} \right| \cong 0$ 
into account.

The above bounds show only the absolute upper-bound 
for each mixing, but are less useful itself. It should better to present them
in the ratios that correspond to the less model-dependent ratios of some
processes. This can be achieved with the help of quadrangle relation in Eq.
(\ref{eq:zv}), and of course, by using the bounds of the first and second
rows matrix elements in Sec. \ref{subsubsec:eccs} and the preceding section.
Our aim is to derive the ratios between the mixings of FCNC and
FCCC. Especially our interest is the ratio
$ \left| {{z_d}^{d_1 d_2}}/{{V_{CKM}}^{td_1} {{V_{CKM}}^{td_2}}^\ast}
\right| $, with $d_i = d,s,b$. The reasons that most of the
FCCC processes are dominated by the contribution of top exchange in the
internal line. As the first step, rewrite Eq. (\ref{eq:zv}),
\begin{equation}
	\frac{{z_d}^{d_1 d_2}}{{V_{CKM}}^{u_2 d_2}}
	= \frac{ {V_{CKM}}^{u_1 d_1} {V_{CKM}}^{u_1 d_2} }{
		{V_{CKM}}^{u_2 d_2} + {V_{CKM}}^{u_2 d_1}}
	+ \frac{{V_{CKM}}^{td_1} {V_{CKM}}^{td_2}}{{V_{CKM}}^{u_2 d_2}} \: ,
\end{equation}
where $u_i = u,c$. Further equation,
\begin{equation}
	\left| \frac{{z_d}^{d_1 d_2}}{{V_{CKM}}^{td_1} {{V_{CKM}}^{td_2}}^\ast}
	\right| = \left| \frac{{z_d}^{d_1 d_2}}{{V_{CKM}}^{u_2 d_2}} \right|
		\times \left| \frac{{V_{CKM}}^{u_2 d_2}}{
		{V_{CKM}}^{td_1} {{V_{CKM}}^{td_2}}^\ast} \right|
\end{equation}
leads to the wanted ratios, since the upper-bounds of each ratio in
r.h.s. have been determined.

Therefore, on using Eqs. (\ref{eq:vud}) $\sim$ (\ref{eq:vub}),
(\ref{eq:zsdus}) and (\ref{eq:zqbcb}), we obtain
\begin{eqnarray}
	\left.
	\begin{array}{c}
		d_1 = d \: , \: d_2 = b \\
		u_1 = u \: , \: u_2 = c
	\end{array}
	\right\} & \longrightarrow &
	\left| \frac{{z_d}^{db}}{{V_{CKM}}^{td} {{V_{CKM}}^{tb}}^\ast}
	\right| \le 2.15 \: ,
	\label{eq:ztdb} \\
	\left.
	\begin{array}{c}
		d_1 = s \: , \: d_2 = b \\
		u_1 = u \: , \: u_2 = c
	\end{array}
	\right\} & \longrightarrow &
	\left| \frac{{z_d}^{sb}}{{V_{CKM}}^{ts} {{V_{CKM}}^{tb}}^\ast}
	\right| \le 0.05 \: ,
	\label{eq:ztsb} \\
	\left.
	\begin{array}{c}
		d_1 = d \: , \: d_2 = s \\
		u_1 = c \: , \: u_2 = u
	\end{array}
	\right\} & \longrightarrow &
	\left| \frac{{z_d}^{ds}}{{V_{CKM}}^{td} {{V_{CKM}}^{ts}}^\ast}
	\right| \le 1.13 \times 10^{-3} \: .
	\label{eq:ztsd}
\end{eqnarray}
Nevertheless, in the $K$ meson system, the contribution of charm quark in
the internal line is not negligible. So we need also
the mixing ratio of ${z_d}^{ds}$ and ${V_{CKM}}^{cd} {{V_{CKM}}^{cs}}^\ast$.
This can be verified easily by using the fact that it is larger than the
top's one. Then from Eq. (\ref{eq:ztsd}),
\begin{equation}
	\left| \frac{{z_d}^{ds}}{{V_{CKM}}^{cd} {{V_{CKM}}^{cs}}^\ast}
	\right| \ll 1.13 \times 10^{-3} \: .
	\label{eq:zcsd}
\end{equation}

According to the above upper-bounds, the sizes 
of tree-level FCNC's are large enough in the $db$ sector, and 
conversely negligible in the $ds$ sector. We will use these 
ratios frequently in the next. Therefore, we can conclude here that
the tree-level FCNC's do not give significant contributions to the $K$ 
meson system. Then the constraints for the CKM matrix from the neutral 
$K$ meson system are same with the SM ones that already done in the preceding 
sections. On the other hand, the FCNC's in the $qb$ sectors 
are large enough to have significant contributions for some processes 
in the neutral $B$ meson system. This will be discussed in the next 
section. 

\subsubsection{Quadrangle relation}
\label{subsubsec:quadrangle}

\begin{figure}[t]
	\begin{center}
		\input{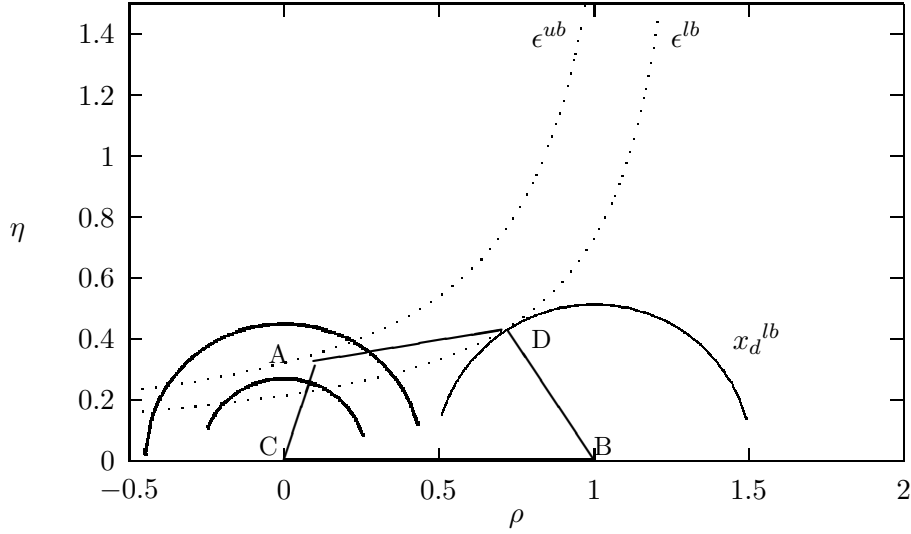}
	\end{center}
	\caption{The ideal quadrangle of the $db$ sector
		with the shortest $\left| {\rm DB} \right|$ 
		in the small FCNC case.}
	\label{fig:trisqm}
\end{figure}

According to the results in the preceding 
section, the tree-level FCNC should not contribute to the
neutral $K$ meson system in a good approximation. It means 
that ${V_{CKM}}^{ts}$, ${V_{CKM}}^{cs}$ and ${V_{CKM}}^{cd}$
are non-zero, and the unitarity in the $ds$ sector is 
conserved.
On the other hand, the matter is different in the $db$
sector, since ${z_d}^{db}$ can be comparable with 
${V_{CKM}}^{td} {{V_{CKM}}^{tb}}^\ast$. So the unitarity 
will be violated, and the relation, 
\begin{equation}
	{V_{CKM}}^{ud} {{V_{CKM}}^{ub}}^\ast + 
	{V_{CKM}}^{cd} {{V_{CKM}}^{cb}}^\ast + 
	{V_{CKM}}^{td} {{V_{CKM}}^{tb}}^\ast 
	= {z_d}^{db} 
\end{equation}
from Eq. (\ref{eq:zv}), is materialized.
This relation changes the unitary triangle of CKM matrix in
the SM to be quadrangle as depicted in Fig. (\ref{fig:trisqm}). 

However, there is also a considerable case when the above
quadrangle relation would be violated in the $db$ sector.
It means that the size of ${z_d}^{db}$ is  
not large enough such that comparable with the other terms in 
the relation. We are showing it now. 
The first task is evaluating the parametrization of the 
modified CKM matrix in Eq. (\ref{eq:ckmodsqm}).
To compare it with the SM one, it should better to adopt the
same way of Wolfenstein parametrization. 

Employ the definitions of ${z_d}^{\alpha \beta}$ in Eq. (\ref{eq:zd}),
the bounds in Eqs. (\ref{eq:ztdb}), (\ref{eq:ztsb}) and 
(\ref{eq:ztsd}), 
\begin{eqnarray}
	\left| {z_d}^{ds} \right| & = & \left| V^{d b^\prime} {V^{s b^\prime}}^{\ast}
		 \right| \: , \nonumber \\
	\left| {z_d}^{sb} \right| & = & \left| V^{s b^\prime} {V^{b b^\prime}}^{\ast}
		 \right| \: , \\
	\left| {z_d}^{db} \right| & = & \left| V^{d b^\prime} {V^{b b^\prime}}^{\ast}
		\right| \: , \nonumber
\end{eqnarray}
that yields, from the bounds in Eqs. (\ref{eq:zdds}), (\ref{eq:zdsb}) 
and (\ref{eq:zddb}), 
\begin{eqnarray}
	\left| \frac{{z_d}^{ds}}{{z_d}^{sb}} \right| & = &  
		\left| \frac{V^{d b^\prime}}{V^{b b^\prime}}
		\right| < 5.5 \times {10}^{-2} \: , 
		\label{eq:vdbb} \\
	\left| \frac{{z_d}^{db}}{{z_d}^{sb}} \right| & = &  
		\left| \frac{V^{d b^\prime}}{V^{s b^\prime}}
		\right| < 6.4 \times {10}^{-1} \: , \\
	\left| \frac{{z_d}^{ds}}{{z_d}^{db}} \right| & = &  
		\left| \frac{V^{s b^\prime}}{V^{b b^\prime}}
		\right| < 8.6 \times {10}^{-2} \: .
		\label{eq:vsbb}
\end{eqnarray}

Now, we are going on parametrizing the CKM matrix in the model.
For simplicity, without loss of generality let us assume that
\begin{enumerate}
	\item $V^{d b^\prime}$, $V^{s b^\prime}$ and $V^{b b^\prime}$ 
		are real.
	\item The $3 \times 3$ part of CKM matrix is generated 
		from the unitary matrix 
		that diagonalizes the up-quark fields and 
		normalized by $V^{b b^\prime}$ 
\end{enumerate}
Under these assumptions, the CKM matrix can be written by 
the unitary matrices $U$ and $V$ as following,
\begin{eqnarray}
	V_{CKM} & = & U \, V^\dagger \nonumber \\
	& \equiv & \left(
	\begin{array}{ccc}
		{V_{CKM}}^{ud} 	& {V_{CKM}}^{us} 	& {V_{CKM}}^{ub}\\
		{V_{CKM}}^{cd} 	& {V_{CKM}}^{cs} 	& {V_{CKM}}^{cb}\\
		{V_{CKM}}^{td} 	& {V_{CKM}}^{ts} 	& {V_{CKM}}^{tb}
	\end{array}
	\right) \times 
	\left(
	\begin{array}{cccc}
		1 & 0 & 0 & {V^{db^\prime}}/{V^{b b^\prime}} \\
		0 & 1 & 0 & {V^{sb^\prime}}/{V^{b b^\prime}} \\
		0 & 0 & 1 & 1
	\end{array}
	\right) \nonumber \\
	& \cong & \left(
	\begin{array}{cccc}
		{V_{CKM}}^{ud} 	& {V_{CKM}}^{us} 	& 
			{V_{CKM}}^{ud} \left( 
			{V^{db^\prime}}/{V^{bb^\prime}} \right)
			+ {V_{CKM}}^{ub} \\
		{V_{CKM}}^{cd} 	& {V_{CKM}}^{cs} 	& 
			{V_{CKM}}^{cs} \left( 
			{V^{sb^\prime}}/{V^{bb^\prime}} \right)
			+ {V_{CKM}}^{cb} \\
		{V_{CKM}}^{td} 	& {V_{CKM}}^{ts} 	& 
			{V_{CKM}}^{tb} \\
	\end{array}
	\right) \: , 
\end{eqnarray}
with $U$ and $V$ are defined in Eq. (\ref{eq:ckmodsqm}). 
We have kept only $3 \times 4$ part of $V^\dagger$.
On using the upper-bounds in Eq. (\ref{eq:vdbb}), the fourth column can be
written in term of $\lambda$ up to order $O({\lambda }^4)$ roughly, 
\begin{equation}
	{V_{CKM}}^{(4)} < \left(
	\begin{array}{c}
		\lambda^3 \\
		\lambda^2 \\
		1
	\end{array}
	\right) \: ,
\end{equation}
from the bounds in Eqs. (\ref{eq:vdbb}) and (\ref{eq:vsbb}).
But, from Eqs. (\ref{eq:ztsb}), (\ref{eq:ztsd}) and (\ref{eq:zcsd}) we hope that, at least,  
$ds$ sector is unitary. So it must have value with order smaller than 
$O(\lambda^4)$ to conserve the unitarity of CKM matrix in
the $ds$ sector. For the other sectors, especially $db$
sector, the unitarity are violated. When we concern a
case where the unitarities in all sectors are conserved, 
i.e. $ {V_{CKM}}^{(4)}_{all} \propto \lambda^4 $, 
we will find a new bound for CKM matrix elements
as done below.

In the case, we can adopt the parametrization in Eq.
(\ref{eq:ckmw}) for the $3 \times 3$ part of the modified CKM 
matrix in the model without any change. Then, the side 
$\left| {\rm AD} \right|$ in Fig. (\ref{fig:trisqm}) must be 
proportional to $\left| {z_d}^{db} \right|$ when 
the quadrangle is conserved. As notations,
put the angles as $\alpha^\prime$, $\beta^\prime$, $\gamma^\prime$ 
in place of $\alpha$, $\beta$ and $\gamma$ in the SM, and call 
the angle between $\vec{AD}$ and $\vec{DB}$ as $\theta_d$. 
Derivation of this angle will be given in Sec. \ref{subsubsec:mixing}.
Further, more precise bound for $\left| {V_{CKM}}^{td} {{V_{CKM}}^{tb}}^\ast \right|$ 
can be found from the figure, that is the shortest line from point B to 
the hyperbola $\epsilon^{lb}$. It gives, 
\begin{equation}
	0.513 \le \left| {\rm DB} \right|^{ODVQM} \le 1.386 \: , 
	\label{eq:dbsqm} 
\end{equation}
and 
\begin{equation}
	0.0041 \le \left| {V_{CKM}}^{td}
	{{V_{CKM}}^{tb}}^\ast \right| \le 0.0109 \: , 
	\label{eq:vtdb} 
\end{equation}
where the upper-bounds is found from ${x_d}^{ub}$ 
when $\left| {z_d}^{db} \right| \cong 0$ as derived in Eq.
(\ref{eq:abtotal}) by using Eqs. (\ref{eq:lambda}) and (\ref{eq:a}).

Hence, under the assumption that the quadrangle relation is
conserved, the size of $\left| {z_d}^{db} \right|$ must have
same order with 
$\left| {V_{CKM}}^{td} {{V_{CKM}}^{tb}}^\ast \right|$. 
However, employing the bound of $\left| {V_{CKM}}^{td} 
{{V_{CKM}}^{tb}}^\ast \right|$ in Eq. (\ref{eq:vtdb}) into
the well-measured mixing size $x_d$ will show that it can
not be realized. This will be done in the next.

\subsubsection{${B_q}^0 - \bar{B_q}^0$ mixing}
\label{subsubsec:mixing}

The new contribution for $x_q$ in the ODVQM is coming from
the tree-level $\Delta B = 2$ diagrams in Fig. (\ref{fig:b0b0}b)
(ref. \cite{morozumi}).
In fact, the $\chi^0$ and $H$ exchange diagrams must be considered
here. But, the diagram will be suppressed by a factor
${m^2}/{M^2}$ with $m$ is external line quark mass, and
$M$ is $H$ or $\chi^0$ mass.

\begin{figure}[t]
	\begin{center}
		\input{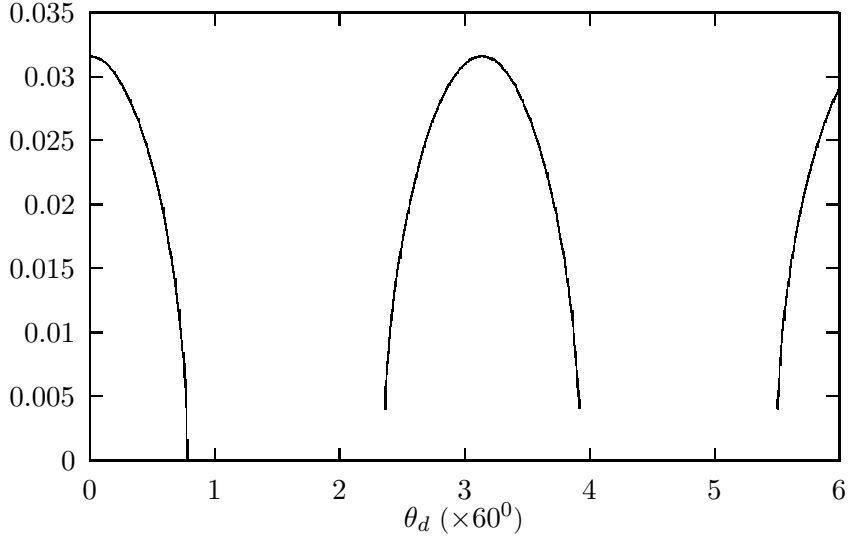}
	\end{center}
	\caption{The upper-bound for the mixing ratio 
		$ \left| {{z_d}^{db}}/{
		{V_{CKM}}^{td} {{V_{CKM}}^{tb}}^\ast} \right| $ 
		from $x_d$ in the conserved unitarity case.}
	\label{fig:mixrat}
\end{figure}
After doing a calculation for the $Z$ tree diagram as done
in Sec. \ref{subsubsec:tlfcnc}, and combining Eqs. (\ref{eq:xqsm})
and (\ref{eq:xqtl}),
\begin{eqnarray}
	{x_q}^{ODVQM} & = & \frac{{G_F}^2}{6 \pi^2} m_{B_q} {M_W}^2
		\tau_{B_q} \eta_{B_q} {f_{B_q}}^2 B_{B_q}
		\left| {V_{CKM}}^{tq} {{V_{CKM}}^{tb}}^\ast \right|^2
		\left| {F_{\Delta B = 2}}^{SM}(x_t) \right|
		\left| \Delta_q \right| \: ,
		\label{eq:xqsqm}
\end{eqnarray}
where $x_q^{ODVQM}$ is normalized by the SM contribution.
Here we adopt same notations as used in ref. \cite{morozumi},
\begin{equation}
	\Delta_q \equiv	1 + r_q e^{2 i \theta_q} \: ,
	\label{eq:deltaq}
\end{equation}
where, $\theta_q$ has been introduced in Sec. \ref{subsubsec:bdgsqm},
and
\begin{equation}
	r_q \equiv \frac{4 \sqrt{2} \pi^2}{G_F {M_W}^2}
		\frac{1}{\left| {F_{\Delta B = 2}}^{SM}(x_t) \right|}
		\left| \frac{{z_d}^{qb}}{{V_{CKM}}^{tq}
		{{V_{CKM}}^{tb}}^\ast} \right|^2 \: .
	\label{eq:rq}
\end{equation}
The absolute value of $\Delta_q$ can be written as,
\begin{equation}
	\left| \Delta_q \right| = \sqrt{1 + {r_q}^2 +
		2 \, r_q \, \cos 2 \theta_q} \: .
	\label{eq:deltaqa}
\end{equation}
Note here that the QCD correction factor $\eta_{B_q}$ is approximated
to be same for the $Z$ contribution, since QCD correction above the
scale of $M_Z$ are negligible.

The appearence of phase $\theta_d$ ($\neq {180}^0$)
reconstructs the unitary triangle 
as depicted in Fig. (\ref{fig:trisqm}). Then, Eq. (\ref{eq:xqsqm}) 
becomes 
\begin{equation}
	{x_d}^{ODVQM} = \left( 2.51 \times 10^5 \right) \times
		{f_{B_d}}^2 B_{B_d} \, A^2 \, \lambda^6 \,
		\left[ \left( 1 - \rho \right)^2 + \eta^2 \right]
		\left( 1 + r_d e^{2 i \theta_d} \right) \: ,
		\label{eq:circlesqm}
\end{equation}
with
\begin{equation}
	r_d \cong 514.48 \left| \frac{{z_d}^{db}}{
		{V_{CKM}}^{td} {{V_{CKM}}^{tb}}^\ast} \right|^2 \: ,
		\label{eq:rd}
\end{equation}
where the same values as in Eq. (\ref{eq:circle}) have been used.
We can see here that Eq. (\ref{eq:circlesqm}) is sensitive for the term
$r_d$. The reason is clear, since contribution of the SM is one order higher
than that of the ODVQM. So the contribution of the ODVQM may
be comparable with the SM one even for small 
$\left| {{z_d}^{qb}}/{{V_{CKM}}^{tq} {{V_{CKM}}^{tb}}^\ast} \right| < O(10^{-2})$.

Finally, in the case when the unitarity is conserved, 
substituting the value of Eq. (\ref{eq:dbsqm}) or 
Eq. (\ref{eq:vtdb}) into Eq. (\ref{eq:circlesqm}) gives 
the upper-bound for the 
mixing ratio. Fig. (\ref{fig:mixrat}) shows the allowed values 
for mixing ratio as a function of $\theta_d$. Then, the upper-bound is, 
\begin{equation}
	\left| \frac{{z_d}^{db}}{
		{V_{CKM}}^{td} {{V_{CKM}}^{tb}}^\ast} \right|
		\le 0.032 \: .
	\label{eq:dblast}
\end{equation}
This bound justifies the result in the preceding section
when the unitarity is nearly conserved. This will be 
used to analyze some processes with small FCNC 
in Sec. \ref{subsec:effect}.

\subsubsection{$S,T,U$ parameters}
\label{subsubsec:stu}

\begin{figure}[t]
	\epsfxsize=16cm
	\epsfysize=3cm
	\epsffile{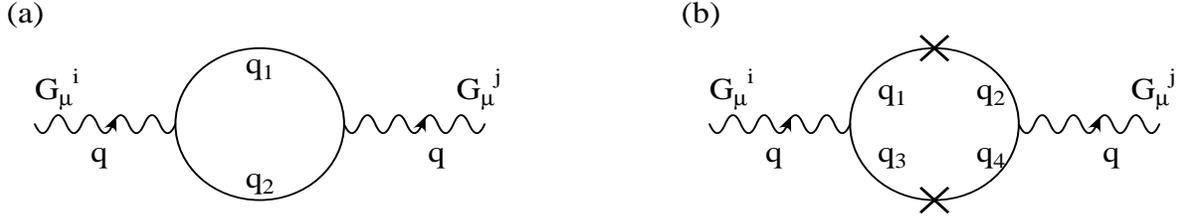}
	\caption{Vacuum polarization diagram
		responsible for the $S,T,U$ parameters,
		with ${G_\mu}^i$ and $q_i$ denote gauge bosons and
		 quarks, and $q$ denotes momentum.}
	\label{fig:stu}
\end{figure}

Before going to predict some processes in the neutral $B$ meson 
system, we will confirm the well-known $S,T,U$ parameters 
with the new effects from the ODVQM (ref. \cite{lavoura}). 
$S,T,U$ parameters indicate the effects of oblique corrections from new physics 
as depicted in Fig. (\ref{fig:stu}).
According to the original definitions in ref. \cite{peskin}, the $S,T,U$
parameters have been defined as linear combinations of gauge bosons,
${W_\mu}^i$ and $B_\mu$ ($i = 1,2,3$). The combinations are,
\begin{eqnarray}
	S & \equiv & -16 \pi \left. \frac{d}{dq^2}
		{\Pi_{3Y}}^{new}(q^2) \right|_{q^2=0} \: , \\
	\alpha T & \equiv & \frac{g^2 + {g^\prime}^2}{{M_Z}^2}
		\left[ {\Pi_{11}}^{new}(0) - {\Pi_{33}}^{new}(0)
		\right] \: , \\
	U & \equiv & 16 \pi \left. \frac{d}{dq^2} \left[
		{\Pi_{11}}^{new}(q^2) - {\Pi_{33}}^{new}(q^2)
		\right] \right|_{q^2=0} \: ,
\end{eqnarray}
where ${\Pi_{ij}}^{new}$ are two point functions of two gauge
bosons, ${G_\mu}^i$ and ${G_\mu}^j$, due to new physics.
Here $g^\prime$ and $g$ are the couplings for $U(1)$ and
$SU(2)$ gauge fields respectively.

Fortunately, in the model there is no interaction between vector-like 
quark and the $SU(2)_L$ gauge bosons, ${W_\mu}^i$. So it is clear that 
without including any mixings in the internal fermions like Fig.
(\ref{fig:stu}b), the $S,T,U$ parameters in the model are
exactly zero. It means that in the model, the $\rho$ parameter is
nearly conserved, since $\alpha T \equiv \rho - 1 \cong 0$. 

However, it may be worthwhile to concern the effects of the mixings as 
verified in ref. \cite{lavoura}. We note here that, the contributions 
of the lowest order of mixings in the internal fermions in 
Fig. (\ref{fig:stu}b) 
would be suppressed, roughly by a factor $({m_{d^i}}/{m_{b^\prime}})^2$.
The reason is simple, since according to Eq. (\ref{eq:dmass}),
the mixings between ordinary quarks $d^i$ and vector-like quark
$b^\prime$ are proportinal to $m_{d^i}$. On the other hand,
a factor $1/{{m_{b^\prime}}^2}$ appears from the heavy
fermion propagators. 
So, we can say that the model would not give significant alteration 
for $S,T,U$ parameters. Then we could not expect more strict bounds 
for the sizes of FCNC's from the parameters. 

\subsection{The effects of tree-level FCNC's on the neutral $B$ meson 
system}
\label{subsec:effect}

The next task after deriving the bounds for each mixing in the preceding 
section, is reevaluating some processes in the neutral $B$ meson system 
in the model. We will give predictions for $B \rightarrow X_d \, \gamma$ 
(refs. \cite{handoko2} and \cite{gautam}), 
$B \rightarrow X_d \, l^+ \, l^- $ and the $CP$ violations in the 
neutral $B$ meson system (ref. \cite{morozumi}).
All of processes has not been measured yet. 

Especially, the $B \rightarrow X_d \, l^+ \, l^- $ and the $CP$ violations 
in the neutral $B$ meson system are appropriate processes to verify the model. 
The reason is, in the processes significant tree-level
FCNC's contributions are expected, since the new
contributions appear as $Z$ exchange tree diagrams. So, 
the contribution will be one order larger 
than the CC mediated diagrams' ones. 
On the other hand, the situation is different in the 
$B \rightarrow X_d \, \gamma$ process, since all of
contributions from 
CC mediated and NC mediated diagrams, appear in the one-loop level. 
Therefore, we need a mixing ratio with order $O(1)$ to make the NC 
contributions to be comparable. But, the $B \rightarrow X_d \, \gamma$ is 
still important to search the mass of down-type vector-like quark, since 
the heavy vector-like quark contributes in the internal line of NC 
mediated diagrams. 
Unfortunately, when the upper-bound for the mixing ratio in the 
$db$ sector is suppressed, i.e. $\sim O({10}^{-2})$ (Eq. (\ref{eq:dblast})), 
the tree-level FCNC contribution should not be visible in the 
$B \rightarrow X_d \, \gamma$.

Experimentally, the rate of $B \rightarrow X_d \, \gamma$ is expected 
to be larger than $B \rightarrow X_d \, l^+ \, l^- $, roughly 
by a factor with order $1/\alpha \cong O(10^2)$. It has consequence that 
$B \rightarrow X_d \, \gamma$ is easier to observe. 
So, for the present we will also give the predictions for
the decay.
\begin{figure}[t]
	\begin{center}
		\input{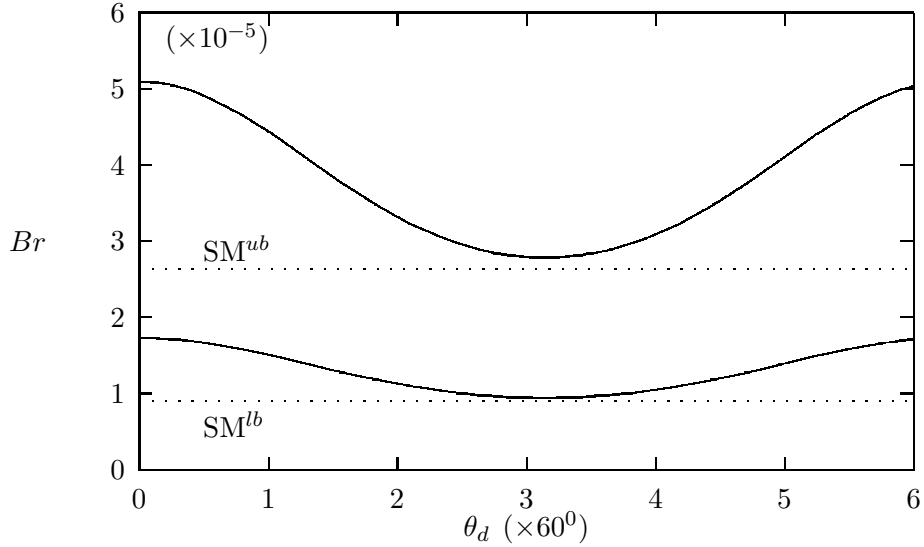}
	\end{center}
	\caption{The branching-ratio for the inclusive 
		$B \rightarrow X_d \, \gamma$ decay as a function
		of $\theta_d$.}
	\label{fig:bdgu}
\end{figure}

\subsubsection{$B \rightarrow X_d \, \gamma$}
\label{subsubsec:bdgsqm}

Similar with the SM case, here we assumpt the quark subprocess
$b \rightarrow d \, \gamma$ decay. The new contributions that appear 
in the ODVQM, are mainly from the $Z$ and $\chi^0$ mediated penguin diagrams 
with down-quarks in the internal line.
\begin{figure}[t]
	\begin{center}
		\input{bdgm.plt}
	\end{center}
	\caption{The branching-ratio for the inclusive 
		$B \rightarrow X_d \, \gamma$ decay as a function
		of $m_{b^\prime}$.}
	\label{fig:bdgm}
\end{figure}

Adopt the result in Appendix \ref{app:bqgsqm}, the branching
ratio is predicted as,
\begin{eqnarray}
	{Br(b \rightarrow d \gamma)}^{ODVQM} & \cong &
		\frac{\alpha {G_F}^2}{128 \pi^4} {m_b}^5 \, \tau_B \,
		{Q_u}^2
		\left| {V_{CKM}}^{td} {{V_{CKM}}^{tb}}^\ast \right|^2 \;
		\left| {F_{bd\gamma}}^{ODVQM}(m_b) \right|^2
		\nonumber \\
	& = &	0.252 \times A^2 \, \lambda^6 \,
		\left[ \left( 1 - \rho \right)^2 + \eta^2 \right]
		\left| {F_{bd\gamma}}^{ODVQM}(m_b) \right|^2 \: ,
\end{eqnarray}
where the QCD corrected function ${F_{bd\gamma}}^{ODVQM} (m_b)$ is given
in Appendix \ref{app:bqgsqm},
\begin{eqnarray}
	{F_{bd\gamma}}^{ODVQM}(m_b) & \equiv & F_L^T(m_b)
		\cong F_R^T(m_b) \nonumber \\
	& \cong & 0.67 \left( F - 0.42 F^g -
		0.88 \right) \: ,
\end{eqnarray}
because approximately the values of $F$ and $F^g$ are the following,
\begin{eqnarray}
	F & \equiv & F_L \cong F_R \nonumber \\
	& \sim & -0.586 1 + \frac{Q_d}{Q_u}
		\left| \frac{{z_d}^{db}}{{V_{CKM}}^{td} {{V_{CKM}}^{tb}}^\ast}
		\right| e^{i \theta_d}
		\left[ 1 + \frac{2}{3} \times 0.234 \times (-1.667)
		\right. \nonumber \\
	& & \left.
		+ \left( {z_d}^{dd} + {z_d}^{bb} \right) 0.333
		- \left| V^{b^\prime b^\prime} \right|^2
		\left( F_2^{NC}(r_{b^\prime},w_{b^\prime}) +
		F_3^{NC}(r_{b^\prime}) \right)	\right] \: ,
		\label{eq:ffin} \\
	F^g & \equiv & F_L^g \cong F_R^g \nonumber \\
		& \sim & -0.097 + \frac{Q_d}{Q_u}
		\left| \frac{{z_d}^{db}}{{V_{CKM}}^{td} {{V_{CKM}}^{tb}}^\ast}
		\right| e^{i \theta_d}
		\left[ \frac{5}{3} + \frac{2}{3} \times 0.234 \times (-1.667)
		\right. \nonumber \\
	& & \left.
		+ \left( {z_d}^{dd} + {z_d}^{bb} \right) 0.333
		- \left| V^{b^\prime b^\prime} \right|^2
		\left( F_2^{NC}(r_{b^\prime},w_{b^\prime}) +
		F_3^{NC}(r_{b^\prime}) \right)	\right] \: ,
		\label{eq:fgfin}
\end{eqnarray}
where $ \theta_d \equiv {\rm arg} \left( {z_d}^{db}/{{V_{CKM}}^{td}
{{V_{CKM}}^{tb}}^\ast} \right)$ and $m_t = 174$(GeV). 
The existence of the mixing ratio
in the NC parts makes the contribution to be suppressed  for
small mixing ratio in Eq. (\ref{eq:dblast}).
We note here that the function ${F_{bd\gamma}}^{ODVQM}$
has dependence on the down-type vector-like quark mass $m_{b^\prime}$ due to
the functions $F_2^{NC}$ and $F_3^{NC}$ in Appendix (\ref{app:bqgsqm}).

We give the figures for the branching-ratio in Fig.
(\ref{fig:bdgu}) for $m_{b^\prime} = 2$(TeV), 
and Fig. (\ref{fig:bdgm}) for $\theta_d = 0^0, 180^0$.
Here, $\left| {z_d}^{d^\alpha d^\alpha} \right| = 1$, 
$m_H = 1$(TeV), mixing ratio $= 2$, and 
$\left| {V_{CKM}}^{td} {{V_{CKM}}^{tb}}^\ast \right|$ as in 
Eq. (\ref{eq:vtdb}).
Anyway, the prediction of the SM for the branching ratio is,
\begin{equation}
	8.971 \times 10^{-6} \le {Br}^{SM} \le 2.641 \times 10^{-5} \: .
\end{equation}

\subsubsection{$B \rightarrow X_d \, l^+ \, l^- $}

\begin{figure}[t]
	\begin{center}
		\input{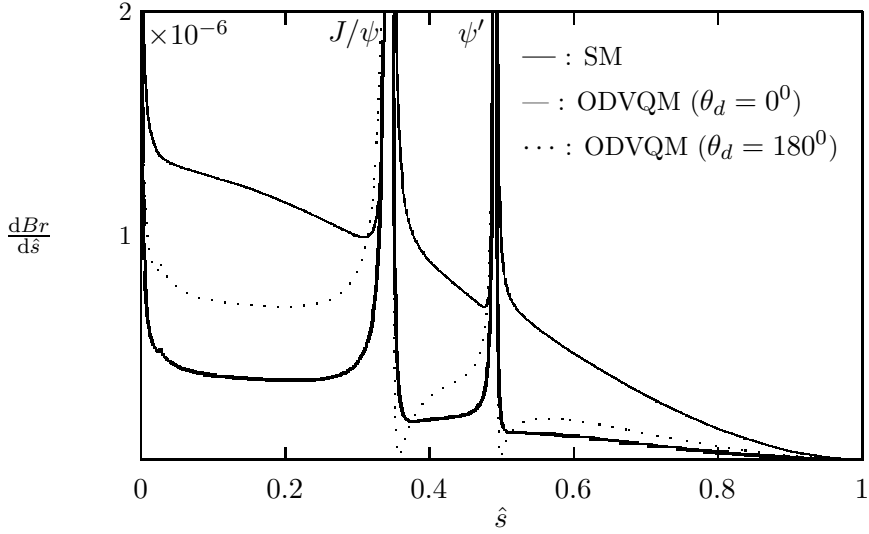}
	\end{center}
`	\caption{The differential branching ratio for 
		$B \rightarrow X_d \, l^+ \, l^- $ as a 
		function of the scaled invariant dilepton mass 
		($\hat{s} \equiv {m^2}/{m_b}^2 $). }
	\label{fig:bdll}
\end{figure}
As stated before, the $B \rightarrow X_q \, l^+ \, l^- $
decay is an appropriate test for the model, even for the tiny 
mixing ratio. Because, in the model the decay also occurs at 
$Z$ mediated tree diagram, as well as one-loop level CC 
diagrams of off-shell $b \rightarrow q \, \gamma$ as explained in 
Appendix \ref{app:bqg}. Note that, 
the one-loop NC diagrams should not give significant
contribution for small mixing ratio as shown in the 
inclusive $B \rightarrow X_d \, \gamma $ decay.

The detailed calculation for the CC diagrams contribution 
can be seen in ref. \cite{mannel}. In the reference,  
the calculation have been done including the effects 
of resonances of $J/\psi$ and $\psi^\prime$. 
In the small FCNC case, we just need to concern the 
contribution from $Z$ 
mediated tree diagram that the result has been given in Eq. 
(\ref{eq:bqmm}). The mixing ratio appears again if we 
combine both contributions, and normalize them by 
the CC contribution.
Since the equation is too complicated,  
we do not write it again in the paper. 
The readers are expected to cite the reference for 
the calculation of CC diagrams. 
Here, we just give a figure that shows the effects of 
tree-level FCNC in the differential branching ratio 
of the $B \rightarrow X_d \, l^+ \, l^- $ decay including 
long distance effects due to $J/\psi$ and $\psi^\prime$ 
resonances. 
We have put $m_t = 174$(GeV), mixing ratio $\sim 0.025$,
and the CKM matrix elements as in Sec. \ref{subsubsec:eccs}
in the figure. 
The values of the other parameters are same with 
ref. \cite{mannel}.

\subsubsection{{\it CP} violations in ${B_q}^0 - \bar{B_q}^0$ mixing}

\begin{figure}[t]
	\begin{center}
		\input{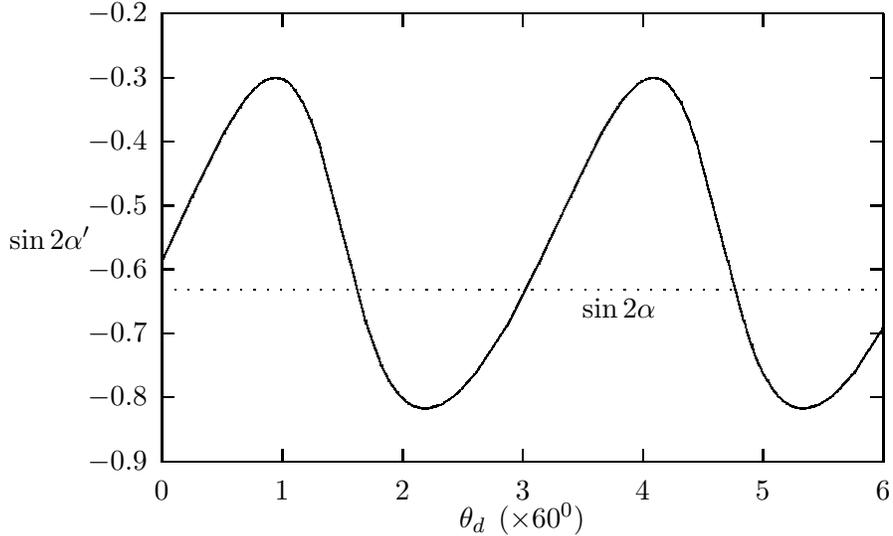}
	\end{center}
	\caption{The dependence of $\sin 2 \alpha^\prime$
		on the $\theta_d$. }
	\label{fig:sinta}
\end{figure}
The effects of vector-like quark in the $CP$ violated decays in
Sec. \ref{subsubsec:ecdbs} appear in the mixing phase
in ${B_q}^0 - \bar{B_q}^0$ mixing. As done in ref. \cite{morozumi},
the tree $Z$ exchange diagram generates new mixing phase from
Eq. (\ref{eq:deltaq}), that is
\begin{equation}
	{\rm arg} \: \Delta_q = {\rm tan}^{-1} \left(
		\frac{r_q \, \sin 2 \theta_q}{1 + r_q \,
		\cos 2 \theta_q} \right) \: .
	\label{eq:argq}
\end{equation}
Therefore, the mixing phase $\phi_f$ in Sec. \ref{subsubsec:ecdbs},
should be changed as,
\begin{equation}
	{\phi_f}^\prime \equiv \phi_f + {\rm arg} \: \Delta_q
	\label{eq:phisqm}
\end{equation}
in the ODVQM, with $r_q$ is given in Eq. (\ref{eq:rq}).

\begin{figure}[t]
	\begin{center}
		\input{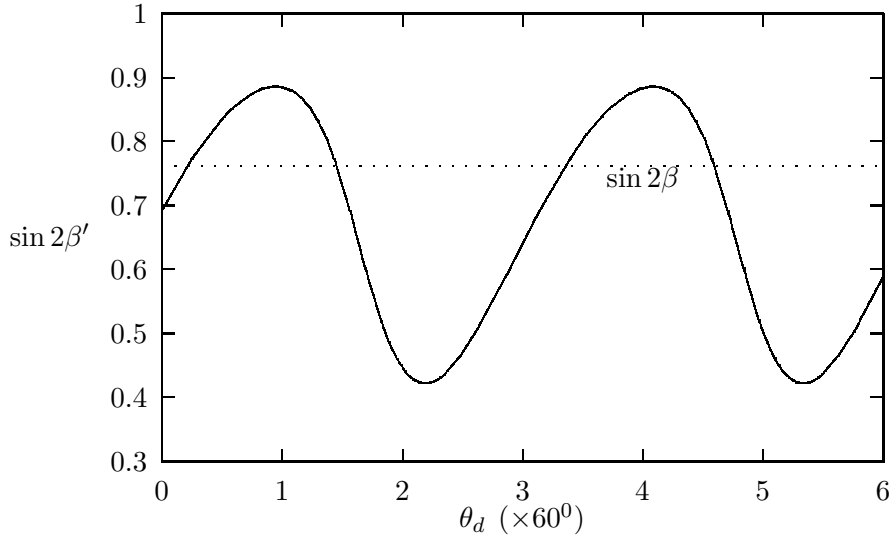}
	\end{center}
	\caption{The dependence of $\sin 2 \beta^\prime$
		on the $\theta_d$. }
	\label{fig:sintb}
\end{figure}
As discussed in the preceding section, the contributions of
FCNC's are significant for the angles, even for the sizes
are small. The reason is clear, since the mixing ratio in
Eq. (\ref{eq:rd}) is multiplicated by a factor with order
$O(10^2)$. The same situation can be said for $q = s$
and its mixing ratio in Eq. (\ref{eq:ztsb}).
The dependence of the angles on the $\theta_d$ has been discussed in
ref. \cite{morozumi}, and the figures are given again in 
Figs. (\ref{fig:sinta}) and (\ref{fig:sintb}). 
The figures are drawn by using Eq. (\ref{eq:argq}). 
In the figures, we have put $r_d \cong 0.32$ for mixing ratio 
$\sim 0.025$. The values of the angles in the SM are 
$\sin 2 \alpha = -0.59$ and $\sin 2 \beta = 0.69$ 
for the most reliable pair of $(\rho,\eta)$, that is $\rho = 0.25$ 
and $\eta = 0.3$. From this equation and the figures, it is clear 
that the contribution of tree-level FCNC may be constructive or
destructive depend on the $\theta_d$. The SM results will be
restored again at $\theta = n \pi $ with $n$ is any arbitrary
integer number.

\subsection{Theoretical studies on the FCNC}
\label{subsec:tsfcnc}

In this subsection, we will show that there are considerable constraints due
to the features of mass matrix in the VQM. The idea started from the
possibility of deriving the mass differences in the quark sector
by introducing the vector-like quarks. The reason that, the sizes of FCNC's,
which are expressed as $z_{u(d)}$, have been related with the unitary
matrices which diagonalize the mass terms in the Yukawa sector.
The relations have been given in Eqs. (\ref{eq:zd}) and (\ref{eq:zu}).
Unfortunately, the form of these unitary matrices are not unique.
Then, the sizes of FCNC's will depend on the assumptions
which have been made in generating the matrices.
Here we will give the theoretical studies for the FCNC's
in the down-quark sector ($z_d$), since our interest in this paper
is the triangle in the $db$ sector. Let the uniqueless of the matrices
as an outstanding problem that must be solved in the future.
However, the studies will support the phenomenological
result in Sec. \ref{subsubsec:quadrangle}, i.e. the case
when the FCNC's are small.

From the results in Appendix \ref{app:mm}, we can divide the constraints
into two classes, that is
\begin{enumerate}
  	\item Theoretical constraints which make the
  		FCNC's to be exactly zero (refs. \cite{handoko2} 
		and \cite{handoko1}).
  	\item Theoretical constraints which give non-zero FCNC's
  		(ref. \cite{handoko3}).
\end{enumerate}
For the detailed derivations, the class (1) is given in the Appendix
\ref{app:nsfcnc}, and Appendix \ref{app:nefcnc} for the class (2).

In order to show that, let us start with the following down-type
mass matrix without loss of generality,
\begin{equation}
	M_d^0  = \left(
	\begin{array}{cccc}
		{m_d}^0		& 0		& 0		& 0 \\
		0		& {m_s}^0	&  0 		& 0 \\
		0		& 0		& {m_b}^0	& 0 \\
		J_d		& J_s		&  J_b		&
			{m_{b^\prime}}^0
	\end{array}
	\right) \: \: ,
\end{equation}
where $m^0$ is the mass parameters in the weak basis.
It can be shown that with an appropriate choice of weak basis, we can always
transform the down-type mass matrix in this particular form as has been proved
in Eq. (\ref{eq:mdo}). To derive the unitary matrix $V$ which diagonalizes
the down-quark mass, we need to analyze the following matrix,
\begin{equation}
	{M_d}^0 {{M_d}^0}^{\dagger} = \left(
	\begin{array}{cccc}
		{{m_d}^0}^2	& 0	& 0	& {m_d}^0 {J_d}^{\ast} \\
		0	& {{m_s}^0}^2	& 0 	& {m_s}^0 {J_s}^{\ast} \\
		0	& 0	& {{m_b}^0}^2	& {m_b}^0 {J_b}^{\ast} \\
		{m_d}^0 J_d	& {m_s}^0 J_s	& {m_b}^0 J_b	& M^2
	\end{array}
	\right) \: ,
	\label{eq:mdmd}
\end{equation}
where, $M \equiv \sqrt{{m_{b^\prime}}^2 + \left| J_d \right|^2 +
\left| J_s \right|^2 + \left| J_b \right|^2 }$.
The necessary condition to have non-zero ${z_d}^{ij}$ is the presence of
$J_i$ and $J_j$. But, this is not sufficient condition here as occured
in the class (1).

\subsubsection{Natural suppression of FCNC's}
\label{subsubsec:nsfcnc}

First, let see the class (1). This case is equivalent with the simplest form
of Eq. (\ref{eq:mdmd}), that is
\begin{equation}
	{M_d}^0 {{M_d}^0}^{\dagger} = \left(
	\begin{array}{cccc}
		{m^0}^2	& 0	& 0	& m^0 J^{\ast} 	\\
		0	& {m_s^0}^2	& 0 	& 0 	\\
		0	& 0	& {m^0}^2	& m^0 J^{\ast} \\
		m^0 J	& 0	& m^0 J	& 2 \left| J \right|^2 + {m^0}^2
	\end{array}
	\right) \: \: ,
	\label{eqn:mmd}
\end{equation}
by supposing that $J_d = J_b \equiv J$, $J_s = 0$ and
${m_d}^0 = {m_b}^0 = {m_{b^\prime}}^0 \equiv m^0$.
In this case, we concentrate on the FCNC between $d$ and $b$ quarks
because of the presence of $J_d$ and $J_b$. Moreover,
at $m^0 \ll J$, we find a unitary matrix $V$ which diagonalizes
Eq. (\ref{eq:mdmd}) from Eq. (\ref{eq:vdmat}),
\begin{equation}
	V \cong \frac{1}{\sqrt{2}} \left(
	\begin{array}{cccc}
		-1 & 0 & -1 &
		\frac{2 m^0 J^{\ast}}{2 \left| J \right|^2} \\
		0 & \sqrt{2} & 0 & 0 \\
		1 & 0 & -1 & 0 \\
		\frac{\sqrt{2} m^0 J}{2 \left| J \right|^2} & 0
		& \frac{\sqrt{2} m^0 J}{2 \left| J \right|^2} & \sqrt{2}
	\end{array}
	\right) .
	\label{eq:vmat}
\end{equation}
Under the above approximation, the mass eigenvalues are found by the
diagonalized ${M_d}^0 {{M_d}^0}^{\dagger}$,
\begin{eqnarray}
	M_d \, {M_d}^{\dagger} & \equiv &
		V \, {M_d}^0 \, {{M_d}^0}^{\dagger} \, V^{\dagger}
		\nonumber \\
	& \equiv & {\rm diag} \left({m_d}^2 \: , \: {m_s}^2 \: ,
		\: {m_b}^2 \: , \: {m_{b^\prime}}^2 \right)
		\nonumber \\
	& \cong & {\rm diag} \left(
		 \frac{{m^0}^4}{2 \left| J \right|^2}
		 \: , \: {{m_s}^0}^2 \: , \: {m^0}^2 \: ,
		 \: 2 \left| J \right|^2 \right) \: .
	\label{eq:diam}
\end{eqnarray}
Here we denote $m_{\alpha}$ ($\alpha = d,s,b,b^\prime$) as the physical mass.

Finally, under the assumption which are considered,
for example the mixing ${z_d}^{db}$ vanishes since from Eqs. (\ref{eq:vmat})
and (\ref{eq:zd}), $ \left| {z_d}^{db} \right| = \left| V^{db^\prime}
{V^{bb^\prime}}^{\ast} \right| = 0 $. However, there are some interesting
points in Eq. (\ref{eq:diam}) ,
\begin{enumerate}
	\item Despite of assuming the same masses for $d$, $b$ and
		$b^\prime$ quarks first, we can derive the mass differences
		of the physical	masses by introducing only two parameters
		$m^0$ and $J$ in the lagrangian.
	\item There is a constraint on the ratio of the mass mixing $J$ and
		the theoretical mass $m^0$, that is
		\begin{equation}
			\frac{2 \left| J \right|^2}{{m^0}^2}
			\cong \frac{{m_b}^2}{{m_d}^2} \sim 2.5 \times 10^{5}
			\: . \label{eq:jm}
		\end{equation}
		Thus, $\left| J \right| \cong 1.77$(TeV).
	\item From the above value of $\left| J \right|$, the mass of the
		down-type vector-like quark can be predicted to be
		$m_{b^\prime} \cong 2.5$(TeV).
\end{enumerate}

\subsubsection{Natural existence of FCNC's}

In this section, we will show the case where theoretically
$z_{u(d)}$ are non-zero and may be enhanced in many cases.
The same approach with Appendix \ref{app:nsfcnc} will be done.
The difference is just the assumptions which have been made to
generate the eigenvalues and eigenvectors.

We start with an assumption that $J_s = 0$. In fact, this
assumption is needed to make the the mass matrix in Eq.
(\ref{eq:mdmd}) to be simply solved, since we do not need
solve a four by four matrix.
\begin{equation}
	{M_d}^0 {{M_d}^0}^{\dagger} = \left(
	\begin{array}{cccc}
		{{m_d}^0}^2	& 0	& 0	& {m_d}^0 {J_d}^{\ast} \\
		0	& {{m_s}^0}^2	& 0 	& 0 \\
		0	& 0	& {{m_b}^0}^2	& {m_b}^0 {J_b}^{\ast} \\
		{m_d}^0 J_d	& 0	& {m_b}^0 J_b	& M^2
	\end{array}
	\right) \: .
\end{equation}
Here, $M^2 \equiv \sqrt{{m_{b^\prime}}^2 + \left| J_d \right|^2
+ \left| J_b \right|^2}$.
However, even for under this assumption we still have difficulty
solving the mass matrix exactly.
Here, further simplification can be made by making the following
assumptions. Suppose that $m_i \ll M$, so the mass matrix can be
expanded in term of ${m_i}/M$. Then, let us consider two radical
cases,
\begin{enumerate}
  	\item ${m_d}^0 \approx {m_b}^0$, and
  	\item ${m_d}^0 \ll {m_b}^0$.
\end{enumerate}

In the case (1), we find the mass eigenvalues from
Appendix \ref{app:nefcnc} as,
\begin{equation}
	M_d \, {M_d}^{\dagger} \cong {\rm diag} \left(
		\frac{{m^0}^2 {m_{b^\prime}}^2}{M^2} \left[
		1 - 2 r \frac{\left| J_b \right|^2 - \left| J_d \right|^2}{
		\left| J_b \right|^2 + \left| J_d \right|^2} \right] \: , \:
		{{m_s}^0}^2 \: , \: {m^0}^2 \left[
		1 - 2 r \frac{\left| J_b \right|^2 - \left| J_d \right|^2}{
		\left| J_b \right|^2 + \left| J_d \right|^2} \right]^{-1}
		\: , \: M^2 \right) \: ,
\end{equation}
where $m \equiv \frac{1}{2} \left( {m_b}^0 + {m_d}^0 \right)$ and
$r \equiv {\left( {m_b}^0 - {m_d}^0 \right)}/{(2 m)}$.
Moreover, from the eigenvectors the mixing in $db$ sector is obtained,
\begin{equation}
	\left| {z_d}^{db} \right| \cong r \:
	\frac{m_d m_b}{{m_{b^\prime}}^2} \frac{J_d J_b}{{m_{b^\prime}}^0 M}
	\frac{2 \left| J_d \right|^2 \left( {m_{b^\prime}}^2 + \left| J_d
	\right|^2 \right) + \left| J_b \right|^2 \left( 2 \left| J_d
	\right|^2 - \left| J_b \right|^2 \right) }{\left| J_d \right|^2
	\left| J_b \right|^2} \: .
	\label{eq:zdbr}
\end{equation}
From Eq. (\ref{eq:zdbr}), it is easily understood that the discussion in the
Sec. \ref{subsubsec:nsfcnc} is contained, since it corresponds to $r = 0$.
Thus, in this case the existence of FCNC has a sensitibility for the size
of $r$.

In the second case, the mass eigenvalues are found as the following,
\begin{equation}
	M_d \, {M_d}^{\dagger} \cong {\rm diag} \left(
	\frac{{{m_d}^0}^2 {{m_{b^\prime}}^0}^2}{
	{{m_{b^\prime}}^0}^2 + \left| J_d \right|^2} \: , \:
	{{m_s}^0}^2 \: , \:
	\frac{{{m_b}^0}^2 \left( {{m_{b^\prime}}^0}^2 + \left| J_d \right|^2
	\right)}{M^2} \: , \:
	M^2 \right) \: .
\end{equation}
The size of FCNC should be,
\begin{equation}
	\left| {z_d}^{db} \right| \cong r \:
	\frac{m_d m_b}{{m_{b^\prime}}^2} \frac{J_d J_b}{{m_{b^\prime}}^0 M}
	\left( 1 + \frac{\left| J_b \right|^2}{{{m_{b^\prime}}^0}^2
	+ \left| J_d \right|^2 } \right) \: .
\end{equation}
Therefore, the size of FCNC could be enhanced by adjusting the parameters
$J_d$, $J_b$ and ${m_{b^\prime}}^0$ under a condition that down-type
vector-like quark mass should be $m_{b^\prime} \ge 46$(GeV)
from $e^+ \, e^-$ collider and $m_{b^\prime} \ge 86$(GeV)
from hadron collider (ref. \cite{data}).

\clearpage
\section{SUMMARY AND CONCLUSIONS}

We have restudied the unitarity of CKM matrix in the SM
by employing profitably the triangle that is consequences
of the orthogonal conditions in Sec. \ref{sec:fpsm}. The study has
been concentrated on the triangle of $db$ sector. To evaluate it 
the main processes and the present results of 
experiments have been used. 

Further evaluation have been done to examine the effects of tree-level 
FCNC. The situation is realized in VQM model in Sec. \ref{subsec:sqm}. 
In relation to the $db$ sector, we have concerned a special case of 
the model, where only one down-type vector-like quark has been added. 
In the model, the unitary triangle has been changed to be quadrangle. 
From the upper-bound of each mixing under the assumption that the tree-level 
FCNC's are dominant, we have found that the neutral $K$ meson processes 
would not be influenced. But the matter is different in the
neutral $B$ meson system.
However, in the case when the FCNC is small, 
more precise bound has been found for the mixing ratio in the $db$ 
sector from the ${B_d}^0 - \bar{B_d}^0$ mixing 
and the lower-bound of $\epsilon$, i.e. mixing ratio $\sim O(10^{-2})$. 
This has consequence that the unitary triangle in the $db$
sector is also nearly conserved, that is
\begin{equation}
	{V_{CKM}}^{ud} {{V_{CKM}}^{ub}}^\ast + 
	{V_{CKM}}^{cd} {{V_{CKM}}^{cb}}^\ast + 
	{V_{CKM}}^{td} {{V_{CKM}}^{tb}}^\ast 
	= {z_d}^{db} \cong 0 \: ,  
\end{equation}
as well as the unitary triangles in the $ds$ and $sb$ sectors.
Additionally, by a rough dimension analysis we found that
$S,T,U$ parameters give no significant constraints 
for the tree-level FCNC's in the model.

In the last, we made some predictions for not yet measured processes 
in the neutral $B$ meson system. On using the allowed values of 
the the mixing ratio in the case when the FCNC is
small, the predictions for 
$B \rightarrow X_d \, l^+ \, l^-$ and $CP$ asymmetries in the 
neutral $B$ meson system have been made. The significant contributions 
on these processes are expected. 
On the other hand, a prediction for inclusive $B \rightarrow X_d \, \gamma$
decay has been made for large FCNC case. The important point is that
this process has dependence on the down-type vector-like 
quark mass. Lastly, it can be said that, the present status of
experiments still approve the possibilities of new physics beyond
the SM, in this case is ODVQM. 

In addition to the experimental constraints, we argued the relations
between the sizes of FCNC's and the down-quark mass matrices. From these 
relations, there are some considerable theoretical constraints depends 
on the form of down-quark mass matrix. As typical examples, we have 
concerned two radical cases, when the FCNC's are exactly vanish, and 
the small FCNC's are exist. In either cases the results
suggest that the sizes of FCNC's must be small.
However, the relations are not unique, so it is less 
powerful to use them seriously when we examine the triangle
in the ODVQM.
On the other hand, the approaches may give an interesting mechanism,
for example to generate the mass differences of fermions.

The present paper, however discuss a part of Yukawa sector. 
Next studies for the other sectors of unitary triangles, especially for 
the up-quarks sector are expected. We are looking forward to 
the progress of experiments on the up-quark contained mesons ($D$ meson, etc), 
since the heavy up-type vector-like quarks will have significant 
contributions on the processes.

\noindent
\bigskip \\
\addcontentsline{toc}{section}{\protect\numberline{}{\bf ACKNOWLEDGMENTS}}
{\large \bf ACKNOWLEDGMENTS}

We would like to thank T. Morozumi for his guidance and collaboration
in this research,
and also to T. Muta for his assistance and useful discussion during the period
of our attendance at Elementary Particle Physics Laboratory -
Hiroshima University. Moreover, we feel grateful to all of our laboratory
members, especialy for J. Kodaira, T. Onogi, T. Goto, S.
Hashimoto, T. Inagaki and
T. Yoshikawa for their private teachings, and E. Nakamoto for her warmth
assistance.
The study of the author is supported by a grant from
Overseas Fellowship Program, BPPT-Indonesia.

\clearpage
\noindent
\addcontentsline{toc}{section}{\protect\numberline{}{\bf APPENDIX}}
{\large \bf APPENDIX}
\appendix

\section{THE INTERACTIONS IN THE SM}
\label{app:sml}

We will not reproduce all of the terms of lagrangian here. The aim of this
appendix is to appreciate how flavor changing transitions
emerge in the VQM. More detailed procedure could be seen in ref.
\cite{cheng} and references therein.

Through the calculations of the lagrangians in the Appendices
(\ref{app:sml}) and (\ref{app:sqml}), we use the following notations,
\begin{eqnarray}
	\psi_L 	& : & {\rm doublet} \: \: {\rm fermions}
		\nonumber \\
	q 	& : & {\rm singlet} \: \: {\rm fermions}
		\nonumber \\
	D_\mu	& \equiv & \partial_\mu - i g T^a {W_\mu}^a
			- i g_s Y B_\mu \: ,
		\label{eq:dc} \\
	{W_{\mu \nu}}^a & \equiv & \partial_\mu {W_\nu}^a
			- \partial_\nu {W_\mu}^a
			+ g \epsilon^{abc} {W_\mu}^b {W_\nu}^c \: ,
	\label{eq:wb} \\
	B_{\mu \nu} & \equiv & \partial_\mu B_\nu - \partial_\nu B_\mu
		\label{eq:u1b} \\
	\tilde{\phi} & \equiv & \left(
		\begin{array}{c}
			\frac{1}{\sqrt{2}} \left( v + {\phi^0}^{\ast}
			\right) \\
			{-}\chi^{-} \\
		\end{array}
		\right) \: ,
	\label{eq:bb} \\
	\phi^0 & \equiv & H + i \, \chi^0 \: .
	\label{eq:p0}
\end{eqnarray}
Here, $T^a \equiv \frac{1}{2} \sigma^a$ are three Pauli matrices
($a = 1,2,3$), $Y$ denotes isospin, $g$ is $SU(2)$ coupling constant,
$g^\prime$ is $U(1)$ coupling constant, and
$e \equiv g \sin \theta_W = g^\prime \cos \theta_W$
with $\theta_W$ is Weinberg angle. Meanwhile,
$\tilde{\phi} \equiv i \, \, T_2 \phi^\ast$ and $\chi^{-} \equiv
\left( \chi^+ \right)^\ast$. Weak eigenstates are denoted by
$\tilde{\psi}, \tilde{q}$ and so on.

\subsection{The fermion and gauge sectors}

The kinetic term for the fermions is,
\begin{equation}
	{\cal L}_F = i \, \bar{\tilde{q}} \, \gamma^\mu \,
	D_\mu \, \tilde{q} \: ,
\end{equation}
where $D_\mu$ is defined in Eq. (\ref{eq:dc}).

After making redefinition of the gauge fields in order to get the
physical masses, the explanation has been omitted here,
it can be rewritten as,
\begin{eqnarray}
	{\cal L}_{NC} & = & e \, J_Q^\mu \, A_\mu +
		\frac{g}{\cos \theta_W} J^\mu Z_\mu \: ,
	\label{eq:nclsm} \\
	{\cal L}_{CC} & = & \frac{g}{\sqrt{2}}
		\left( {J^\mu}^+ {W_\mu}^+ + {J^\mu}^- {W_\mu}^- \right)
		+ h.c. \: .
	\label{eq:cclsm}
\end{eqnarray}
Here, ${J_Q}^\mu$, $J^\mu$ and ${J^\mu}^\pm$ are currents for neutral
and charge change interactions, $Q \equiv T_3 + Y $, and
\begin{eqnarray}
	{W_\mu}^\pm & \equiv & \frac{1}{\sqrt{2}} \left(
		{W_\mu}^1 \mp i {W_\mu}^2 \right) \: , \\
	A_\mu	& \equiv & \sin \theta_W \, {W_\mu}^3 +
		\cos \theta_W \, B_\mu \: , \\
	Z_\mu	& \equiv & \cos \theta_W \, {W_\mu}^3 -
		\cos \theta_W \, B_\mu \: , \\
	{J_Q}^\mu & \equiv & Q_q \bar{\tilde{q}} \, \gamma^\mu \, \tilde{q}
		\: , \label{eq:asm} \\
	J^\mu	& \equiv & \bar{\tilde{q}} \, \gamma^\mu \left[
		\left( T_3 - Q_q \sin^2 \theta_W \right) L
		- Q_q \sin^2 \theta_w R \right] \, \tilde{q} \: ,
		\label{eq:zsm} \\
	{J^\mu}^\pm & \equiv & \bar{\tilde{\psi}} \, \gamma^\mu \left(
		T_1 \pm i T_2 \right) L \, \tilde{\psi} \: .
\end{eqnarray}

In the SM, there are two kinds of gauge boson. One is belong to
$SU(2)$, and the other one is $U(1)$ gauge group. The kinetic terms
for them are,
\begin{equation}
	{\cal L}_G = -\frac{1}{4} \left(
		{W_{\mu \nu}}^a \, {W^{\mu \nu}}^a
		+ B_{\mu \nu} \, B^{\mu \nu} \right) \: .
\end{equation}
Substituting Eqs. (\ref{eq:dc}), (\ref{eq:wb}) and (\ref{eq:u1b}),
\begin{eqnarray}
	{\cal L}_{W^{\pm} W^{\mp} A} & = & i e \, \left[
		\left( \partial^\mu A^\nu - \partial^\nu A^\mu \right)
		{W_\mu}^+ {W_\nu}^- + A^\nu \left(
		\partial^\mu {W^\nu}^+ - \partial^\nu {W^\mu}^+ \right)
		{W_\mu}^- \right. \nonumber \\
	& & 	\left. - A^\nu \left( \partial_\mu {W_\nu}^-
		- \partial_\nu {w_\mu}^- \right) {W^\mu}^+
		\right] \: .
\end{eqnarray}
Note here that the others $W^{\pm} W^{\mp} A$ terms also appear in
the gauge fixing lagrangian in Eq. (\ref{eq:gf}).

\subsection{The Higgs sector}

For the Higgs sector, we can write it as below,
\begin{equation}
	{\cal L}_Y = \left( D_\mu \phi \right)^\dagger
		\left( D^\mu \phi \right) - V(\phi) \: ,
\end{equation}
where, $V(\phi)$ is the potential term,
$V(\phi) \equiv -\mu^2 \phi^\dagger \phi +
\lambda \left( \phi^\dagger \phi \right)^2 $. This potential term will
give the mass of the physical Higgs.
By substituting Eqs. (\ref{eq:scalar}), (\ref{eq:dc}) and (\ref{eq:bb}),
\begin{eqnarray}
	{\cal L}_Y & = & \left\{ \left[ \partial_\mu +
		i g \sin \theta_W A_\mu + \frac{i g}{2 \cos \theta_W}
		\left( \cos^2 \theta_W - \sin^2 \theta_W \right) Z_\mu
		\right] \chi^- \right.
		\nonumber \\
	& &	\left. + i M_W {W_\mu}^-
		+ \frac{i g}{2} \left( H - i \chi^0 \right) {W_\mu}^-
		\right\}
		\nonumber \\
	& & 	\times \left\{ \left[ \partial^\mu - i g \sin \theta_W A^\mu
		- \frac{i g}{2 \cos \theta_W}
		\left( \cos^2 \theta_W - \sin^2 \theta_W \right) Z^\mu
		\right] \chi^+ \right.
		\nonumber \\
	& &	\left. - i M_W {W_\mu}^+
		- \frac{i g}{2} \left( H + i \chi^0 \right) {W_\mu}^+
		\right\}
		\nonumber \\
	& &	+ \frac{1}{2} \left[ {W_\mu}^+ \chi^- - i M_Z Z_\mu
		+ \left( \partial_\mu - \frac{i g}{2 \cos \theta_W} Z_\mu
		\right) \left(H - i \chi^0 \right) \right]
		\nonumber \\
	& &	\times \left[ {W^\mu}^- \chi^+ + i M_Z Z^\mu
		+ \left( \partial^\mu + \frac{i g}{2 \cos \theta_W} Z^\mu
		\right) \left(H + i \chi^0 \right) \right]
		- V(\phi)  \: .
\end{eqnarray}
Here, $W^\pm$ and $Z$ bosons masses are defined as,
\begin{equation}
	{M_W}^2 \equiv \frac{g^2 v^2}{4} \: \: \: , \: \: \:
	{M_Z}^2 \equiv \frac{g^2 v^2}{8 \cos^2 \theta_W} \: ,
\end{equation}
where $v$ is the vacuum-expectation-value (VEV) of Higgs,
$ \langle \phi \rangle \equiv v $.

\subsection{The Yukawa and mass sectors}
\label{app:yukawasm}

The most important term here is the lagrangian of Yukawa sector. This
is the standard mechanism in the SM to generate the fermion masses.
The lagrangian for Yukawa sector with up and down fermions is,
\begin{equation}
	{\cal L}_Y = -{f_d}^{ij} {\bar{\tilde{\psi}}_L}^i \, \phi
		\, {\tilde{d}_R}^j - {f_u}^{ij} {\bar{\tilde{\psi}}_L}^i
		\phi {\tilde{u}_R}^j + h.c. \: .
	\label{eq:higgssm}
\end{equation}
Here, ${f_d}^{i \alpha}$ and ${f_u}^{i \alpha}$ are the Yukawa couplings
which give the masses of fermions through the vacuum expectation value
of Higgs doublet.

Substitute the scalar (Eqs. (\ref{eq:scalar}) and (\ref{eq:p0})), and
pick up the VEV terms, the would-be mass terms are presented as,
\begin{equation}
	{\cal L}_{mass} = -{f_d}^{ij} {\bar{\tilde{\psi}}_L}^i \,
		\frac{v}{\sqrt{2}} \, {\tilde{d}_R}^j - {f_u}^{ij}
		{\bar{\tilde{\psi}}_L}^i \, \frac{v}{\sqrt{2}} \,
		{\tilde{u}_R}^j + h.c. \: .
	\label{eq:masssm}
\end{equation}
The fermion fields should be diagonalized as,
\begin{equation}
	{f_d}^{ij} \frac{v}{\sqrt{2}} \equiv
		\left( {V_L}^{\dagger} m_d V_R \right)^{ij}
	\: \: , \: \:
	{f_u}^{ij} \frac{v}{\sqrt{2}} \equiv
		\left( {U_L}^{\dagger} m_u U_R \right)^{ij} \: ,
\end{equation}
where $V$ and $U$ are $3 \times 3$ unitary matrices which relate the
weak-eigenstates to mass-eigenstates as,
\begin{eqnarray}
	{d_L}^{i} & \equiv & V^{ij} {\tilde{d}_L}^j \: ,
	\label{eq:ddcsm} \\
	{u_L}^{i} & \equiv & U^{ij} {\tilde{u}_L}^j \: .
	\label{eq:uucsm}
\end{eqnarray}

Redefinitions of the fermion fields in Eqs. (\ref{eq:ddcsm}) and
(\ref{eq:uucsm}) make a principal change in the CC lagrangian
in Eq. (\ref{eq:cclsm}), without changing the NC one.
The effect of the diagonalizations is usually expressed as (Eq. (\ref{eq:ckmd})),
\begin{equation}
	{V_{CKM}}^{ij} \equiv \sum_{k=1}^{3} U^{ki} {V^{kj}}^{\ast} \: ,
	\label{eq:ckmsm}
\end{equation}
and called as the {\it Cabibbo-Kobayashi-Maskawa matrix} (CKM matrix).
Through this CKM matrix, flavor changing in the CC interactions
in the SM has occured.

\subsection{The gauge-fixing term}

About the gauge-fixing term, we employ the procedure of ref. \cite{grig},
\begin{equation}
	{\cal L}_{GF} \equiv -\frac{1}{\xi} \left(f_W {f_W}^{\dag}
		+ {f_Z}^2 \right) ,
	\label{eq:gf}
\end{equation}
where,
\begin{eqnarray}
	f_W & \equiv & \partial_{\mu}{W^{\mu}}^-
		+ i e A_{\mu} {W^{\mu}}^- + i \xi M_W \chi^- ,  \\
	f_Z & \equiv & \frac{1}{\sqrt{2}} \left(\partial_{\mu}Z^{\mu}
		+ \xi M_Z \chi^0\right)  .
\end{eqnarray}
According to ref. \cite{grig}, the advantage of this gauge-fixing
is the absence of $W^{\pm} \chi^{\mp} A$ interaction in the full lagrangian.
Therefore the number of diagrams is reduced without changing the propagator
of each field, for example in the $b \rightarrow s(d) \gamma$ process.

\section{THE INTERACTIONS IN THE VQM}
\label{app:sqml}

In this appendix, mainly we will give the lagrangian for the
$Z$ (ref. \cite{branco}) and the neutral Higgs (ref. \cite{handoko2})
sector, and show how the tree-level FCNC's comes out. We do similar procedures
as in Appendix \ref{app:sml}, and gauge fixing lagrangian is as
Eq. (\ref{eq:gf}).

\subsection{The Yukawa sector}

In principle, the Yukawa sector is same with Eq. (\ref{eq:higgssm}).
The problem is how to generate the mass of vector-like quarks. In general,
there are two solutions for the problem, that is
\begin{enumerate}
	\item Introducing one singlet Higgs, and
	\item Putting a bare mass for each vector-like quark in the lagrangian.
\end{enumerate}
Here, we adopt the second solution. However the discussion of our interest
is not altered by the choice.

Adding the bare mass terms into the Yukawa sector lagrangian in
Eq. (\ref{eq:higgssm}),
\begin{equation}
	{\cal L}_Y = -{f_d}^{i \alpha} {\bar{\tilde{\psi}}_L}^i \,
		\phi \, {\tilde{d}_R}^{\alpha}
	- {f_d}^{4 \alpha} {\bar{\tilde{d}}_L}^4 \,
		{\tilde{d}_R}^{\alpha} \, v^{\prime}
	- {f_u}^{i \alpha} {\bar{\tilde{\psi}}_L}^i \phi {\tilde{u}_R}^{\alpha}
	- {f_u}^{4 \alpha} {\bar{\tilde{u}}_L}^4 \,
		{\tilde{u}_R}^{\alpha} \, v^{\prime}
	+ h.c. \: .
	\label{eq:higgssqm}
\end{equation}
The terms which are proportionated to
${f_d}^{4 \alpha}$ and ${f_u}^{4 \alpha}$ are bare mass terms.
These terms contain both diagonal couplings for vector-like quarks
($t^\prime$, $b^\prime$), and off-diagonal couplings between
left-handed $SU(2)$ vector-like quarks (${t_L}^\prime$, ${b_L}^\prime$)
and right-handed vector-like quarks
(${u_L}^i$, ${d_L}^i$). We diagonalize the Yukawa couplings as,
\begin{eqnarray}
	{f_d}^{i \alpha} \frac{v}{\sqrt{2}} \equiv
		\left( {V_L}^{\dagger} m_d V_R \right)^{i \alpha} & , &
	{f_d}^{4 \alpha} v^{\prime} \equiv
		\left( {V_L}^{\dagger} m_d V_R \right)^{4 \alpha} \: ,
	\label{eq:fd} \\
	{f_u}^{i \alpha} \frac{v}{\sqrt{2}} \equiv
		\left( {U_L}^{\dagger} m_u U_R \right)^{i \alpha} & , &
	{f_u}^{4 \alpha} v^{\prime} \equiv
		\left( {U_L}^{\dagger} m_u U_R \right)^{4 \alpha} \: ,
	\label{eq:fu}
\end{eqnarray}
where $V$ and $U$ are $4 \times 4$ unitary matrices which relate the
weak-eigenstates to mass-eigenstates and generate CKM matrix as,
\begin{eqnarray}
	{d_L}^{\beta} & \equiv & V^{\beta\alpha} {\tilde{d}_L}^{\alpha},
	\label{eq:ddqf} \: , \\
	{u_L}^{\beta} & \equiv & U^{\beta\alpha} {\tilde{u}_L}^{\alpha},
	\label{eq:duqf} \: , \\
	{V_{CKM}}^{\alpha\beta} & \equiv & \sum_{i=1}^{3}
		U^{i\alpha} {V^{i\beta}}^{\ast} .
	\label{eq:ckmsqm}
\end{eqnarray}
Substituting the above relations into the lagrangian
in Eq. (\ref{eq:higgssqm}), the neutral Higgs sector becomes,
\begin{eqnarray}
	{\cal L}_{\phi^0} & = & -\frac{1}{v} {m_d}^{\beta} \,
		V^{\alpha i} {V^{\beta i}}^{\ast}
		{\bar{d}_L}^{\alpha} \, {{d}_R}^{\beta} \, \phi^0
	- \frac{1}{v} {m_u}^{\beta} \, U^{\alpha i} {U^{\beta i}}^{\ast}
		{\bar{u}_L}^{\alpha} \, {{u}_R}^{\beta} \, {\phi^0}^{\ast}
		+ h.c. \: .
	\label{eq:nhs}
\end{eqnarray}
With the following definitions,
\begin{eqnarray}
	{z_d}^{\alpha\beta} & \equiv & \sum_{i=1}^{3} V^{\alpha i}
		{V^{\beta i}}^{\ast} = \delta^{\alpha\beta} -
		V^{\alpha 4} {V^{\beta 4}}^{\ast} ,
	\label{eq:w4} \\
	{z_u}^{\alpha\beta} & \equiv & \sum_{i=1}^{3} U^{\alpha i}
		{U^{\beta i}}^{\ast} = \delta^{\alpha\beta} -
		U^{\alpha 4} {U^{\beta 4}}^{\ast} ,
	\label{eq:w5}
\end{eqnarray}
the neutral Higgs sector (Eq. \ref{eq:nhs}) can be written as,
\begin{eqnarray}
	{\cal L}_{\phi^0} & = & -\frac{1}{v} {m_d}^{\beta} \,
		{z_d}^{\alpha \beta}
		{\bar{d}_L}^{\alpha} \, {{d}_R}^{\beta} \, H
	- \frac{1}{v} {m_u}^{\beta} \, {z_u}^{\alpha \beta}
		{\bar{u}_L}^{\alpha} \, {{u}_R}^{\beta} \, H
	\nonumber \\
	&  & -\frac{i}{v} {m_d}^{\beta} \,
		{z_d}^{\alpha \beta}
		{\bar{d}_L}^{\alpha} \, {{d}_R}^{\beta} \, \chi^0
	+ \frac{i}{v} {m_u}^{\beta} \, {z_u}^{\alpha \beta}
		{\bar{u}_L}^{\alpha} \, {{u}_R}^{\beta} \, \chi^0
		+ h.c. \: .
\end{eqnarray}

It is apparent that if there are non-zero off-diagonal elements in
${z_d}^{\alpha \beta}$ and ${z_u}^{\alpha \beta}$, FCNC's arise in the neutral
Higgs sector,
\begin{eqnarray}
	{\cal L}_H & = & \frac{-g}{2 M_W} \left[
		{z_u}^{\alpha\beta} \bar{u}^{\alpha}
		\left(m_{u^{\alpha}} L + m_{u^{\beta}} R\right) u^{\beta}
		+ {z_d}^{\alpha\beta} \bar{d}^{\alpha}
		\left(m_{d^{\alpha}} L + m_{d^{\beta}} R\right) d^{\beta}
		\right] \, H \: ,  \\
	{\cal L}_{\chi^0} & = & \frac{-ig}{2 M_W} \left[
		{z_u}^{\alpha\beta} \bar{u}^{\alpha}
		\left(m_{u^{\alpha}} L - m_{u^{\beta}} R\right) u^{\beta}
		- {z_d}^{\alpha\beta} \bar{d}^{\alpha}
		\left(m_{d^{\alpha}} L - m_{d^{\beta}} R\right) d^{\beta}
		\right] \, \chi^0 \: .
\end{eqnarray}

\subsection{The neutral gauge bosons sector}

The lagrangian in these sectors can be found easily. By using
Eqs. (\ref{eq:nclsm}), (\ref{eq:asm}) and (\ref{eq:zsm}), and
inserting the new vector-like quarks, the current for $A$ sector
is,
\begin{eqnarray}
	{J_Q}^\mu & \equiv &
	Q_q \bar{\tilde{q}} \, \gamma^\mu \, \tilde{q} +
	Q_{b^\prime} \bar{\tilde{b^\prime}} \, \gamma^\mu \,
	\tilde{b^\prime} +
	Q_{t^\prime} \bar{\tilde{t^\prime}} \, \gamma^\mu \,
	\tilde{t^\prime} \nonumber \\
	& = & \frac{1}{3} \left(
		2 \bar{u}^{\alpha} \gamma^{\mu} u^{\alpha}
		- \bar{d}^{\alpha} \gamma^{\mu} d^{\alpha}
		\right) \, A_{\mu} \: ,
\end{eqnarray}
after making diagonalization of the quark fields through
Eqs. (\ref{eq:ddqf}) and (\ref{eq:duqf}). 

For the $Z$ sector, we obtain,
\begin{equation}
	J^\mu \equiv \bar{\tilde{\psi}_L}^i \, \gamma^\mu T_3 \,
		{\tilde{\psi}_L}^i - \sin^2 \theta_W \left( Q_q \,
		\bar{\tilde{q}}^i \, \gamma^\mu \, \tilde{q}^i
		+ Q_{b^\prime} \bar{\tilde{b^\prime}} \, \gamma^\mu \,
		\tilde{b^\prime}
		+ Q_{t^\prime} \bar{\tilde{t^\prime}} \, \gamma^\mu \,
		\tilde{t^\prime} \right) \: .
\end{equation}
Diagonalizing the quarks field by Eqs. (\ref{eq:ddqf}) and (\ref{eq:duqf}),
\begin{eqnarray}
	J^\mu & = & \frac{1}{2}
		\left\{ \bar{u}^{\alpha} \gamma^{\mu} \left[\left(
		{z_u}^{\alpha\beta} - \frac{4}{3} \sin^2\theta_W
		\delta^{\alpha\beta}
		\right) L - \frac{4}{3} \sin^2\theta_W \delta^{\alpha\beta} R
		\right] u^{\beta} \right. \nonumber \\
	& &	+ \left. \bar{d}^{\alpha} \gamma^{\mu}
		\left[\left( \frac{2}{3} \sin^2\theta_W \delta^{\alpha\beta} -
		{z_d}^{\alpha\beta}\right) L
		+ \frac{2}{3}\sin^2\theta_W \delta^{\alpha\beta} R\right] 
		d^{\beta}\right\} \: ,
\end{eqnarray}
by using definitions in Eqs. (\ref{eq:w4}) and (\ref{eq:w5}).

\section{CKM MATRIX IN THE SM}

In this appendix, we give two usually employed parametrizations. 
By these parametrizations, it will also be shown 
that there is only one remaining phase in the CKM matrix.

In the Yukawa sector lagrangian in Eq. (\ref{eq:higgssm}),
suppose that the up-quark sector is diagonalized first,
\begin{equation}
	{{U_L}^\ast}^{ij} {f_u}^{jk} \frac{v}{\sqrt{2}}
	{U_R}^{kl} \equiv {m_u}^{il} \: .
\end{equation}
Generally, when the matrices $U_L$ diagonalizes the mass matrix in
the up-quark sector, the mass matrix in the down-quark sector is not
automatically diagonalized, that is
\begin{equation}
	{\cal L}_{d-{\rm mass}} = {{U_L}^\ast}^{ij} {f_d}^{jk}
		\frac{v}{\sqrt{2}}
		{V_R}^{kl} \, {\bar{\tilde{d}_L}}^i \, {\tilde{d}_R}^l \: ,
\end{equation}
is not diagonalized. So, in order to make it diagonalized, the down-quark
fields should be redefined again,
\begin{eqnarray}
	{m_d}^{il} & \equiv & {{U^\prime_L}^\dagger}^{in} \,
		\left( {{U_L}^\dagger}^{nm} \, {f_d}^{jk} \frac{v}{\sqrt{2}} \,
		{V_R}^{kl} \right) \nonumber \\
	& = &	{\left( U_L \, U^\prime_L \right)^\dagger}^{im}
		{f_d}^{mk} \frac{v}{\sqrt{2}} \, {V_R}^{kl} \: ,
\end{eqnarray}
where $V_L$ is newly introduced unitary matrix. Thus, the relation between
weak-eigenstate and mass-eigenstate of $SU(2)$ doublet $d$ field becomes,
\begin{equation}
	{d_L}^i \equiv \left( U_L \, U^\prime_L \right)^{il}
	{d^\prime_L}^l \: ,
	\label{eq:ckmd}
\end{equation}
where $d^\prime_L$ is the mass eigenstate of down-quark.
Write $U_L \, U^\prime_L \equiv V_L$, we obtain the CKM matrix as Eq. (\ref{eq:ckmsm})
in the CC interactions.

\subsection{Kobayashi-Maskawa parametrization}
\label{app:kobayashi}

This Appendix is based on refs. \cite{kobayashi} and
\cite{georgi}. The task is to parametrize the CKM matrix which has
been defined in Eq. (\ref{eq:ckmsm}). The CKM matrix is expressed as
a multiplication of three unitary matrices. Since a unitary matrix
could be considered as a rotation without loss of generality, we can
suppose the matrices as rotations with three different angles in
three dimensions space,
\begin{equation}
	R \left( \theta_1 \, \theta_2 \, \theta_3 \right) =
		R_y(\theta_3) \, R_x(\theta_2) \, R_z(\theta_1) \: ,
\end{equation}
where $R_i(\theta_j)$ is a rotation about $i$ with angle $\theta_j$.
For convenience, it can be expressed as,
\begin{equation}
	R \left( \theta_1 \, \theta_2 \, \theta_3 \right) =
		R_z(\theta_3) \, R_y(\theta_2) \, R_z(\theta_1) \, .
\end{equation}

Since as stated above, the matrices can be rewritten as rotation,
let us express the matrices like below,
\begin{eqnarray}
	R_z(\theta_{1(3)}) & \equiv & \left(
		\begin{array}{ccc}
			c_{1(3)} & -s_{1(3)} e^{-i \varphi_{1(3)}} & 0 \\
			s_{1(3)} e^{i \varphi_{1(3)}}	& c_{1(3)} & 0 \\
			0			& 0	& 1
		\end{array}
		\right) \: , \\
	R_y(\theta_2) & \equiv & \left(
		\begin{array}{ccc}
			c_2 	& 0		& -s_2 e^{-i \varphi_2} \\
			0			& 1		& 0 \\
			s_2 e^{i \varphi_2}	& 0		& c_2
		\end{array}
		\right) \: .
\end{eqnarray}
Here, $c_i \equiv \cos \theta_i$ and $s_i \equiv \sin \theta_i$.
Then, we can make $R_z(\theta_{1(3)})$ to be real, since for example,
\begin{equation}
	\left(
	\begin{array}{cc}
		1	& 0 			\\
		0	& e^{-i \varphi}
	\end{array}
	\right) \times
	\left(
	\begin{array}{cc}
		c			& -s e^{-i \varphi} 	\\
		s_ e^{i \varphi}	& c
	\end{array}
	\right) \times
	\left(
	\begin{array}{cc}
		e^{i \varphi^\prime}	& 0 				\\
		0			& e^{i (\varphi + \varphi^\prime)}
	\end{array}
	\right) \propto
	\left(
	\begin{array}{cc}
		c	& -s \\
		s	& c
	\end{array}
	\right) \: : \: {\rm real} \: .
	\label{eq:real}
\end{equation}
Thus, we can redefine
\begin{eqnarray}
	R^\prime \left( \theta_1 \, \theta_2 \, \theta_3 \right) & \equiv &
		U \: R \left( \theta_1 \, \theta_2 \, \theta_3 \right) U^\prime
		\nonumber \\
	& = &	U \, R_z(\theta_3) \, R_y(\theta_2) \, R_z(\theta_1)\, U^\prime
		\nonumber \\
	& = & 	\left( U \, R_z(\theta_3) \, {U^{\prime \prime}}^\ast \right)
		\left( U^{\prime \prime} \, R_y(\theta_2) \,
		{U^{\prime \prime \prime}}^\ast \right)
		\left( U^{\prime \prime \prime}  \, R_z(\theta_1) \: U^\prime
		\right)
		\nonumber \\
	& \equiv & R^\prime_z(\theta_3) \, R^\prime_y(\theta_2) \,
		R^\prime_z(\theta_1) \: ,
\end{eqnarray}
with $U$, $U^\prime$, $U^{\prime \prime }$, $U^{\prime\prime \prime }$
are diagonal unitary matrices like the matrices in Eq. (\ref{eq:real}), and
\begin{equation}
	R^\prime_z(\theta_{1(3)}) \equiv \left(
	\begin{array}{ccc}
		c_{1(3)} 	& -s_{1(3)} 	& 0 \\
		s_{1(3)}	& c_{1(3)}	& 0 \\
		0		& 0		& 1
	\end{array}
	\right) \: .
	\label{eq:rz13}
\end{equation}

Further redefinition for $R^\prime_y(\theta_2)$ is done by
multiplying on either of $R^\prime \left( \theta_1 \,
\theta_2 \, \theta_3 \right)$ by a matrix of the form,
\begin{equation}
	U^{\prime \prime \prime \prime} \equiv
	\left(
	\begin{array}{ccc}
		\beta 	& 0		& 0 \\
		0	& \alpha	& 0 \\
		0	& 0		& \beta
	\end{array}
	\right) \: , \\
\end{equation}
where $U^{\prime \prime \prime \prime}$ commutes with
$R^\prime_z(\theta_{1(3)})$ and
$\left| \alpha \right| = \left| \beta \right| \equiv 1$.
Then it yields,
\begin{eqnarray}
	R^{\prime \prime} \left( \theta_1 \, \theta_2 \, \theta_3 \right)
		& \equiv &
		 U^{\prime \prime \prime \prime} \:
		 R^\prime \left( \theta_1 \, \theta_2 \, \theta_3 \right)
		{U^{\prime \prime \prime \prime}}^\ast	\nonumber \\
	& = & R^\prime_z(\theta_3) \left(
		 U^{\prime \prime \prime \prime} \, R^\prime_y(\theta_2) \,
		{U^{\prime \prime \prime \prime}}^\ast	\right)
		R^\prime_z(\theta_1) \nonumber \\
	& \equiv & R^\prime_z(\theta_3) \, R^{\prime \prime}_y(\theta_2)
		\, R^\prime_z(\theta_1) \: ,
\end{eqnarray}
where
\begin{equation}
	R^{\prime \prime}_y(\theta_2) \equiv \left(
	\begin{array}{ccc}
		c_2 		& 0			& -s_2 \\
		0		&  e^{i \delta}		& 0 \\
		s_2		& 0			& c_2
	\end{array}
	\right) \: .
	\label{eq:ry2}
\end{equation}
It is clear that just only one phase $\delta$ has been remained in the
CKM matrix. Finally, the CKM matrix in the Kobayashi-Maskawa parametrization
is given by multiplying the matrices $R^\prime_z(\theta_{1(3)})$ and
$R^{\prime \prime}_y(\theta_2)$ of Eqs. (\ref{eq:rz13}) and (\ref{eq:ry2}),
\begin{equation}
	V_{CKM} = \left(
	\begin{array}{ccc}
		c_1 c_2 c_3 - s_1 s_3 e^{i \delta} & -s_1 c_2 c_3 - c_1 s_3
			e^{i \delta} & -s_2 c_3 \\
		c_1 c_2 s_3 + s_1 c_3 e^{i \delta} & -c_1 c_2 s_3 + c_1 c_3
			e^{i \delta} & -s_2 s_3 \\
		c_1 s_2		& -s_1 s_2	& c_2
	\end{array}
	\right) \: .
\end{equation}

\subsection{Wolfenstein parametrization}
\label{app:wolfenstein}

This Appendix is based on ref. \cite{wolfenstein}. This parametrization is
based on the phenomenological approximation. The main principle is
the fact that $\left| {V_{CKM}}^{us}\right| \cong \sin \theta_C \cong 0.22$
is quite well determined, where $\theta_C$ is Cabibbo angle.
Then, by defining $\left| {V_{CKM}}^{us} \right| \equiv \lambda$,
it could be considered an expansion of $V_{CKM}$ in powers of $\lambda$.
The experiment bound for $\left| {V_{CKM}}^{cb} \right|$ in Eq. (\ref{eq:vcb})
suggests that $\left| {V_{CKM}}^{cb} \right|$ is order of $\lambda^2$ rather
than $\lambda$, so it should be parametrized as $\left|
{V_{CKM}}^{cb} \right| \equiv A \lambda^2 $ with $A$ is a
constanta which is determined by experiment bound on
$\left| {V_{CKM}}^{cb} \right|$. Note here that the elements
of ${V_{CKM}}^{ub}$ and
${V_{CKM}}^{td}$ are zero in order $O(\lambda^2)$.

It leads that we have to make an approximation up to order $O(\lambda^3)$.
Unitarity of CKM matrix then describes the following form,
\begin{equation}
	V_{CKM} \cong \left(
	\begin{array}{ccc}
		1 - \frac{1}{2} \lambda^2 	& \lambda	&
			A \lambda^3 (\rho - i \eta) \\
		- \lambda	& 1 - \frac{1}{2} \lambda^2	&
			A \lambda^2 \\
		A \lambda^3 (1 - \rho - i \eta)	& -A \lambda^2 	& 1
	\end{array}
	\right) + O(\lambda^4) \: .
\end{equation}
Here, two new parameters $\rho$ and $\eta$ must be introduced to satisfy the
unitary condition. The detailed prove must be seen in ref. \cite{wolfenstein}.

However, the parametrization up to $O(\lambda^2)$ is too rough to
describe the $CP$ violation. The reason is that, generally
all $CP$ violating quantities are proportional to,
\begin{equation}
	J_{CP} = {\rm Im} \left( {V_{CKM}}^{cb} {V_{CKM}}^{us}
	{{V_{CKM}}^{ub}}^\ast {{V_{CKM}}^{cs}}^\ast \right)
\end{equation}
as derived in ref. \cite{jarlskog}. Then, to get non-zero
result, one additionally needs the leading $CP$ violating
pieces of ${V_{CKM}}^{cs}$ and ${V_{CKM}}^{cb}$. For this,
the unitarity up to $O(\lambda^5)$ in the imaginary parts
has to be satisfied.
The Wolfenstein parametrization up to this order
is then given as Eq. (\ref{eq:ckmw}).

\section{MASS MATRIX IN THE ODVQM}
\label{app:mm}

Let us start from the Yukawa sector lagrangian in Eq. (\ref{eq:higgssqm})
and the diagonalizations of Yukawa couplings in Eqs. (\ref{eq:fd}) and
(\ref{eq:fu}).
The first task is to derive the quark mass matrices.
Pick up the mass terms of down-quark sector in Eq. (\ref{eq:higgssqm}),
since the same procedure can be done in the up-quark sector.
\begin{eqnarray}
	{\cal L}_{d-{\rm mass}} & = &
		-\frac{v}{2} \, {f_d}^{i \alpha} \, {\bar{\tilde{d}}_L}^i
		\, {\tilde{d}_R}^{\alpha}
		- v^{\prime} \, {f_d}^{4 \alpha} {\bar{\tilde{d}}_L}^4 \,
		{\tilde{d}_R}^{\alpha} \nonumber \\
	& = & -\left(
		\begin{array}{cc}
			{\bar{\tilde{d}}_L}^i & {\bar{\tilde{d}}_L}^4
		\end{array}
		\right) \:
		\left(
		\begin{array}{c}
			\frac{v}{2} \, {f_d}^{i \alpha} \\
			v^{\prime} \, {f_d}^{4 \alpha}
		\end{array}
		\right) \:
		{\tilde{d}_R}^{\alpha} \: .
	\label{eq:ldm}
\end{eqnarray}
Redefine each field by the following transformations,
\begin{eqnarray}
	{\tilde{\psi}_L}^i & \equiv & {A_L}^{ij} \, {\psi_L}^i
		\: \: \longrightarrow
		{\tilde{d}_L}^i \equiv {A_L}^{ij} \, {d_L}^i \: , \\
	{\tilde{d}_L}^4 & \equiv & e^{-i \theta_d} \, {d_L}^4 \: , \\
	{\tilde{d}_R}^\alpha & \equiv & {B_R}^{\alpha \beta} \, {d_R}^\beta
		\: ,
\end{eqnarray}
where $A_L$ and $B_R$ are arbitrary unitary matrices.
Substituting them into Eq. (\ref{eq:ldm}),
\begin{eqnarray}
	{\cal L}_{d-{\rm mass}} & = &
		-\left(
		\begin{array}{cc}
			{\bar{d}_L}^j & {\bar{d}_L}^4
		\end{array}
		\right) \:
		\left(
		\begin{array}{cc}
			{A_L}^{ji}	& 0  	\\
			0 		& e^{-i \theta_d}
		\end{array}
		\right) \:
		\left(
		\begin{array}{c}
			\frac{v}{2} \, {f_d}^{i \alpha} \\
			v^{\prime} \, {f_d}^{4 \alpha}
		\end{array}
		\right) \:
		{B_R}^{\alpha \beta} \: {d_R}^{\alpha}
		\nonumber \\
	& \equiv &
		-\left(
		\begin{array}{cc}
			{\bar{d}_L}^j & {\bar{d}_L}^4
		\end{array}
		\right) \:
		\left(
		\begin{array}{c}
			{m^0}^{i \beta} \\
			{m^0}^{4 \beta}
		\end{array}
		\right) \:
		{d_R}^{\alpha}
	\label{eq:dmass}
\end{eqnarray}
where,
\begin{eqnarray}
	{m^0}^{i \beta} & \equiv & {A_L}^{ji} \, \frac{v}{2} \,
		{f_d}^{i \alpha} \, {B_R}^{\alpha \beta}
		\equiv \left(
		\begin{array}{cccc}
			{\rm Re} \left[ {m_d}^0 \right] & 0 & 0 & 0 \\
			0 & {\rm Re} \left[ {m_s}^0 \right] & 0 & 0 \\
			0 & 0 & {\rm Re} \left[ {m_b}^0 \right] & 0
		\end{array}
		\right) \: , \\
	{m^0}^{4 \beta} & \equiv & e^{-i \theta_d} \, v^\prime \,
		{f_d}^{4 \alpha} \, {B_R}^{\alpha \beta}
		\equiv e^{-i \theta_d} \, \left(
		\begin{array}{cccc}
			J_d	& J_s	& J_b	& m_{b^\prime}
		\end{array}
		\right) \nonumber \\
	& = & \left(
		\begin{array}{cccc}
			J_d	& J_s	& J_b
			& {\rm Re} \left[ m_{b^\prime} \right]
		\end{array}
		\right) \: .
\end{eqnarray}
Here, $J_i$ are generally complex numbers. Phases in the imaginary
number $f_d$ are cancelled by the phases contained in the unitary
matrices which have been multiplyed on either. Therefore, we
obtain the mass matrix of down-quarks as,
\begin{eqnarray}
	M_d^0 & \equiv & \left(
	\begin{array}{c}
		{m^0}^{i \beta} \\
		{m^0}^{4 \beta}
	\end{array}
	\right) \nonumber \\
	& = & \left(
	\begin{array}{cccc}
		{m_d}^0		& 0		& 0		& 0 \\
		0		& {m_s}^0	&  0 		& 0 \\
		0		& 0		& {m_b}^0	& 0 \\
		J_d		& J_s		& J_b	& {m_{b^\prime}}^0
	\end{array}
	\right) \: \: ,
	\label{eq:mdo}
\end{eqnarray}
where we leave out the symbols {\rm Re} and {\rm Im} as
knowledgment, i.e. ${m_\alpha}^0$ are real and $J_i$ are
complex quantities.

The same procedure can be done in the up-quark sector respectively.
Strictly, the mass matrix for up-quark sector becomes,
\begin{eqnarray}
	M_u^0 & = & \left(
	\begin{array}{cccc}
		{m_u}^0		& 0		& 0		& 0 \\
		0		& {m_c}^0	&  0 		& 0 \\
		0		& 0		& {m_t}^0	& 0 \\
		J_u		& J_c		& J_t	& {m_{t^\prime}}^0
	\end{array}
	\right) \: \: ,
	\label{eq:muo}
\end{eqnarray}

\subsection{Natural suppression of tree-level FCNC's}
\label{app:nsfcnc}

With the procedure in refs. \cite{handoko2} and \cite{handoko1}, we try
to derive the unitary matrix which diagonalize Eq. (\ref{eq:mdmd})
in the case (1). When we concentrate on the mixing ${z_d}^{db}$,
we can make an assumption that $J_d = J_b \equiv J$, $J_s = 0$ and
${m_d}^0 = {m_b}^0 = {m_{b^\prime}}^0 \equiv m^0$.
Under these assumptions, the discussion becomes simpler since it
does not need to calculate the eigenvalues and eigenvectors of
four by four matrix. Further, it can be reduced to the $2 \times 2$
matrix problem, that is
\begin{equation}
	{M_d}^0 {{M_d}^0}^{\dagger} = \left(
	\begin{array}{cc}
		{m^0}^2	& m^0 J^{\ast} \\
		m^0 J	& 2 \left| J \right|^2
	\end{array}
	\right) \: .
\end{equation}

The problems are to find the mass eigenvalues $M_d {M_d}^{\dagger}$,
with $M_d {M_d}^{\dagger} \equiv V \, {M_d}^0 {{M_d}^0}^{\dagger} \, V^\dagger$.
So we have to calculate the eigenvalues and eigenvectors of
this matrix. From the definition of $M_d {M_d}^{\dagger}$, the eigenfunction
is written as,
\begin{equation}
	\left( {M_d}^0 {{M_d}^0}^{\dagger} \right) {V_i}^\dagger =
	\lambda_i \, {V_i}^\dagger \: ,
\end{equation}
with $\lambda_i \equiv \left( M_d {M_d}^{\dagger} \right)_i $.
After any calculations, the eigenvalues are,
\begin{eqnarray}
	\lambda_1 & = & \frac{{m_o}^4}{2 \left| J \right|^2} \equiv {m_d}^2
		\: , \\
	\lambda_2 & = & m^2 \equiv {m_b}^2 \: , \\
	\lambda_3 & = & 2 \left| J \right|^2 -
		\frac{{m_o}^4}{2 \left| J \right|^2} \equiv {m_{b^\prime}}^2
		\: ,
\end{eqnarray}
and ${{m_s}^0}^2 \equiv {m_s}^2$ respectively.
Meanwhile, the eigenvectors correspond to the eigenvalues can strictly be
found,
\begin{equation}
	V^\dagger \cong \frac{1}{\sqrt{2}} \left(
	\begin{array}{cccc}
		-1 & 0 & 1 &
		\frac{\sqrt{2} m^0 J}{2 \left| J \right|^2} \\
		0 & \sqrt{2} & 0 & 0 \\
		-1 & 0 & -1 &
		\frac{\sqrt{2} m^0 J}{2 \left| J \right|^2} \\
		\frac{2 m^0 J^{\ast}}{2 \left| J \right|^2}
		& 0 & 0 & \sqrt{2}
	\end{array}
	\right) \: .
	\label{eq:vdmat}
\end{equation}

\subsection{Natural existence of tree-level FCNC's}
\label{app:nefcnc}

In this appendix, we will show the case where theoretically
$z_{u(d)}$ are non-zero as done in ref. \cite{handoko3}.
The same approach with Appendix \ref{app:nsfcnc} will be done,
by changing the assumption, that is $J_s = 0$.
This assumption is needed to avoid solving $4 \times 4$ mass
matrix of Eq. (\ref{eq:mdmd}).
\begin{equation}
	{M_d}^0 {{M_d}^0}^{\dagger} = \left(
	\begin{array}{cccc}
		{{m_d}^0}^2	& 0	& 0	& {m_d}^0 {J_d}^{\ast} \\
		0	& {{m_s}^0}^2	& 0 	& 0 \\
		0	& 0	& {{m_b}^0}^2	& {m_b}^0 {J_b}^{\ast} \\
		{m_d}^0 J_d	& 0	& {m_b}^0 J_b	& M^2
	\end{array}
	\right) \: \: \longrightarrow \: \:
	\left(
	\begin{array}{ccc}
		{{m_d}^0}^2	& 0		& {m_d}^0 {J_d}^{\ast} \\
		0		& {{m_b}^0}^2	& {m_b}^0 {J_b}^{\ast} \\
		{m_d}^0 J_d	& {m_b}^0 J_b	& M^2
	\end{array}
	\right) \: .
\end{equation}

Further assumption is $m_i \ll M$, then the matrix can be expanded
in term of ${m_i}/M$,
\begin{equation}
	{M_d}^0 {{M_d}^0}^{\dagger} = M^2 \: \left(
	\begin{array}{ccc}
		\frac{{{m_d}^0}^2}{M^2}	 & 0
			& \frac{{m_d}^0 {J_d}^{\ast}}{M^2} \\
		0			 & \frac{{{m_b}^0}^2}{M^2}
			& \frac{{m_b}^0 {J_b}^{\ast}}{M^2} \\
		\frac{{m_d}^0 J_d}{M^2}	 & \frac{{m_b}^0 J_b}{M^2} 	& 1
	\end{array}
	\right) \: ,
\end{equation}
where, $M \equiv \sqrt{{m_{b^\prime}}^2 + \left| J_d \right|^2 +
\left| J_s \right|^2 + \left| J_b \right|^2 }$.
By solving the eigenvalues in order to order,
\begin{equation}
	\left( {M_d}^0 {{M_d}^0}^{\dagger} \right)^{(j)}
		{{V^\dagger}_i}^{(j)} = {\lambda_i}^{(j)}
		\, {{V^\dagger}_i}^{(j)} \: ,
\end{equation}
with $i = 1,2,3,4$ are the column numbers and $j = 0,1,2$ are the
order indices. Since the calculation is too long and complicated,
we do not give it here. Note that, the calculation to first order
is sufficient to derive the mixing, but up to second order
calculation is required to derive the mass eigenvalues.
The fourth eigenvector, which shoud be used to get ${z_d}^{db}$,
is found as,
\begin{equation}
	{V^\dagger}_{4} \cong \frac{1}{M^2} \left(
	\begin{array}{c}
		{m_d}^0 {J_d}^\ast \\
		{m_b}^0 {J_d}^\ast \\
		{\rm x}
	\end{array}
	\right) \: ,
\end{equation}
where {\rm x} is an undetermined number in the calculation to first order.
So, it yields
\begin{equation}
	\left| {z_d}^{db} \right| \cong \frac{{m_d}^0 {m_b}^0 {J_d}^\ast {J_d}^\ast}{M^4}
	\neq 0 \: .
\end{equation}
Of course, this result has not finished yet, since the mass eigenvalues
must be substituted into the equation. However, by this equation, at least,
it is clear that non-zero mixing is possible for finite values of
ordinary quarks masses.

\section{$b \rightarrow q \, \gamma$}
\label{app:bqg}

\begin{figure}[t]
	\epsfxsize=15cm
	\epsfysize=10cm
	\epsffile{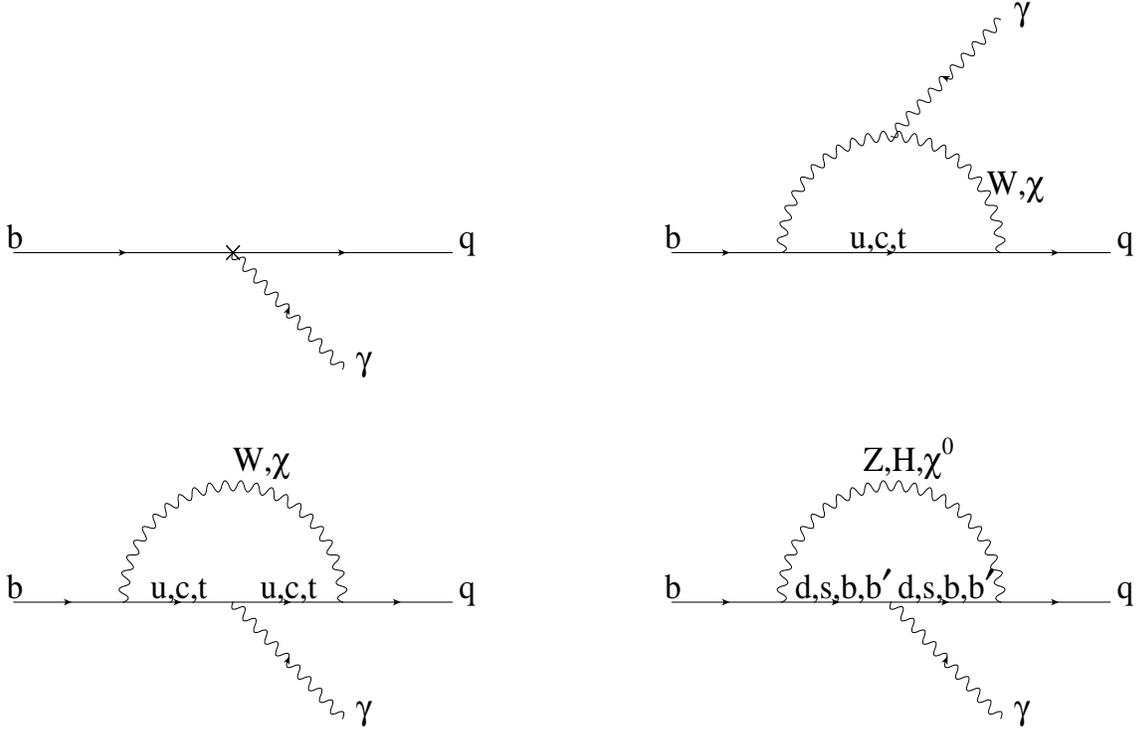}
	\caption{Diagrams related to $b \rightarrow q \, \gamma$ in the
		SM and ODVQM.}
	\label{fig:bqg}
\end{figure}

We will give a detail calculation for both on-shell and
off-shel $b \rightarrow q \, \gamma$ processes. The 
on-shell process is responsible for 
the inclusive $B \rightarrow X_q \, \gamma$ decay, while 
for the $b \rightarrow q \, l^+ \, l^- $ decay both of them
are responsible.
As stated in Appendix \ref{app:sml}, the diagrams that responsible for
the $b \rightarrow q \, \gamma$ ($q = d,s$) decays are one-loop
penguin diagrams as depicted in Fig. (\ref{fig:bqg}). The $W^\pm X^\mp A$ 
vertices contained penguin diagrams disappear due to the gauge fixing in 
Eq. (\ref{eq:gf}). The $b \rightarrow q \, g$ diagrams, which are needed in 
the calculation of QCD correction, are realized by the third and fourth 
diagrams with replacing the external photon lines by the gluon lines.

Technically, the calculation have been done in on-shell renormalization
up to second order in the external momenta.
The counter terms are realized by $W^\pm$, $\chi^\pm$, $Z$, $H$ and $\chi^0$
exchange quark self-energy diagrams, and symbolically depicted in the
first diagram in Fig. (\ref{fig:bqg}). More detailed calculations can
be seen in refs. \cite{grin}, \cite{grig} and \cite{inami}.

\subsection{In the SM}
\label{app:bqgsm}

In the SM, the must-be calculated diagrams are the CC diagrams, i.e.
the second and third diagrams in Fig. (\ref{fig:bqg}). The content of
the would-be counter term diagrams are $W^\pm$ and $\chi^\pm$ exchange
quark self-energy diagrams.

After making a straightforward calculation, the effective 
hamiltonian for $b \rightarrow q \, \gamma$ is given as the following,
\begin{eqnarray}
	H_{eff}^{SM} & = & \frac{G_F e}{8 \sqrt{2} \pi^2} Q_u
		{V_{CKM}}^{tq} {{V_{CKM}}^{tb}}^{\ast} 
		\left\{ F_1^{CC}(x_t) 
		\left[\gamma_{\mu},\not{q}\right]
		\left(m_q L + m_b R \right) \right.
		\nonumber \\
	& & \left. + F_3^{CC}(x_t) \left( q_\mu \not{q}    
		- q^2 \gamma_\mu  \right) L \right\} \: .  
\end{eqnarray}
The functions are,  
\begin{eqnarray}
	F_1^{CC}(x_t) & = & x_t
		\frac{-7 + 5 x_t + 8 {x_t}^2}
		{8 (1 - x_t)^3}
		- {3 x_t}^2 \frac{(2 - 3 x_t) \ln x_t}{4(1 - x_t)^4}
		\: , \\
	F_3^{CC}(x_t) & = & \frac{2}{3} \ln x_t + 
		{x_t}^2 \frac{25 - 19 x_t}{24 (1 - x_t)^3} -
		{x_t}^2 \frac{5 {x_t}^2 - 2 x_t - 6}{18 (1 - x_t)^4}
		\ln x_t \: , 
\end{eqnarray}
with $x_t \equiv \left( m_t/{M_W} \right)^2$ and $Q_u = 2/3$.
We note here that, the cancellation of divergences are independent on
the GIM mechanism. The GIM mechanism just works for making simplification
of the light internal quarks (up and charm) terms,
since $m_u$ and $m_c$ are lighter than $M_W$, then approximately
$x_u \cong x_c \cong 0$. On using orthogonal condition in
Eq. (\ref{eq:occol}),
\begin{equation}
	{V_{CKM}}^{uq} {{V_{CKM}}^{ub}}^{\ast}
		+ {V_{CKM}}^{cq} {{V_{CKM}}^{cb}}^{\ast}
		= -{V_{CKM}}^{tq} {{V_{CKM}}^{tb}}^{\ast} \: ,
	\label{eq:canckm}
\end{equation}
the up and charm quarks terms can be rewritten in term of
${V_{CKM}}^{tq} {{V_{CKM}}^{tb}}^{\ast}$.

In order to complete the calculation, let us consider the QCD correction
up to leading logarithm (refs. \cite{grin} and \cite{grig}).
The QCD corrected amplitude for on-shell $b \rightarrow q \, \gamma$ is,
\begin{equation}
	T^{SM} = \frac{G_F e}{8 \sqrt{2} \pi^2} Q_u \,
		{V_{CKM}}^{tq} {{V_{CKM}}^{tb}}^{\ast} \,
		{F_{bq\gamma}}^{SM}(m_b) \; \bar{q}(p^{\prime})
		\left[\gamma_{\mu},\not{q}\right]
		\left(m_q L + m_b R \right) \; b(p) \; \epsilon^{\mu} \, ,
\end{equation}
with,
\begin{equation}
	{F_{bq\gamma}}^{SM}(m_b) \cong 0.67
		\left[ F_1^{CC} - 0.42 F_2^{CC} - 0.88 \right] .
\end{equation}
Here we suppose $\Lambda_{\rm QCD} = 150$(MeV) for flavor
number $n_f = 5$, $\mu_0 = M_W$
and $\mu = m_b$. The functions $F_2^{CC}$ are induced by $b \rightarrow q \, g$
diagrams, that is
\begin{equation}
	F_2^{CC}(x_t) = -x_t \frac{2 + 5 x_t - {x_t}^2}{8(1 - x_t)^3} -
		\frac{3 {x_t}^2 \ln x_t}{4(1 - x_t)^4}\: .
\end{equation}
Note here that QCD 'correction' in the $B \rightarrow X_q \,
\gamma$ decay is large. There is an enhancement for about
one order of the branching ratio.
But the enhancement will be smaller as top-quark mass is larger. The detailed
behaviour of the QCD correction in these processes can be seen in
refs. \cite{grin} and \cite{grig}.

\subsection{In the ODVQM}
\label{app:bqgsqm}

\begin{figure}[t]
	\begin{center}
		\input{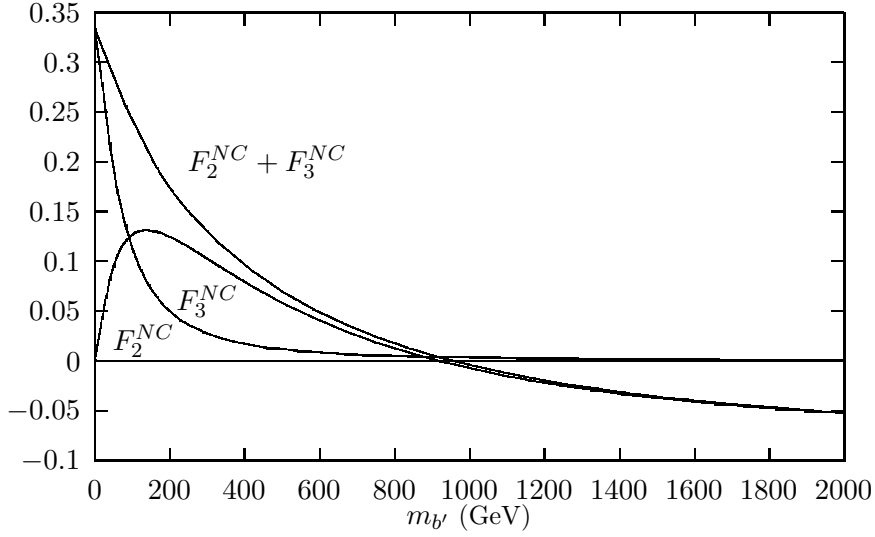}
	\end{center}
	\caption{The dependence of NC contribution functions
		on $m_{b^\prime}$.}
	\label{fig:f23}
\end{figure}
The calculation of the decay in the ODVQM is similar with in the SM
as done in refs. \cite{handoko2} and \cite{gautam}.
However, the problem is coming from the unitarity violation of CKM matrix
as written in Eq. (\ref{eq:zdbv}). Then in place of Eq. (\ref{eq:canckm}),
we have
\begin{equation}
	{V_{CKM}}^{uq} {{V_{CKM}}^{ub}}^{\ast}
		+ {V_{CKM}}^{cq} {{V_{CKM}}^{cb}}^{\ast}
		= -{V_{CKM}}^{tq} {{V_{CKM}}^{tb}}^{\ast}
		+ {z_d}^{qb} \: .
	\label{eq:canckmsqm}
\end{equation}
The effective hamiltonian for $b \rightarrow q \, \gamma$ in
the model is given as the following,
\begin{eqnarray}
	H_{eff}^{ODVQM} & = & \frac{G_F e}{8 \sqrt{2} \pi^2} Q_u
		{V_{CKM}}^{tq} {{V_{CKM}}^{tb}}^{\ast} 
		\left\{ \left[\gamma_{\mu},\not{q}\right]
		\left( F_L \, m_q L + F_R \, m_b R \right) \right.
		\nonumber \\
	& & \left. + F \left( q_\mu \not{q}    
		- q^2 \gamma_\mu  \right) L \right\} \: .  
\end{eqnarray}
where we normalized it by the CC contribution.
The contained functions are, 
\begin{eqnarray}
	F_L & \equiv & \left( F_1^{CC}(x_t) \right)^{ODVQM}
		+ \frac{Q_d}{Q_u} \frac{{z_d}^{qb}}{
		{V_{CKM}}^{tq} {{V_{CKM}}^{tb}}^{\ast}}
		\left[ \frac{2}{3} \sin^2\theta_W F_1^{NC}(r_q)
		\right. \nonumber \\
	& & \left. + {z_d}^{qq} \left( F_2^{NC}(r_q,w_q)
		+ F_3^{NC}(r_q) \right) + {z_d}^{bb} \left(
		F_2^{NC}(r_b,w_b)
		+ F_3^{NC}(r_b) \right) \right. \nonumber \\
	& & \left. - \left| V^{b^\prime b^\prime} \right|^2 
		\left( F_2^{NC}(r_{b^\prime},w_{b^\prime})
		+ F_3^{NC}(r_4) \right) \right] \: , \\
	F_R & \equiv & \left( F_1^{CC}(x_t) \right)^{ODVQM}
		+ \frac{Q_d}{Q_u} \frac{{z_d}^{qb}}{
		{V_{CKM}}^{tq} {{V_{CKM}}^{tb}}^{\ast}}
		\left[ \frac{2}{3} \sin^2\theta_W F_1^{NC}(r_b)
		\right. \nonumber \\
	& & \left. + {z_d}^{qq} \left( F_2^{NC}(r_q,w_q)
		+ F_3^{NC}(r_q) \right) + {z_d}^{bb} \left(
		F_2^{NC}(r_b,w_b)
		+ F_3^{NC}(r_b) \right) \right. \nonumber \\
	& & \left. - \left| V^{b^\prime b^\prime} 
		\right|^2 \left( F_2^{NC}(r_{b^\prime},w_{b^\prime})
		+ F_3^{NC}(r_4) \right) \right] \: , \\
	F & = & \left( F_3^{CC}(x_t) \right)^{ODVQM} +
		\frac{Q_d}{Q_u} \frac{{z_d}^{qb}}{
		{V_{CKM}}^{tq} {{V_{CKM}}^{tb}}^{\ast}}
		\left[ \frac{2}{3} \sin^2\theta_W 
		\left( F_4^{NC}(r_q) + F_4^{NC}(r_b) \right)
		\right. \nonumber \\
	& &	\left. + \left| V^{b^\prime b^\prime} \right|^2
		\left( F_4^{NC}(r_\alpha) -  
		r_\alpha \, F_5^{NC}(r_\alpha) - 
		w_\alpha \, F_5^{NC}(w_\alpha) \right) \right]
		\: .
\end{eqnarray}
These equations are obtained by using the upper-bounds in Sec.
\ref{subsubsec:tlfcnc}. The contributions of $\alpha = d$ for $q = s$
and $\alpha = s$ for $q = d$ can be neglected, since
$\left| {z_d}^{ds} {z_d}^{db} \right|$ and
$\left| {z_d}^{ds} {z_d}^{sb} \right|$ are tiny. Hence, the remaining
mixings are ${z_d}^{qq} {z_d}^{qb}$, ${z_d}^{qb} {z_d}^{bb}$ and
${z_d}^{qb^\prime} {z_d}^{b^\prime b}$.
Furthermore, rewriting the mixing between vector-like quark and 
ordinary quarks,
\begin{eqnarray}
	{z_d}^{qb^\prime} {z_d}^{b^\prime b} & = & V^{qb^\prime}
		{V^{b^\prime b^\prime}}^{\ast} \;
		V^{b^\prime b^\prime} {V^{bb^\prime}}^{\ast} \nonumber \\
	& = & -{z_d}^{qb} \; \left| V^{b^\prime b^\prime} \right|^2 \: ,
\end{eqnarray}
gives simplification of functions as in the above equations.

On using Eq. (\ref{eq:canckmsqm}), the modified functions
$F_i^{CC}$ from CC contributions are,
\begin{eqnarray}
	\left( F_1^{CC}(x_t) \right)^{ODVQM} & \equiv & x_t
		\frac{-7 + 5 x_t + 8 {x_t}^2}{8 (1 - x_t)^3}
		- {3 x_t}^2 \frac{(2 - 3 x_t) \ln x_t}{4(1 - x_t)^4}
		\nonumber \\
	& & 	- \frac{1}{2} \frac{{z_d}^{qb}}{
		{V_{CKM}}^{tq} {{V_{CKM}}^{tb}}^{\ast}} \: , \\
	\left( F_2^{CC}(x_t) \right)^{ODVQM} &  = &
		-x_t \frac{2 + 5 x_t - {x_t}^2}{8(1 - x_t)^3} -
		\frac{3 {x_t}^2 \ln x_t}{4(1 - x_t)^4}
		- \frac{5}{6} \frac{{z_d}^{qb}}{
		{V_{CKM}}^{tq} {{V_{CKM}}^{tb}}^{\ast}} \: ,\\
	\left( F_3^{CC}(x_t) \right)^{ODVQM} & =   & \frac{2}{3} \ln x_t + 
		{x_t}^2 \frac{25 - 19 x_t}{24 (1 - x_t)^3} -
		{x_t}^2 \frac{5 {x_t}^2 - 2 x_t - 6}{18 (1 - x_t)^4}
		\ln x_t  \nonumber \\
	& &	- \frac{33}{18} \frac{{z_d}^{qb}}{
		{V_{CKM}}^{tq} {{V_{CKM}}^{tb}}^{\ast}} \: . 
\end{eqnarray}
On the other hand, the NC contribution functions are,
\begin{eqnarray}
	F_1^{NC}(r_{\alpha}) & = & \frac{-10 + 15 r_{\alpha}
		+ 6 {r_{\alpha}}^2 - 11 {r_{\alpha}}^3 - 6 r_{\alpha}
		\left( 3 - 4 r_{\alpha} \right) \ln r_{\alpha}}
		{6 (1 - r_{\alpha})^4} \: , \\
	F_2^{NC}(r_{\alpha},w_{\alpha}) & = & r_{\alpha}
		\frac{-20 + 39 r_{\alpha} - {24 r_{\alpha}}^2
		+ 5 {r_{\alpha}}^3 - 6(2 - r_{\alpha}) \ln r_{\alpha}}
		{24 (1 - r_{\alpha})^4} \nonumber \\
	& &	- w_{\alpha} \frac{-16 + 45 w_{\alpha} - 36 {w_{\alpha}}^2
		+ 7 {w_{\alpha}}^3 - 6\left(2
		- 3 w_{\alpha}\right) \ln w_{\alpha}}
		{24 (1 - w_{\alpha})^4} \: , \\
	F_3^{NC}(r_{\alpha}) & = & -\frac{
		-4 + 9 r_\alpha - 5 {r_\alpha}^3 
		- 6 \left( 1 - 2 r_\alpha \right) \ln r_\alpha }{
		12 (1 - r_{\alpha})^4} \: , \\
	F_4^{NC}(r_{\alpha}) & = & \frac{
		2 + 27 r_\alpha - 54 {r_\alpha}^2 + 25 {r_\alpha}^3 
	 	- \left( 12 - 54 r_\alpha + 36 {r_\alpha }^2
		\right) \ln r_\alpha}{36 (1 - r_{\alpha})^4} \: , \\
	F_5^{NC}(r_{\alpha}) & = & \frac{
		-28 + 27 r_\alpha + {r_\alpha}^3 - 
		6 \left( 2 - 3 r_\alpha \right) \ln r_\alpha}{
		36 (1 - r_{\alpha})^4} \: .
\end{eqnarray}
We denote $Q_u = 2/3$, $Q_d = -1/3$,
$x_{\alpha} \equiv {m_{\alpha}}^2/{{M_W}^2}$, $r_{\alpha} \equiv
{m_{\alpha}}^2/{{M_Z}^2}$ and $w_{\alpha} \equiv {m_{\alpha}}^2/{{M_H}^2}$.
Note that, $F_1^{CC}$ and $F_3^{CC}$ are $W^\pm$ and $\chi^\pm$ exchange;
$F_1^{NC}$, $F_3^{NC}$ and $F_4^{NC}$ are $Z$ exchange; 
$F_2^{NC}$ and $F_5^{NC}$ are $\chi^0$ and $H$ exchange diagrams
contribution. In $F_2^{NC}$, the first term comes from
$\chi^0$ exchange, while the second term comes from $H$ exchange diagram.
The dependence on the vector-like quark mass in the
inclusive $B \rightarrow X_q \, \gamma$ decay occur in the
functions $F_2^{NC}$ and $F_3^{NC}$. The dependence is
depicted in Fig. (\ref{fig:f23}) with $m_H = 750$(GeV).

Finally, the QCD corrected amplitude for 
on-shell $b \rightarrow q \, \gamma$ in the ODVQM becomes,
\begin{equation}
	T^{ODVQM} = \frac{G_F e}{8 \sqrt{2} \pi^2} Q_u
		{V_{CKM}}^{tq} {{V_{CKM}}^{tb}}^{\ast}
		\bar{q}(p^{\prime}) \left[ \gamma_{\mu},\not{q}\right]
		\left(F_{bq\gamma}^L(m_b) m_q L + F_{bq\gamma}^R(m_b) m_b R \right)
		\; b(p) \; \epsilon^{\mu},
	\label{eqn:dwqcd}
\end{equation}
where $F_L^T$ and $F_R^T$ are defined as,
\begin{eqnarray}
	F_{bq\gamma}^L(m_b) & \cong & 0.67
		\left[ F_L - 0.42 F_L^g - 0.88 \right] , \\
	F_{bq\gamma}^R(m_b) & \cong & 0.67
		\left[ F_R - 0.42 F_R^g - 0.88 \right] .
\end{eqnarray}
The functions $F_{L(R)}^g $ are given as the following,
\begin{eqnarray}
	F_L^g & \equiv & \left( F_2^{CC}(x_t) \right)^{ODVQM}
		+ \frac{Q_d}{Q_u} \frac{{z_d}^{qb}}{
		{V_{CKM}}^{tq} {{V_{CKM}}^{tb}}^{\ast}}
		\left[ \frac{2}{3} \sin^2\theta_W F_1^{NC}(r_q)
		\right. \nonumber \\
	& & \left. + {z_d}^{qq} \left( F_2^{NC}(r_q,w_q)
		+ F_3^{NC}(r_q) \right) + {z_d}^{bb} \left(
		F_2^{NC}(r_b,w_b)
		+ F_3^{NC}(r_b) \right) \right. \nonumber \\
	& & \left. - \left| V^{b^\prime b^\prime} \right|^2 \left( 
		F_2^{NC}(r_{b^\prime},w_{b^\prime})
		+ F_3^{NC}(r_4) \right) \right] \: , \\
	F_R^g & \equiv & \left( F_2^{CC}(x_t) \right)^{ODVQM}
		+ \frac{Q_d}{Q_u} \frac{{z_d}^{qb}}{
		{V_{CKM}}^{tq} {{V_{CKM}}^{tb}}^{\ast}}
		\left[ \frac{2}{3} \sin^2\theta_W F_1^{NC}(r_b)
		\right. \nonumber \\
	& & \left. + {z_d}^{qq} \left( F_2^{NC}(r_q,w_q)
		+ F_3^{NC}(r_q) \right) + {z_d}^{bb} \left(
		F_2^{NC}(r_b,w_b)
		+ F_3^{NC}(m_b) \right) \right.  \nonumber \\
	& & \left. - \left| V^{b^\prime b^\prime} \right|^2
		\left( F_2^{NC}(r_{b^\prime},w_{b^\prime})
		+ F_3^{NC}(r_4) \right) \right] \: .
\end{eqnarray}

\section{NEUTRAL MESONS MIXING AND {\it CP} VIOLATIONS}

In the first part of this appendix, we shall give a common
explanation that hold for either mixings in the neutral $K$ and $B$
mesons system. The discussion is based on the refs. \cite{sanda},
\cite{nir}, \cite{bigi} and \cite{kurimoto}.

Consider an arbitrary neutral meson $P^0$ and its antiparticle $\bar{P}^0$.
The neutral $P$ meson state could be expressed as a linear combination
of them as,
\begin{equation}
	\left| P \rangle \right. \equiv a \left| P^0 \rangle \right. +
		b \left| \bar{P}^0 \rangle \right. \: ,
	\label{eq:p}
\end{equation}
and is governed by the time-dependent Schrodinger equation
\begin{eqnarray}
	H \left| P \rangle \right. & = & E \left| P \rangle \right.
		\nonumber \\
	& \equiv & \left( M - \frac{i}{2} \Gamma \right)
		\left| P \rangle \right. \: .
	\label{eq:se}
\end{eqnarray}
Here, $M$ and $\Gamma$ are $2 \times 2$ matrices. $M$ is the Hermitian
part of $H$ and would-be a mass matrix. $\Gamma$ is anti-Hermitian part
of $H$, and describes the exponential decay of the $P$ meson.
On using Eqs. (\ref{eq:p}) and (\ref{eq:se}), and multiplying the
l.h.s. by $\left. \langle P \right|$,
\begin{equation}
	\left. \langle \bar{P} \right| H \left| P \rangle \right. =
		E \left. \langle \bar{P} \right| P \rangle \: .
\end{equation}
Generally, the states $\left| P^0 \rangle \right. $ and
$\left| \bar{P}^0 \rangle \right. $ are orthogonal. This orthogonality
causes,
\begin{eqnarray}
	E \left(
	\begin{array}{c}
		a \\
		b
	\end{array}
	\right) & = &
	\left(
	\begin{array}{cc}
		\langle P^0 \left| H \right| P^0 \rangle &
		\langle P^0 \left| H \right| \bar{P}^0 \rangle \\
		\langle \bar{P}^0 \left| H \right| P^0 \rangle &
		\langle \bar{P}^0 \left| H \right| \bar{P}^0 \rangle
	\end{array}
	\right) \left(
	\begin{array}{c}
		a \\
		b
	\end{array}
	\right) \nonumber \\
	& = & \left(
	\begin{array}{cc}
		M_0 - \frac{i}{2} \Gamma_0 &
		M_{12} - \frac{i}{2} \Gamma_{12} \\
		{M_{12}}^\ast - \frac{i}{2} {\Gamma_{12}}^\ast &
		M_0 - \frac{i}{2} \Gamma_0
	\end{array}
	\right) \left(
	\begin{array}{c}
		a \\
		b
	\end{array}
	\right) \: ,
	\label{eq:mpp}
\end{eqnarray}
since $CPT$ invariance, $\langle P^0 \left| H \right| P^0 \rangle
= \langle \bar{P}^0 \left| H \right| \bar{P}^0 \rangle $, guarantees
$M_{11} = M_{22} \equiv M_0$ and
$\Gamma_{11} = \Gamma_{22} \equiv \Gamma_0$.
From the above equation, the eigenvalues and eigenvectors
are found as,
\begin{eqnarray}
	E_\pm & = & M_0 - \frac{i}{2} \Gamma_0 \pm
		\sqrt{\left( M_{12} - \frac{i}{2} \Gamma_{12} \right)
			\left( {M_{12}}^\ast - \frac{i}{2} {\Gamma_{12}}^\ast
			\right)} \: , \nonumber \\
	\left(
	\begin{array}{c}
		a \\
		b
	\end{array}
	\right) & = &
	\frac{1}{\sqrt{{M_{12}}^\ast - \frac{i}{2} {\Gamma_{12}}^\ast}}
	\left(
	\begin{array}{c}
		\sqrt{M_{12} - \frac{i}{2} \Gamma_{12}} \\
		\pm \sqrt{{M_{12}}^\ast - \frac{i}{2} {\Gamma_{12}}^\ast}
	\end{array}
	\right) \: .
	\label{eq:ev}
\end{eqnarray}

On using Eq. (\ref{eq:ev}), we can define the eigenstates for
the neutral $P$ meson mass matrix as,
\begin{equation}
	\left| P_\pm \rangle \right. \equiv p \left| P^0 \rangle \right.
		\pm q \left| \bar{P}^0 \rangle \right. \: ,
	\label{eq:ppm}
\end{equation}
with eigenvalues $E_\pm \equiv M_\pm - \frac{i}{2} \Gamma_\pm $.
Further, the following definition is usually used in place of $E_\pm$,
\begin{eqnarray}
	\Delta E & \equiv & E_- - E_+ \nonumber \\
	& \equiv & \Delta M - \frac{i}{2} \Delta \Gamma
		\nonumber \\
	& = & 2 \sqrt{\left( M_{12} - \frac{i}{2} \Gamma_{12}
		\right) \left( {M_{12}}^\ast - \frac{i}{2}
		{\Gamma_{12}}^\ast \right)} \: .
		\label{eq:e}
\end{eqnarray}
It causes,
\begin{eqnarray}
	\Delta M & = & 2 \: {\rm Re} \sqrt{\left( M_{12} - \frac{i}{2}
		\Gamma_{12} \right) \left( {M_{12}}^\ast - \frac{i}{2}
		{\Gamma_{12}}^\ast \right)} \: ,
		\label{eq:dm} \\
	\Delta \Gamma & = & -4 \: {\rm Im} \sqrt{\left( M_{12} - \frac{i}{2}
		\Gamma_{12} \right) \left( {M_{12}}^\ast - \frac{i}{2}
		{\Gamma_{12}}^\ast \right)}
		\nonumber \\
	& = &	\frac{4}{\Delta M} {\rm Re} \left( M_{12} {\Gamma_{12}}^\ast
		\right)	\: , \label{eq:dg} \\
	\frac{q}{p} & = & \sqrt{\frac{{M_{12}}^\ast - \frac{i}{2}
		{\Gamma_{12}}^\ast}{M_{12} - \frac{i}{2} \Gamma_{12}}}
		\equiv \frac{1 - \epsilon_t}{1 + \epsilon_t}
		\: . \label{eq:qp}
\end{eqnarray}
The parameter $\epsilon_t$ in Eq. (\ref{eq:qp}) indicates the mixing
that occurs due to the off-diagonal elements in the mass matrix
(Eq. (\ref{eq:mpp})), even for all $\Gamma$'s were zero.
By definition in Eq. (\ref{eq:qp}), Eq. (\ref{eq:ppm}) can be
rewritten again,
\begin{equation}
	\left| P_\pm \rangle \right. = \frac{1}{
		\sqrt{2 \left( 1 + \left| \epsilon_t \right|^2 \right)}}
		\left[ \left( 1 + \epsilon_t \right) \left| P^0 \rangle \right.
		\pm \left( 1 - \epsilon_t \right) \left| \bar{P}^0 \rangle
		\right.	\right] \: .
	\label{eq:cppm}
\end{equation}

Let us make a convention for charge conjugation as below,
\begin{equation}
	C \left| P^0 \rangle \right. = -\left| \bar{P}^0 \rangle \right.
	\: \: \: , \: \: \:
	C \left| \bar{P}^0 \rangle \right. = -\left| P^0 \rangle \right. \: .
\end{equation}
It yields under $CP$ transformation,
\begin{equation}
	CP \left| P^0 \rangle \right. = \left| \bar{P}^0 \rangle \right.
	\: \: \: , \: \: \:
	CP \left| \bar{P}^0 \rangle \right. = \left| P^0 \rangle \right. \: .
\end{equation}
because the $P$'s are pseudoscalars. Then, the parameter
$\epsilon_t$ in Eq. (\ref{eq:qp}) indicates the $CP$ violation in
the neutral $P$ meson system. In particular, the case when
$\epsilon_t \cong 0$, or no relative phase between $M_{12}$ and
$\Gamma_{12}$, means the absence of the $CP$ violation, and
Eq. (\ref{eq:cppm}) would be $CP$ odd and even
states respectively,
\begin{eqnarray}
	\left| P_+ \rangle \right. & = & \frac{1}{\sqrt{2}}
		\left( \left| P^0 \rangle \right. +
		\left| \bar{P}^0 \rangle \right. \right)
	\longrightarrow CP : + \: ,  \\
	\left| P_- \rangle \right. & = & \frac{1}{\sqrt{2}}
		\left( \left| P^0 \rangle \right. -
		\left| \bar{P}^0 \rangle \right. \right)
	\longrightarrow CP : - \: .
\end{eqnarray}

\subsection{Neutral $K$ meson}
\label{app:k0k0sm}

We make further discussion in the case of the neutral $K$ meson
by replacing the above $P$ with $K$.
The description of mixing in the $K^0 - \bar{K}^0$ mixing is simplified by
the fact that it is accounted for, to a good approximation, by physics
of two generations of quark. So the direct $CP$ violation can be
ignored, and the mass eigenstates coincide with the $CP$ eigenstates.
Because of $2 \pi$ is $CP$ even while $3 \pi$ is $CP$ odd, only
\begin{eqnarray}
	K_+ & \rightarrow & \pi^0 \, \pi^0 \: \: , \: \:
		\pi^+ \, \pi^- \: , \\
	K_- & \rightarrow & \pi^0 \, \pi^0 \, \pi^0
		\: \: , \: \: \pi^+ \, \pi^- \, \pi^0
\end{eqnarray}
decays can be realized when the $CP$ is conserved.
Conversely, a small $K_- \rightarrow 2 \pi$ decay will
provide $CP$ violation in the $K$ meson system.
It is customary to call $K_+$ as $K_S$ and $K_-$ as $K_L$,
since two-pion final state has much larger phase than the
three-pion final state. Then, $K_+$ that decays into $2 \pi$,
decays much faster than $K_-$. Experimentally,
\begin{eqnarray}
	\tau_S & = & 0.9 \times 10^{-10} {\rm (s)} \: , \\
	\tau_L & = & 5.2 \times 10^{-8} {\rm (s)} \: , \\
	{\Delta M_K}^{exp} & \equiv &
		M_{K_L} - M_{K_S} \cong 3.5 \times 10^{-12} {\rm (MeV)}
		\: .
\end{eqnarray}
Thus, the relation
\begin{equation}
	\Delta \Gamma \cong -2 \Delta M
\end{equation}
is found, since
\begin{equation}
	{\Gamma_S}^{exp} = 7.4 \times 10^{-12} ({\rm MeV}) \: \: , \: \:
	{\Gamma_L}^{exp} = 1.3 \times 10^{-14} ({\rm MeV}) \: ,
\end{equation}
then, $\Delta \Gamma \equiv \Gamma_L - \Gamma_S \cong -\Gamma_S$.
\begin{figure}[t]
	\epsfxsize=15cm
	\epsfysize=7cm
	\epsffile{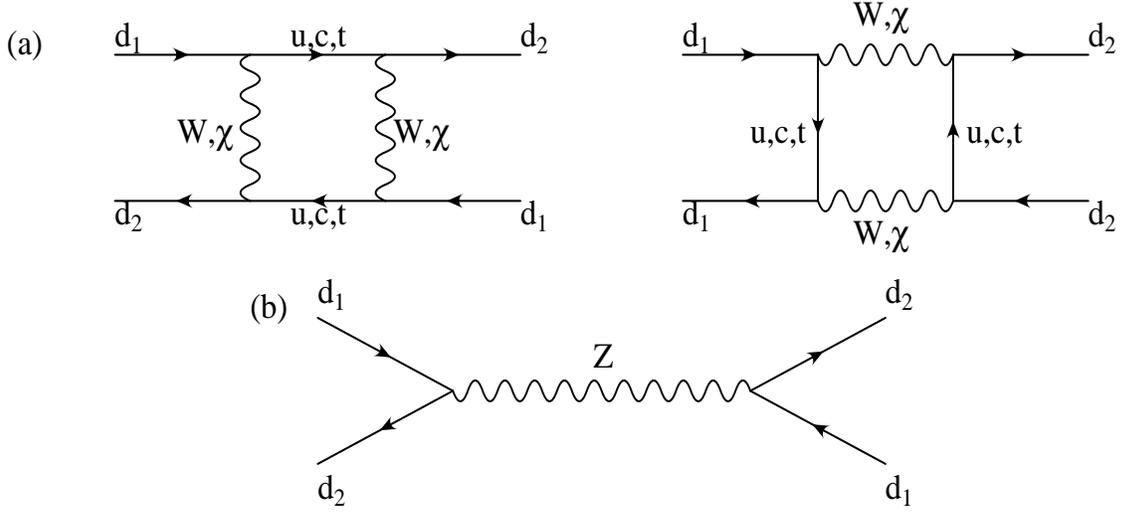}
	\caption{$\Delta P = 2$ diagrams for $P^0 - \bar{P}^0$ mixing
		in the (a) SM and (b) ODVQM, with $P = B,K$.}
	\label{fig:b0b0}
\end{figure}

Now, we are ready to evaluate the indirect $CP$ violation in the neutral
$K$ meson system. This can be described by rewriting Eq. (\ref{eq:cppm}) as,
\begin{eqnarray}
	\left| K_S \rangle \right. & = & \frac{1}{
		\sqrt{\left( 1 + \left| \epsilon_t \right|^2 \right)}}
		\left( \left| K_1 \rangle \right. +
		\epsilon_t \left| K_2 \rangle \right. \right) \: , \\
	\left| K_L \rangle \right. & = & \frac{1}{
		\sqrt{\left( 1 + \left| \epsilon_t \right|^2 \right)}}
		\left( \left| K_2 \rangle \right. +
		\epsilon_t \left| K_1 \rangle \right. \right) \: ,
\end{eqnarray}
where
\begin{equation}
	\left| K_{1(2)} \rangle \right. \equiv \frac{1}{\sqrt{2}} \left(
		\left| K^0 \rangle \right. \pm
		\left| \bar{K}^0 \rangle \right. \right)
	\label{eq:k12}
\end{equation}
are $CP$ eigenstates. Thus, since $K_S$ and $K_L$ are not $CP$
eigenstates, they may decay into $3 \pi$ and $2 \pi$.

Experimentally measured parameter of $CP$ violation in the neutral
$K$ meson system is defined as,
\begin{equation}
	\epsilon \equiv \frac{\langle (2 \pi)_{I=0} \left| H_{\Delta S=2}
		\right| K_L \rangle}{\langle (2 \pi)_{I=0} \left|
		H_{\Delta S=2} \right| K_S \rangle} \: .
	\label{eq:epsilon}
\end{equation}
This quantity is different with $\epsilon_t$ in Eq. (\ref{eq:qp}), that
is independent of the phase convention. In the SM, $M_{12}$ and
$\Gamma_{12}$ are nearly real, then
\begin{eqnarray}
	{\Delta M}^{SM} & \cong & 2 {\rm Re} \, {M_{12}}^{SM} \: , \\
	{\Delta \Gamma}^{SM} & \cong & 2 {\rm Re} \, {\Gamma_{12}}^{SM}
		\: , \\
	\epsilon_t & \cong & \frac{i {\rm Im} \, {M_{12}}^{SM} +
		\frac{1}{2} {\rm Im} \, {\Gamma_{12}}^{SM}}{
		{\Delta M}^{SM} - \frac{i}{2} {\Delta \Gamma}^{SM}}
		\cong e^{i \pi/4} \frac{{\rm Im} \, {M_{12}}^{SM}}{
		\sqrt{2} {\Delta M}^{SM}} \: .
\end{eqnarray}
Therefore, using Eqs. (\ref{eq:k12}) and (\ref{eq:epsilon}),
\begin{eqnarray}
	\epsilon & = & \frac{i {\rm Im} \, A_0 + \epsilon_t {\rm Re} \, A_0}{
		{\rm Re} \, A_0 + \epsilon_t i {\rm Im} \, A_0} \nonumber \\
	& \cong & \epsilon_t + i \frac{{\rm Im} \, A_0}{{\rm
		Re} \, A_0} \nonumber \\
	& \cong & \frac{e^{i \pi/4}}{\sqrt{2} {\Delta M}^{SM}} \left(
		{\rm Im} \, {M_{12}}^{SM} +
		2 \frac{{\rm Im} \, A_0}{{\rm Re} \, A_0}
		{\rm Re} \, {M_{12}}^{SM} \right) \: ,
	\label{eq:epsilonsm}
\end{eqnarray}
where $A_i \equiv \langle (2 \pi)_{I=i} \left| H_{\Delta S=2} \right| K^0
\rangle$, and use $\left| \epsilon_t {\rm Im} \, A_0 \right| \ll \left|
{\rm Re} \, A_0 \right| $, ${\rm Im} \, {\Gamma_{12}}^{SM} \propto
-2 ({\rm Re} \, A_0) \, ({\rm Im} \, A_0)$, and $(1/2) {\Delta \Gamma}^{SM}
\cong {\rm Re} \, {\Gamma_{12}}^{SM} \propto \left( {\rm Re} \, A_0 \right)^2$.

We shall give $M_{12}$ in the SM which can be found by calculating the matrix
elements of the $\Delta S = 2$ effective Hamiltonian as,
\begin{equation}
	{M_{12}}^{SM} = \langle K^0 \left| {H_{\Delta S=2}}^{SM} \right|
		\bar{K}^0 \rangle + {\rm (long \: distance
		\: effects)} \: ,
\end{equation}
where, ${H_{\Delta S=2}}^{SM}$ is found from the calculation of the diagrams
in Fig. (\ref{fig:b0b0} a) with $d_1 = d$ and $d_2 = s$,
\begin{equation}
	{H_{\Delta S=2}}^{SM} = \frac{{G_F}^2 {M_W}^2}{4 \pi^2} \sum_{i,j=c,t}
		\lambda_i \lambda_j \, \eta_{ij} \,
		{F_{\Delta S=2}}^{SM}(x_i,x_j)
		\left( \bar{s} \, \gamma_\mu \, L \, d \right)^2 + h.c. \: ,
\end{equation}
where the $u$ quark and the external momenta have been ignored, since
$m_u \ll M_W$.
Here, $\eta_{ij}$ is the QCD correction in the diagrams,
$\lambda_i \equiv {V_{CKM}}^{id} {{V_{CKM}}^{is}}^\ast$, the function
$ {F_{\Delta S=2}}^{SM}$'s are (ref. \cite{inami}),
\begin{eqnarray}
	{F_{\Delta S=2}}^{SM}(x,y) & = & \left[ \frac{1}{4} +
		\frac{3}{2(1 - x)} - \frac{3}{4(1 - x)^2} \right]
		\frac{x y \ln x}{x - y} + (x \leftrightarrow y)
		- \frac{3 x y}{4(1 - x)(1 - y)} \: ,
		\label{eq:fs21sm} \\
	{F_{\Delta S=2}}^{SM}(x_i) & = & x_i \left[
		\frac{1}{4} + \frac{9}{4 (1 - x_t)} -
		\frac{3}{2 (1 - x_t)^2} \right] +
		\frac{3}{2} \left[ \frac{x_t}{x_t - 1} \right]^3
		\ln x_t \: ,
		\label{eq:fs22sm}
\end{eqnarray}
The puzzle of hadron matrix element here is evaluated by PCAC, and
for simplicity ignore the long distance effects,
\begin{equation}
	\langle K^0 \left| \left( \bar{s} \, \gamma_\mu \, L \, d \right)^2
		\right| \bar{K}^0 \rangle \cong
		\frac{1}{3} m_K \, {f_K}^2 B_K \: ,
	\label{eq:pcac}
\end{equation}
where $f_K$ and $B_K$ are the decay constant of $K$ meson and
the bag parameter. Lastly, combining all of the above results,
\begin{eqnarray}
	{M_{12}}^{SM} & = & \frac{{G_F}^2}{12 \pi^2} m_K \,
		{f_K}^2 B_K \, {M_W}^2
		\nonumber \\
	& &	\left[ {\lambda_c}^2 \eta_c {F_{\Delta S=2}}^{SM}(x_c)
		+ {\lambda_t}^2 \eta_t {F_{\Delta S=2}}^{SM}(x_t) +
		2 \lambda_c \lambda_t \eta_{ct} {F_{\Delta S=2}}^{SM}(x_c,x_t)
		\right] \: .
		\label{eq:m12sm}
\end{eqnarray}
Therefore, substituting this equation into Eq. (\ref{eq:epsilonsm})
will give a constraint for the contained CKM matrix elements.

On the other hand, we also need the quantity in Eq. (\ref{eq:qp}) which is
frequently used in the calculation of $CP$ violations in the neutral $B$ meson
system. Moreover, $t$ quark and mixing $c-t$ quarks term in
Eq. (\ref{eq:m12sm}) can be ignored too, because of suppression due to
$\left| {V_{CKM}}^{td} {{V_{CKM}}^{ts}}^\ast \right| \cong O(\lambda^5)$.
Therefore, from Eq. (\ref{eq:qp}) approximately,
\begin{equation}
	\left( \frac{q}{p} \right)_{K^0}^{SM} =
	\frac{\left( {V_{CKM}}^{cd} {{V_{CKM}}^{cs}}^\ast \right)^\ast}{
		{V_{CKM}}^{cd} {{V_{CKM}}^{cs}}^\ast} \: .
	\label{eq:qpk0}
\end{equation}

\subsection{Neutral $B$ meson}
\label{app:b0b0}

In principle, the description of mixing in the neutral $B_q$ meson system is
similar with the $K$'s one. We just change $s$ by $b$ and $d$ by $q$, or
$d_1 = q$ and $d_2 = b$ in Fig. (\ref{fig:b0b0}). The main purpose here is
to derive the $CP$ asymmetries when the neutral $B_q$'s decay into
arbitrary final state $f$.

In the SM framework, the important differences with the $K$ meson are that
the contribution in the internal line is just coming from top quark only,
because of the small masses of up and charm quarks, and
the CKM matrix elements of up and charm quarks
have same order with the top one. The second one is,
\begin{equation}
	\left| {\Gamma_{12}}^{SM} \right| \ll \left|
		{M_{12}}^{SM} \right| \: ,
	\label{eq:gm12}
\end{equation}
which can be strictly showed by calculating the real and imaginary parts
of the diagrams in Fig. (\ref{fig:b0b0} a), i.e.
\begin{equation}
	\left| \frac{\Gamma_{12}}{M_{12}} \right|^{SM} \cong
	\left| \frac{\pi}{3 {F_{\Delta B=2}}^{SM}(x_t)} \right|
	\frac{{m_b}^2}{{m_t}^2} \ll 1 \: ,
\end{equation}
where ${F_{\Delta B=2}}^{SM}(x_t) \equiv {{F_{\Delta S=2}}^{SM}(x_t)}/{x_t}$,
and ${F_{\Delta S=2}}^{SM}(x_t)$ has been given in Eq. (\ref{eq:fs22sm}).
However, when $m_u \neq m_c$, a small $CP$ violation in the
${B_q}^0 - {\bar{B}_q}^0$ will occur. It can be seen in the
calculation of imaginary part without ignoring the light
quarks, that is
\begin{equation}
	{\Gamma_{12}}^{SM} \cong \frac{{G_F}^2 {m_b}^3
		{f_{B_q}}^2 B_{B_q}}{36 \pi}
		\left| {V_{CKM}}^{tq} {{V_{CKM}}^{tb}}^\ast \right|^2
		\left[ 1 + \frac{8}{3} \frac{{m_c}^2}{{m_b}^2}
		\frac{{V_{CKM}}^{cq} {{V_{CKM}}^{cb}}^\ast}{
		{V_{CKM}}^{tq} {{V_{CKM}}^{tb}}^\ast}
		\right] \: ,
\end{equation}
by using GIM mechanism and the fact that $m_u \ll m_c \ll m_t$.
Since the real part is
\begin{equation}
	{M_{12}}^{SM} \cong \frac{{G_F}^2 {m_t}^2 m_b
		{f_{B_q}}^2 B_{B_q}}{12 \pi^2}
		\left| {V_{CKM}}^{tq} {{V_{CKM}}^{tb}}^\ast \right|^2
		\left| {F_{\Delta B=2}}^{SM}(x_t) \right| \ ,
\end{equation}
one finds their ratio as,
\begin{equation}
	\left( \frac{\Gamma_{12}}{M_{12}} \right)^{SM} \cong
		\frac{\pi}{3} \frac{1}{
		\left| {F_{\Delta B=2}}^{SM}(x_t) \right|}
		\frac{{m_b}^2}{{m_t}^2}
		\left[ 1 + \frac{8}{3} \frac{{m_c}^2}{{m_b}^2}
		\frac{{V_{CKM}}^{cq} {{V_{CKM}}^{cb}}^\ast}{
		{V_{CKM}}^{tq} {{V_{CKM}}^{tb}}^\ast}
		\right] \: .
\end{equation}
The second term indicates the small $CP$ violation if
$m_c \neq m_b$. However, we are not interested in this small
$CP$ violation, and the detailed discussion can be seen in ref.
\cite{bigi}. From Eqs. (\ref{eq:dm}) and (\ref{eq:gm12}), the
mass splitting approximately becomes
$\Delta {M_q}^{SM} \cong 2 \left| {M_{12}}^{SM} \right|$.
This has consequences for the size of mixing
in the neutral $B_q$ meson to be written as,
\begin{eqnarray}
	{x_q}^{SM} & \equiv & \frac{\Delta {M_q}^{SM}}{\Gamma_B}
		\nonumber \\
	& = & \frac{{G_F}^2}{6 \pi^2} m_{B_q} {M_W}^2 \tau_{B_q}
		\eta_{B_q} {f_{B_q}}^2 B_{B_q}
		\left| {V_{CKM}}^{tq} {{V_{CKM}}^{tb}}^\ast \right|^2
		\left| {F_{\Delta B=2}}^{SM}(x_t) \right| \: .
	\label{eq:xq}
\end{eqnarray}

Now, we are going on evaluating the $CP$ violations in the neutral $B_q$
meson system. First, the time-evolution neutral $B$ states can be written
by using Eqs. (\ref{eq:ppm}), (\ref{eq:dg}) and (\ref{eq:dm}) as,
\begin{eqnarray}
	\left| {B_q}^0(t) \rangle \right. & = & g_+(t)
		\left| {B_q}^0 \rangle \right. + \frac{q}{p} g_-(t)
		\left| \bar{B_q}^0 \rangle \right. \: ,
		\label{eq:bq0} \\
	\left| \bar{B_q}^0(t) \rangle \right. & = & \frac{p}{q} g_-(t)
		\left| {Bq}^0 \rangle \right. + g_+(t)
		\left| \bar{B_q}^0 \rangle \right. \: ,
		\label{eq:bq0b}
\end{eqnarray}
where, the initial condition is, at $t = 0$, ${B_q}^0(t)$
($\bar{B_q}^0(t)$) would be pure ${B_q}^0$ ($\bar{B_q}^0$),
and
\begin{equation}
	g_\pm(t) \equiv \frac{1}{2} e^{-i (M_L - i/2 \Gamma_L)t}
		\left( 1 \pm e^{-i (\Delta M - i/2 \Delta \Gamma)t} \right)
		\: .
		\label{eq:gpm}
\end{equation}

The next step is defining the quantity that measures the $CP$ violations.
This can be done by concerning the asymmetries when the ${B_q}^0$'s decay
into any final state $f$. It is defined as,
\begin{equation}
	a_f(t) \equiv \frac{
		\Gamma\left( {B_q}^0(t) \rightarrow f \right) -
		\Gamma\left( \bar{B_q}^0(t) \rightarrow f \right) }{
		\Gamma\left( {B_q}^0(t) \rightarrow f \right) +
		\Gamma\left( \bar{B_q}^0(t) \rightarrow f \right)} \: .
		\label{eq:asb}
\end{equation}
The next task is then calculating the amplitude of the decays. On using
Eqs. (\ref{eq:gm12}), (\ref{eq:bq0}), (\ref{eq:bq0b}) and
(\ref{eq:gpm}), one finds
\begin{eqnarray}
	\left| A_f(t) \right|^2
	& \pm &
	\left| \bar{A}_f(t) \right|^2
	\nonumber \\
	& \propto &
	\left( 1 \pm \left| \frac{p}{q} \right|^2 \right)
	\left[ \frac{1}{2} \left( 1 + \left| \frac{q}{p}
	\frac{\bar{A_f}}{A_f}\right|^2 \right) + 2 {\rm Im}
	\left( \frac{q}{p} \frac{\bar{A_f}}{A_f} \right)
	\sin (\Delta Mt) e^{-1/2 \Delta \Gamma t} \right]
	\nonumber \\
	& & + \left( 1 \mp \left| \frac{p}{q} \right|^2 \right)
	\cos (\Delta Mt) \left( 1 - \left| \frac{q}{p}
	\frac{\bar{A_f}}{A_f}\right|^2 \right) e^{-1/2 \Delta \Gamma t}
	\: ,
\end{eqnarray}
with $A_f \equiv \langle f \left| H \right| {B_q}^0 \rangle$ and
$\bar{A}_f \equiv \langle f \left| H \right| \bar{B_q}^0 \rangle$.
Substituting them into Eq. (\ref{eq:asb}) yields
\begin{equation}
	a_f(t) \cong - \sin (\Delta M t) \, \sin \phi_f \: .
\end{equation}
Here,
\begin{eqnarray}
	\sin \phi_f & \equiv & {\rm Im} \left[
		\frac{q}{p} \left( \frac{\bar{A}}{A} \right)_f \right] \: ,
	\label{eq:sinp} \\
	\left( \frac{\bar{A}}{A} \right)_f & = &
	\frac{\langle f \left| H \right| \bar{B_q}^0 \rangle }{
		\langle f \left| H \right| {B_q}^0 \rangle } \: .
	\label{eq:aa}
\end{eqnarray}

Nevertheless, in the neutral $B$ meson system, the quantity in Eq.
(\ref{eq:qp}) can be found by calculating the real part of Fig.
(\ref{fig:b0b0}) same with the neutral $K$ meson. But, here the top
quark dominance approximation can be done by the GIM without any
ambiguities. Thus,
\begin{equation}
	\left( \frac{q}{p} \right)_{{B_q}^0}^{SM} =
	\frac{\left( {V_{CKM}}^{tq} {{V_{CKM}}^{tb}}^\ast \right)^\ast}{
		{V_{CKM}}^{tq} {{V_{CKM}}^{tb}}^\ast} \: .
	\label{eq:qpbq}
\end{equation}

\clearpage
\addcontentsline{toc}{section}{\protect\numberline{}{\bf References}}


\begin{thebibliography}{1}
	\bibitem{nelson}
		A. Nelson,
		{\it Phys. Lett.} {\bf 136B} (1984) 387.
	\bibitem{barr}
		S.M. Barr,
		{\it Phys. Rev.} {\bf D30} (1984) 1805.
	\bibitem{branco}
		G.C. Branco and L. Lavoura,
		{\it Nucl. Phys.} {\bf B278} (1986) 738.
	\bibitem{handoko2}
		L.T. Handoko and T. Morozumi,
		to be published in {\it Mod. Phys. Lett.} {\bf A} (1995)
	\bibitem{handoko3}
		L.T. Handoko and T. Morozumi,
		talk given at 
		{\it International Workshop on $B$ Physics '94},
		to appear in the proceeding (World Scientific, 1995).
	\bibitem{lavoura}
		L. Lavoura and J.P. Silva,
		{\it Phys. Rev.} {\bf D47} (1993) 2046.
	\bibitem{peskin}
		M.E. Peskin and T. Takeuchi,
		{\it Phys. Rev.} {\bf D46} (1992) 381.
	\bibitem{cheng}
		T.P. Cheng and L.F. Li,
		{\it Gauge Theory of Elementary Physics}
		(Oxford Science Publications, 1984).
	\bibitem{muta}
		T. Muta,
		{\it Foundations of Quantum Chromodynamics}
		(World Scientific, Singapore, 1987).
	\bibitem{data}
		Particle Data Group,
		{\it Phys. Rev.} {\bf D50} (1994) 1.
	\bibitem{sanda}
		A.I. Sanda, Short Lecturer given at Hiroshima University,
		December 1994.
	\bibitem{cdf}
		F. Abe, et. al.,
		{\it Phys. Rev.} {\bf D50} (1994) 2966.
	\bibitem{ali}
		A. Ali,
		to be published in {\it B Decays} (2nd edition),
		edited by S. Stone
		(World Scientific, 1994).
	\bibitem{nir2}
		Y. Nir,
		{\it Boulder TASI} (1991) 339.
	\bibitem{buras}
		A.J. Buras and M.K. Harlander,
		in {\it Heavy Flavours}, p. 58, edited by A.J. Buras
			and M. Lindner
		(World Scientific, 1992).
	\bibitem{grin}
		B. Grinstein, R. Springer and M.B. Wise
		{\it Nuc. Phys.} {\bf B339}, 628 (1990).
	\bibitem{payne}
		R. Poling,
		talk given at 
		{\it International Workshop on $B$ Physics '94},
		to appear in the proceeding (World Scientific, 1995).
	\bibitem{mannel}
		A. Ali, T. Mannel and T. Morozumi,
		{\it Phys. Lett.} {\bf B273} (1991) 505.
	\bibitem{morozumi}
		G.C. Branco, T. Morozumi, P.A. Parada and M.N. Rebelo,
		{\it Phys. Rev.} {\bf D48} (1993) 1167.
	\bibitem{nir}
		Y. Nir and D. Silverman,
		{\it Phys. Rev.} {\bf D42} (1990) 1477.
	\bibitem{handoko1}
		L.T. Handoko,
		{\it Soryushikenkyuu} {\bf 2} (1995).
	\bibitem{grig}
		R. Grigjanis, P.J. O'Donnell, Mark Sutherland and
		Henri Navalet,
		{\it Phys. Rep.} {\bf 228} (1993) 93.
	\bibitem{kobayashi}
		M. Kobayashi and K. Maskawa,
		{\it Prog. Theor. Phys.} {\bf 49} (1973) 652.
	\bibitem{wolfenstein}
		L. Wolfenstein,
		{\it Phys. Rev. Lett.} {\bf 51} (1983) 1945.
	\bibitem{inami}
		T. Inami and C.S. Lim,
		{\it Prog. Theor. Phys.} {\bf 65} (1981) 297.
	\bibitem{georgi}
		H. Georgi,
		{\it Weak Interaction and Modern Particle Theory}
		(Benjamin Cummings Publishing, 1984)
	\bibitem{gautam}
		G. Bhattacharyya, G.C. Branco and D. Choudhury,
		{\it Phys. Lett.} {\bf B336} (1994) 487.
	\bibitem{bigi}
		I.I. Bigi, V.A. Khoze, N.G. Uraltsev and A.I. Sanda,
		in {\it CP Violation}, p. 175, edited by C. Jarlskog
		(World Scientific, 1989).
	\bibitem{kurimoto}
		T. Kurimoto,
		in {\it Proceeding of the KEK Summer Institute on High
		Energy Phenomenology}, p. 5, edited by T. Hikasa (1991).
	\bibitem{jarlskog}
		C. Jarlskog,
		{\it Phys. Rev. Lett.} {\bf 55} (1985) 1039.
\end{thebibliography}
\end{document}